\documentclass[useAMS,usenatbib,onecolumn]{mn2e}
\usepackage{aas_macros}
\usepackage{graphicx}

\title[Relativistic Self-similar Expansion]{Relativistic Expansion of
Magnetic Loops at the Self-similar Stage}
\author[Hiroyuki R. Takahashi, Eiji Asano and Ryoji Matsumoto]
{Hiroyuki R. Takahashi$^{1}$\thanks{E-mail:takahasi@astro.s.chiba-u.ac.jp}, Eiji Asano$^{2}$ and Ryoji Matsumoto$^{3}$
\\
$^{1}$Graduate School of Science and Technology, Chiba University,
1-33 Yayoi-cho, Inage-ku, Chiba 263-8522, Japan \\
$^{2}$Kwasan and Hida Observatories, Kyoto University,
17 Ohmine-cho, Kita Kazan, Yamashina-ku, Kyoto 607-8471, Japan\\
$^{3}$Department of Physics, Graduate School of Science, Chiba University,
1-33 Yayoi-cho, Inage-ku, Chiba 263-8522, Japan}

\begin{document}

 \date{Accepted 2007 December 15. Received 2007 December 14; in
 original form 2007 October 11}

\pagerange{\pageref{firstpage}--\pageref{lastpage}} \pubyear{2008}

\maketitle
\label{firstpage}

\begin{abstract}
We obtained self-similar solutions of relativistically expanding
magnetic loops taking into account the azimuthal magnetic fields. We
neglect stellar rotation and assume axisymmetry and a purely radial
flow. 
As the magnetic loops expand, the initial dipole magnetic field is
stretched into the radial direction. 
When the expansion speed approaches the light speed, the displacement
current reduces the toroidal current and modifies the distribution of
the plasma lifted up from the central star.
Since these self-similar solutions describe the free expansion of the
magnetic loops, i.e., $Dv/Dt=0$, the equations of motion are similar to
those of the static relativistic magnetohydrodynamics. 
This allows us to estimate the total energy stored in the magnetic loops
by applying the virial theorem. This energy is comparable to that of the
giant flares observed in magnetars.
\end{abstract}


\section{Introduction}\label{intro}
Soft gamma-ray repeaters (SGRs) are believed to be a
young neutron star with strong magnetic fields ($\sim 10^{15}~\mathrm{G}$),
namely the magnetar (see, e.g., \citealt{2006csxs.book..547W};
\citealt{2008A&ARv..15..225M}, for review). The magnetic fields inside
the magnetar 
are amplified by the dynamo mechanism at the birth of
the neutron star. The Lorentz force stressing the crust of the magnetar
balances with the rigidity of the crust.
When the critical twist is accumulated, the magnetic twist injected
into the magnetar magnetosphere will trigger the
expansion of magnetic loops \citep{2001ApJ...561..980T}. The magnetic
reconnection taking place
inside the expanding magnetic loops can be responsible for SGR
flares \citep{2001ApJ...552..748W, 2006MNRAS.367.1594L}.

Recently, relativistic simulations have been performed to study
the dynamics of the magnetospheres of neutron stars
(\citealt{2002MNRAS.336..759K};
\citealt{2005PASJ...57..409A}; \citealt{2006ApJ...648L..51S};
\citealt{2006MNRAS.367...19K}). \citet{2005KITP...CONF..HP} reported the
results of 2-dimensional relativistic force-free
simulations of the magnetar flares triggered by the injection of the
magnetic twists at the footpoints of the loops. 
When the critical twist is accumulated, the magnetic loops expand
relativistically. \citet{2007Chiba...1..1} carried out 2-dimensional
relativistic force-free simulations of expanding magnetic loops and
showed that the Lorentz factor defined by the
drift velocity $\bmath v_d = c(\bmath E \times \bmath B)/B^2$ exceeds 10
\citep[see,][for the definition of the drift
velocity]{1997PhRvE..56.2181U}. These simulations indicate that the
magnetic loops expand self-similarly.

Assuming relativistic force-free dynamics, 
\citet{2003astro.ph.12347L} obtained self-similar solutions of the
spherically expanding magnetic shell. 
\citet{2005MNRAS.359..725P} found
self-similar solutions of the relativistic force-free field. In
these studies of force-free dynamics, gas pressure and inertial terms
are neglected.
In the framework of the relativistic magnetohydrodynamics (MHD),
\citet{2002PhFl...14..963L}
found self-similar solutions of the spherically expanding magnetic
shells. \cite{1982ApJ...261..351L} obtained non-relativistic
self-similar MHD solutions
of the expanding magnetic loops in solar flares or supernovae
explosion by assuming axisymmetry. 
Subsequently, \citet{1984ApJ...281..392L} extended his model
to the case including toroidal magnetic fields and applied it to solar
coronal mass ejections (CMEs). The latter model was employed by
\cite{1992ApJ...388..415S} as a test problem to check the validity
and accuracy of axisymmetric MHD codes.
In magnetar flares, the magnetic loops may be twisted by the
shear motion at the footpoints of the loops. The shear motion generates
Alfv\'en waves propagating along the field lines.
Such twisted magnetic loops expand by the enhanced magnetic
pressure by the toroidal magnetic fields.
Thus we should include the toroidal magnetic field to study the
evolution of magnetic loops during magnetar flares. Also the
relativistic effects should be included. The characteristic wave speed
in the magnetar magnetosphere approaches the light speed because of the
strong magnetic fields.
 Thus our aim is to obtain relativistic self-similar MHD
solutions of expanding magnetic loops taking into account the toroidal
magnetic fields by extending the non-relativistic solutions found by
\cite{1982ApJ...261..351L}. 

This paper is organized as follows; in $\S$ \ref{ssmhd}, we present the
relativistic ideal MHD equations and introduce a
self-similar parameter which depends on both radial distance from the
centre of the star and time. In $\S$ \ref{sss}, we obtain self-similar
solutions. The physical properties of these solutions are discussed in
$\S$ \ref{physprop}.  We summarize the results in $\S$ \ref{summary}.

\section[]{Self-similar MHD Equations}\label{ssmhd}
In the following, we take the light speed as unity. 
The complete set of relativistic ideal MHD equations is
\begin{equation}
 \frac{\partial }{\partial t}(\gamma \rho)+\nabla\cdot(\gamma\rho \bmath
  v)=0, \label{eq:MHDcont}
\end{equation}
\begin{equation}
 \rho\gamma\left[\frac{\partial}{\partial t}+(\bmath v \cdot \nabla)
						\right](\xi\gamma\bmath
 v)= -\nabla p
 +\rho_e \bmath E+\bmath j \times \bmath B-\frac{G M \rho \gamma }{r^2} \bmath e_r,\label{eq:MHDeom}
\end{equation}
\begin{equation}
 \left[\frac{\partial}{\partial t}+(\bmath v \cdot
 \nabla)\right]\left(\ln\frac{p}{\rho^\Gamma}\right)=0,
 \label{eq:MHDentropy}
\end{equation}
\begin{equation}
 \nabla \cdot \bmath E=4\pi \rho_e, \label{eq:Gauss}
\end{equation}
\begin{equation}
 \nabla \cdot \bmath B = 0, \label{eq:nomonopole}
\end{equation}
\begin{equation}
 \frac{\partial \bmath B}{\partial t}+\nabla \times \bmath E=0,\label{eq:Faraday}
\end{equation}
\begin{equation}
 \frac{\partial \bmath E}{\partial t}=\nabla \times \bmath B-4\pi\bmath j,\label{eq:Ampere}
\end{equation}
\begin{equation}
 \bmath E=-\bmath v\times \bmath B,\label{eq:mhdcond}
\end{equation}
where $\bmath E, \bmath B,\bmath j, \bmath v,\gamma,\rho_e, \rho,
p,\Gamma $ are the electric field, the magnetic field, the current density,
the velocity, the Lorentz factor, the charge density, the mass density,
the pressure and the specific heat ratio, respectively. The vector
$\bmath e_r$ is a unit vector in the radial direction. We include the
gravity by a point mass
$M$ as an external force. Here $G$ is the gravitational constant, and
$r$ is the distance from the centre of the star.
The relativistic specific enthalpy $\xi$ is defined as
\begin{equation}
 \xi=\frac{\epsilon+p}{\rho}=1+\frac{\Gamma}{\Gamma-1}\frac{p}{\rho},
\end{equation}
where $\epsilon$ is the energy density of matter including the
photon energy coupled with the plasma. In SGR outbursts, since the
luminosity much exceeds the Eddington luminosity, radiation energy
density can exceed the thermal energy of the plasma. In the following
pressure $p$ includes the contribution from the radiation pressure.

In this paper, we consider relativistic self-similar expansions of
magnetic loops which started expansion at $t=t_\mathrm{s}$ by loss of
dynamical equilibrium and entered into a self-similar stage at $t=t_0 >
t_\mathrm{s}$. We do not consider the evolution of the loops before $t=t_0$.

For simplicity, we ignore the stellar rotation and assume axisymmetry.
We can express the axisymmetric magnetic field in terms of two scalar
functions $\tilde{A}$ and $B$ as
\begin{equation}
 \bmath B=\frac{1}{r \sin\theta}\left(\frac{1}{r}\frac{\partial
				  \tilde{A}}{\partial \theta}, -\frac{\partial
				  \tilde{A}}{\partial r}, B\right),\label{def:A}
\end{equation}
in the polar coordinates $(r,\theta,\phi)$.
The scalar function $\tilde{A}(t,r,\theta)$ denotes the magnetic flux,
whose contours coincide with magnetic field lines projected on to the
$r-\theta$ plane.

We further assume that the fluid flow is purely radial;
\begin{equation}
 \bmath v=v(t,r,\theta)\,\bmath e_r. \label{def:v}
\end{equation}
Equations (\ref{eq:MHDcont}), (\ref{eq:MHDeom}),
(\ref{eq:MHDentropy}), and (\ref{eq:Faraday}) are then expressed as
\begin{equation}
 \frac{\partial (\rho\gamma)}{\partial t}+\frac{1}{r^2}\frac{\partial (r^2 \rho\gamma
  v)}{\partial r}=0,\label{eq:cont}
\end{equation}
\begin{equation}
 \rho\gamma\left[\frac{\partial}{\partial t}+v\frac{\partial }{\partial
   r}\right](\xi \gamma v)= -\frac{\partial p}{\partial
 r}-\frac{1}{4\pi r^2\sin^2\theta}\left\{ \frac{\partial \tilde{A}}{\partial
				  r}\left[\left(\hat{\mathcal{L}}_{(r,\theta)}\tilde{A}\right)
				     +\frac{\partial}{\partial
				  t}\left(v\frac{\partial \tilde{A}}{\partial
				     r}\right)\right]+B\left[\frac{\partial B}{\partial
				  r}+\frac{\partial (vB)}{\partial
				  t}\right]\right\}-\frac{G M
 \rho\gamma}{r^2},\label{eq:eomr}
\end{equation}
\begin{equation}
 4\pi r^2 \sin^2\theta\frac{\partial p }{\partial
 \theta}+(1-v^2)B\frac{\partial B}{\partial\theta}+\frac{\partial
 \tilde{A}}{\partial
 \theta}\left[\left(\hat{\mathcal{L}}_{(r,\theta)}\tilde{A}\right)+\frac{\partial}{\partial
	 t}\left(v\frac{\partial \tilde{A}}{\partial r}\right)\right]-vB^2\frac{\partial
 v}{\partial \theta}=0,\label{eq:eomtheta}
\end{equation}
\begin{equation}
 (1-v^2)\frac{\partial \tilde{A}}{\partial r}\frac{\partial
  B}{\partial \theta}-\frac{\partial \tilde{A}}{\partial \theta}\left[\frac{\partial
  B}{\partial r}+\frac{\partial (vB)}{\partial t}\right]-vB\frac{\partial
  \tilde{A}}{\partial r}\frac{\partial v}{\partial \theta}=0,\label{eq:eomphi}
\end{equation}
\begin{equation}
 \left[\frac{\partial}{\partial t}+v \frac{\partial}{\partial
  r}\right]\left(\ln\frac{p}{\rho^\Gamma}\right)=0,
 \label{eq:MHDentropy2}
\end{equation}
\begin{equation}
 \frac{\partial \tilde{A}}{\partial t}+v\frac{\partial \tilde{A}}{\partial
  r}=0,\label{eq:A}
\end{equation}
\begin{equation}
 \frac{\partial B}{\partial t}+\frac{\partial (vB)}{\partial
  r}=0,\label{eq:B}
\end{equation}
where we used the MHD condition given by (\ref{eq:mhdcond}) and
introduced the operator
\begin{equation}
 \hat{\mathcal{L}}_{(r,\theta)}\equiv \frac{\partial^2}{\partial
  r^2}+\frac{\sin\theta}{r^2}\frac{\partial}{\partial\theta}\left(\frac{1}{\sin\theta}\frac{\partial}{\partial\theta}\right).
\end{equation}
Since our aim is to obtain self-similar solutions of these relativistic MHD
equations, we assume that the time evolution is governed by the
self-similar variable:
\begin{equation}
 \eta=\frac{r}{Z(t)},\label{def:eta}
\end{equation}
where $Z(t)$ is an arbitrary function of time.
We further assume that the flux function $\tilde{A}$ depends on time $t$
and the radial distance $r$ through the self-similar variable $\eta$, as
\begin{equation}
 \tilde{A}(t,r,\theta)=\tilde{A}(\eta,\theta). \label{def:Aself}
\end{equation}
When equation (\ref{def:Aself}) is satisfied, the radial velocity $v$
has a form 
\begin{equation}
 v=\eta \dot Z, \label{def:sv}
\end{equation}
from equation (\ref{eq:A}). Here dot denotes the time derivative. Equation
(\ref{def:sv}) implies
that the radial velocity $v$ does not depend on the polar angle $\theta$.
It then follows from equations (\ref{eq:cont}) and (\ref{eq:B}) that
\begin{equation}
 \rho(t,r,\theta)\gamma(t,r)=Z^{-3}(t) D(\eta,\theta),\label{eq:DZ}
\end{equation}
\begin{equation}
 B(t,r,\theta)=Z^{-1}(t) Q(\eta,\theta),\label{eq:QZ}
\end{equation}
where $Q$ and $D$ are arbitrary functions of $\eta$ and $\theta$. These
relations indicate that the magnetic flux and the total mass are
conserved. Next we take the pressure $p$ as
$p(t,r,\theta)=Z^{l}P(\eta,\theta)$. Substituting this equation into
equations (\ref{eq:eomtheta}) and (\ref{eq:MHDentropy2}), we obtain
\begin{equation}
 4\pi\eta^2Z^{l+4}\sin^2\theta\frac{\partial P}{\partial
  \theta}+\frac{\partial \tilde{A}}{\partial
  \theta}\left[\hat{\mathcal{L}}_{(\eta,\theta)}\tilde{A}+\left(\eta Z \ddot
						   Z-2\eta\dot{Z}^2\right)
	  \frac{\partial \tilde{A}}{\partial
	  \eta}-\eta^2\dot{Z}^2\frac{\partial^2 \tilde{A}}{\partial
	  \eta^2}\right]+(1-\eta^2
  \dot{Z}^2)Q\frac{\partial Q}{\partial\theta}=0,\label{eq:wk1}\\
\end{equation}
\begin{equation}
 \frac{\Gamma \eta^2 Z \ddot Z }{1-\eta^2\dot
  Z^2}+(3\Gamma+l)=0,\label{eq:MHDentropy3}
\end{equation}
where we introduced an operator $\hat{\mathcal{L}}_{(\eta,\theta)}$:
\begin{equation}
 \hat{\mathcal{L}}_{(\eta,\theta)}\equiv \frac{\partial^2
  }{\partial\eta^2}+\frac{\sin\theta}{\eta^2}\frac{\partial}{\partial\theta}\left(\frac{1}{\sin\theta}\frac{\partial}{\partial\theta}\right)
  =\frac{1}{Z^2}\hat{\mathcal{L}}_{(r,\theta)}.
\end{equation}
To satisfy these equations, $p$, $Z$ and $\Gamma$ should have forms
\begin{equation}
 p(t,r,\theta)=Z^{-4}P(\eta,\theta),\label{eq:PZ}
\end{equation}
\begin{equation}
 Z(t)=t, \label{sol:Z}
\end{equation}
and 
\begin{equation}
 \Gamma = \frac{4}{3}.\label{sol:shr}\\
\end{equation}
This adiabatic index corresponds to the radiation pressure dominant
  plasma. Thus our model can describe the evolution of a fireball
  confined by magnetic fields.

  Equations (\ref{eq:DZ}) and (\ref{eq:PZ}) indicate that the magnetic loops
expand adiabatically. By using equations (\ref{def:sv}), (\ref{eq:DZ}),
(\ref{eq:QZ}), (\ref{eq:PZ}), (\ref{sol:Z}) and (\ref{sol:shr}),
equations (\ref{eq:eomr}), (\ref{eq:eomphi}) and (\ref{eq:wk1}) are 
expressed as 
\begin{equation}
 D(\eta,\theta)=\frac{\eta^2}{GM}\left\{\frac{4\eta
				  P}{1-\eta^2}-\frac{\partial
				  P}{\partial \eta}\right. 
 \left.-\frac{1}{4\pi \eta^2
				  \sin^2\theta}\left[\frac{\partial
						\tilde{A}}{\partial
 \eta}
 \left(\hat{\mathcal{L}}_{(\eta,\theta)}\tilde{A}
  -\frac{\partial}{\partial \eta} \left(\eta^2 \frac{\partial\tilde{A}}{\partial \eta}\right)\right)
 +Q \frac{\partial}{\partial \eta}\left(Q(1-\eta^2) \right)\right]\right\},\label{eq:eomr2}
\end{equation}
\begin{equation}
(1-\eta^2)\frac{\partial \tilde{A}}{\partial
  \eta}\frac{\partial Q}{\partial \theta}-\frac{\partial
  \tilde{A}}{\partial\theta}\frac{\partial}{\partial
  \eta}\left[(1-\eta^2)Q\right]=0,\label{eq:eomphi2}
\end{equation}
\begin{equation}
 4\pi\eta^2\sin^2\theta\frac{\partial P}{\partial\theta}+\frac{\partial
  \tilde{A}}{\partial
  \theta}\left[\hat{\mathcal{L}}_{(\eta,\theta)}\tilde{A}
	  -\frac{\partial}{\partial
	  \eta}\left(\eta^2\frac{\partial \tilde{A}}{\partial \eta}\right)
	 \right]+(1-\eta^2)Q\frac{\partial
  Q}{\partial\theta}=0.\label{eq:eomtheta2}
\end{equation}
From equation (\ref{eq:eomphi2}), a formal solution of  $Q$ is
obtained as
\begin{equation}
 Q(\eta,\theta)=\frac{\mathcal{G}(\tilde{A})}{1-\eta^2},\label{sol:Qformal}
\end{equation}
where $\mathcal{G}$ is an arbitrary function.

Self-similar solutions can be constructed as follows. First we prescribe
an arbitrary function $\tilde{A}(\eta,\theta)$ (or
$Q(\eta,\theta)$). Then, equation (\ref{eq:eomphi2}) determines the function
$Q(\eta,\theta)$ (or $\tilde{A}(\eta,\theta)$). Functions $\tilde{A}$ and $Q$ determine
the pressure $P(\eta,\theta)$ according to equation
(\ref{eq:eomtheta2}). Finally, the density function $D(\eta,\theta)$ is
obtained by equation (\ref{eq:eomr2}).

Note that from the equation (\ref{def:sv}) and (\ref{sol:Z}), the radial
velocity has a simple form as
\begin{equation}
 v=\frac{r}{t}. \label{sol:v}
\end{equation}
Since the time derivative of the velocity becomes zero, i.e.,
$D\bmath v/D t=0$, equations (\ref{eq:eomr2})-(\ref{eq:eomtheta2})
describe the freely expanding solution. This means that there is a
reference frame that all forces balance.
By substituting equations (\ref{eq:DZ}), (\ref{eq:QZ}), (\ref{eq:PZ}),
(\ref{sol:Z}), and (\ref{sol:v}) into the
equations of motion (\ref{eq:MHDeom}), we obtain 
\begin{equation}
 \frac{\Gamma}{\Gamma-1}\frac{\gamma^2 v^2 p}{r}\bmath e_r-\nabla
  p+\rho_e \bmath E+\bmath j \times \bmath
  B-\frac{GM\rho\gamma}{r^2}\bmath e_r=0.\label{eq:reducedeom}
\end{equation}
The first term on the left hand side comes from the inertia. For
convenience, we call this term as a thermal
inertial term throughout this paper. When we neglect the terms of order
$\left(v/c\right)^2$,
equation (\ref{eq:reducedeom}) reduces to the equations of the force
balance in non-relativistic MHD. 

\section{Self-similar solutions}\label{sss}
In the previous section, we derived relativistic self-similar MHD
equations, (\ref{eq:eomr2}), (\ref{eq:eomphi2}) and (\ref{eq:eomtheta2}).
In this section, we obtain solutions of these equations by imposing
appropriate boundary conditions. As mentioned in the previous section,
the toroidal magnetic field, the pressure, and the gas density are
calculated by assigning the flux function $\tilde{A}(\eta, \theta)$. In
the following, we introduce
three kinds of flux functions and obtain explicit forms of other
variables.

\subsection{Construction of Solutions}\label{sss0}
\begin{figure}
 \begin{center}
  \includegraphics[width=10cm]{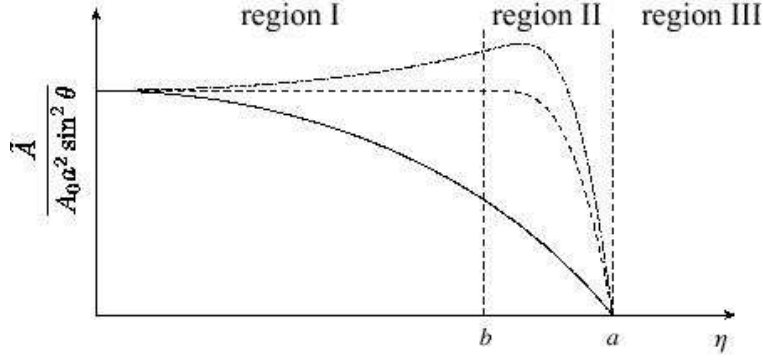}
  \caption{The flux function $\tilde{A}$ as a function
  of $\eta$ for the dipolar solution (solid curve), the shell solution
  (dashed curve), and the flux rope solution (dot-dashed curve).
  The parameters $a$ and $b$ denote the outer and inner boundaries of
  the shell, respectively.}
  \label{fig:shellflux}
 \end{center}
\end{figure}
We assume that the expanding magnetic loops have a spherical outer
boundary at $r=R(t)$.

A simple solution of the expanding magnetic loops is that the poloidal
magnetic field is dipolar near the surface of the star
\citep{1982ApJ...261..351L}.  The
magnetic field should be tangential to the
spherical surface $r=R(t)$ at all time. Such a solution can be
constructed by \begin{equation}
 \tilde{A}(\eta,\theta)=A_0
  \frac{a^2-\eta^2}{\sqrt{1-\eta^2}}\sin^2\theta,\label{sol1:A}
\end{equation}
where $A_0$ and $a$ are constants. The radius $R(t)$ where
$\tilde{A}=0$ is given by
\begin{equation}
 R(t)=at. \label{sol1:radius}
\end{equation}
We hereafter call the solution constructed from equation (\ref{sol1:A})
as {\it dipolar solution}.

\begin{figure}
 \begin{tabular}{ccc}
  \begin{minipage}{0.33\hsize}
   \includegraphics[width=4.8cm]{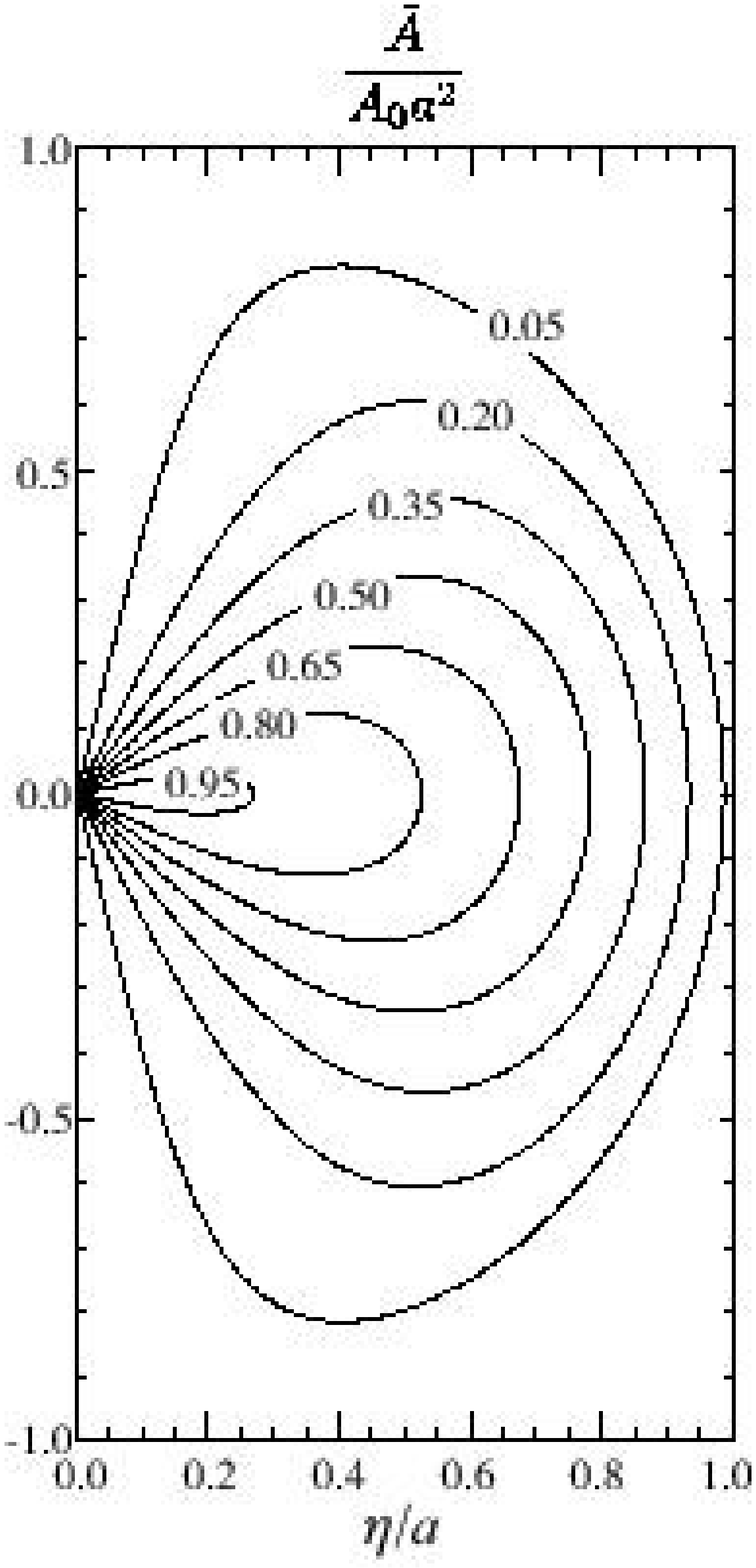}
  \end{minipage}
  \hspace*{1mm}
  \begin{minipage}{0.33\hsize}
   \includegraphics[width=4.8cm]{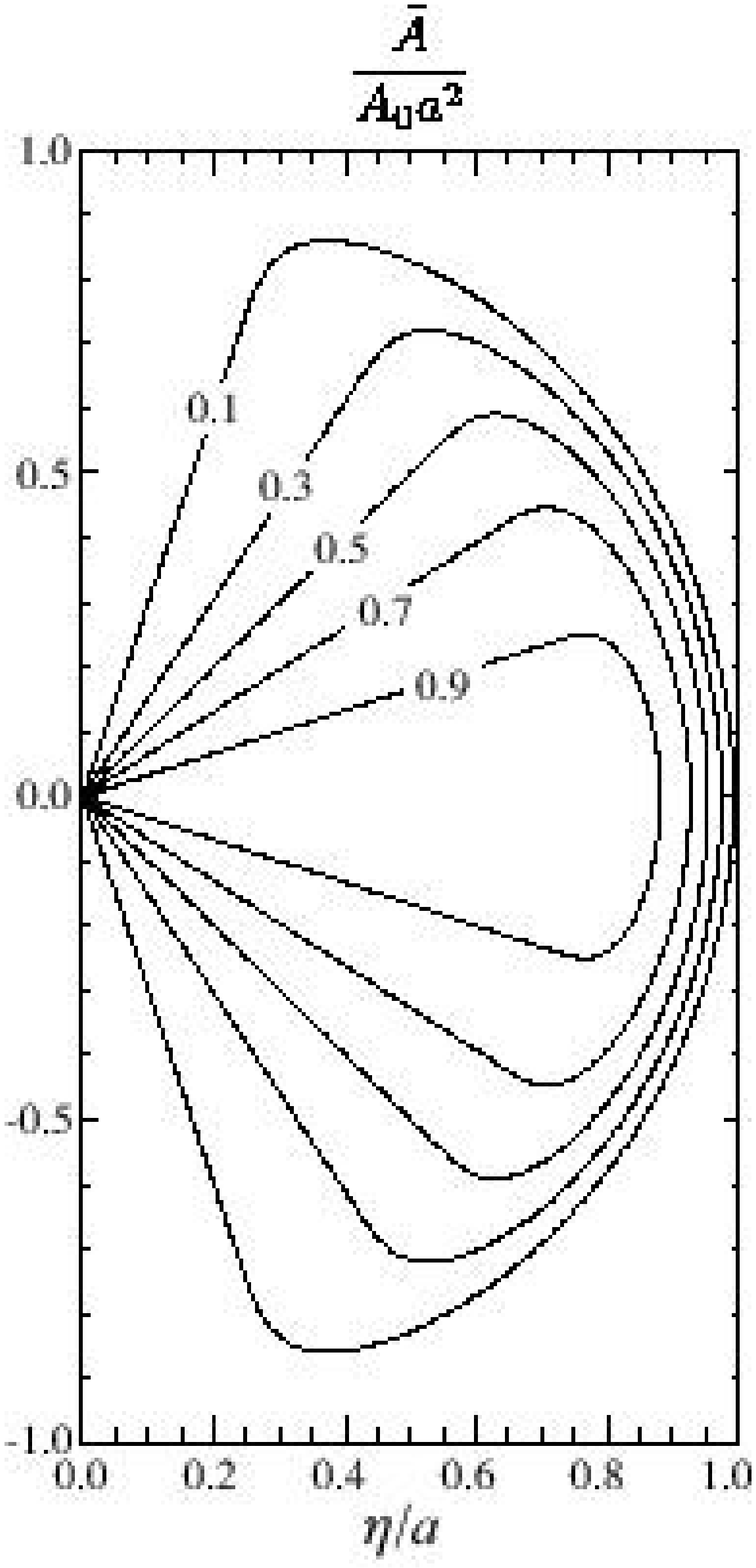}
  \end{minipage}
  \hspace*{1mm}
  \begin{minipage}{0.33\hsize}
  \includegraphics[width=4.8cm]{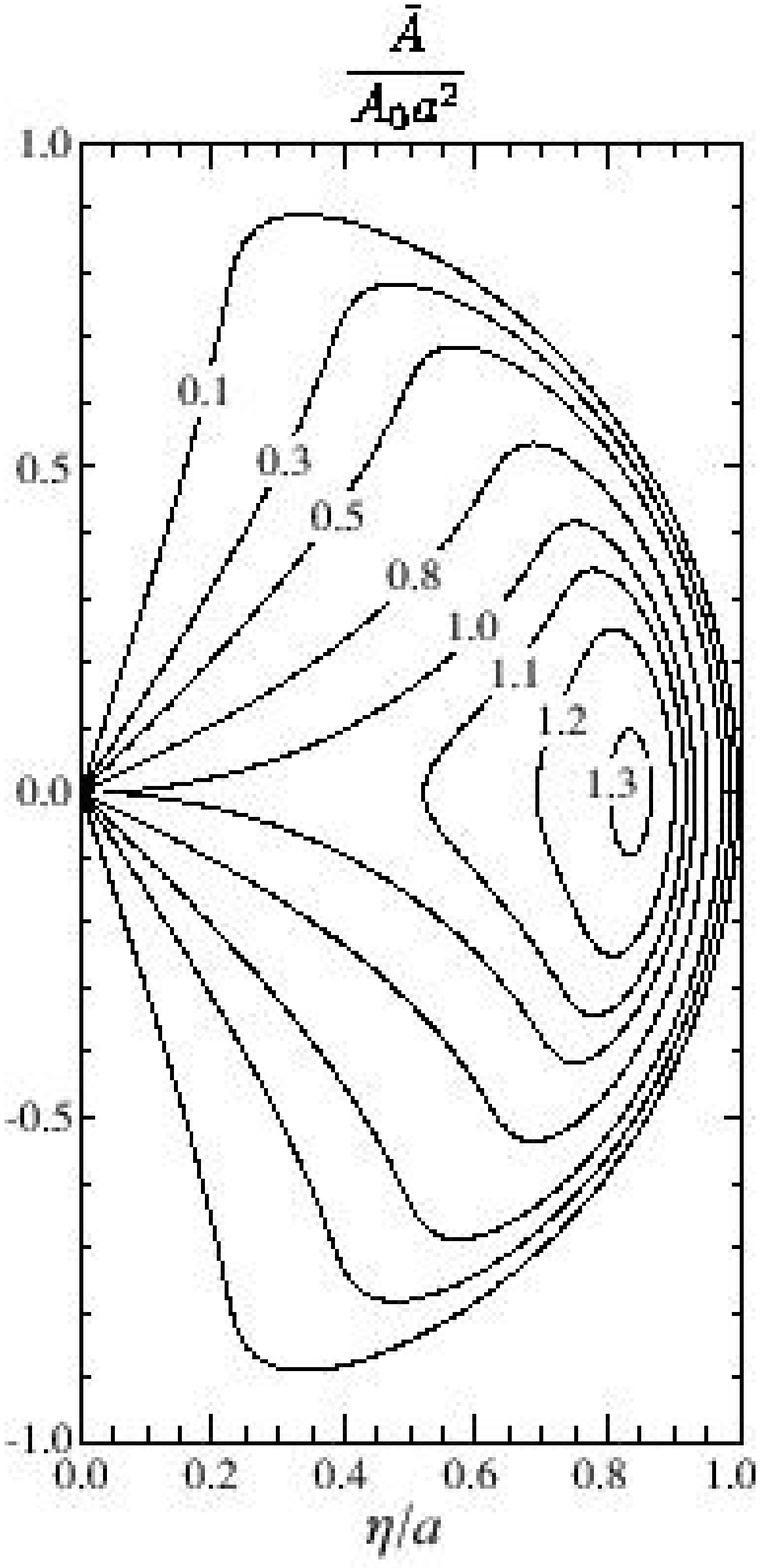}
  \end{minipage}
 \end{tabular}
 \caption{Contour plots of the magnetic flux $\tilde{A}$ which
 constructs the dipolar, shell, and flux rope solutions from left to
 right, respectively. The parameters are taken as $a=0.8$, $b/a=0.7$,
 and $k=\pi/[4(a-b)]$.}
 \label{fig:flux}
\end{figure}
Solid curve in Fig.~\ref{fig:shellflux} shows the flux function
$\tilde{A}$ as a function of $\eta$ for the dipolar solution. Contour
plots of $\tilde{A}$ for dipolar solution is shown in the left panel of
Fig.~\ref{fig:flux}.

When the flux function is given by equation (\ref{sol1:A}),
the magnetic flux crossing the annulus at the equatorial plane
$\theta=\pi/2$ decreases with radius (see Fig. \ref{fig:shellflux}).
In actual MHD explosion, the magnetic flux can be swept up into a thin
shell just behind the loop top. The shell boundaries are
assumed to be at $r=bt$ and $r=at$ (region II, see
Fig. \ref{fig:shellflux}).
Such a self-similar field can be constructed by
\begin{eqnarray}
 \tilde{A}(\eta,\theta)=\left\{\begin{array}{lll}
		 {\displaystyle A_0a^2\sin^2\theta}, &(\rm{region\ I:\
		  }\eta\lid \textit{b}), \\
			{\displaystyle
			 A_0a^2\Lambda(\eta)\sin^2\theta},  
			 &(\rm{region\ II:\ }\textit{b}< \eta \lid
			 \textit{a}),
		\end{array}\right. \label{sol2:A}
\end{eqnarray}
where
\begin{equation}
 \Lambda(\eta)=1-\frac{\sin^4T(\eta)}{\sin^4T(a)}, \label{sol2:Lambda}
\end{equation}
\begin{equation}
 T(\eta)=k(\eta-b),\label{func:T} 
\end{equation}
and $a$, $b$ and $k$ are constants \citep{1982ApJ...261..351L}.  The
flux functions in region I ($\eta \lid b$) and region II ($b < \eta
\lid a$) are connected smoothly at
$\eta=b$. The loop boundary locates at $r=at$, where $\tilde{A}=0$. 

The flux function for this solution is shown by a dashed curve in
Fig.~\ref{fig:shellflux}. 
It can be easily shown that the magnetic field lines projected on to the
$r-\theta$ plane are all radial in region I.
We call the solution constructed from equation (\ref{sol2:A}) as {\it shell
solution}. 
The middle panel of Fig.~\ref{fig:flux} shows the contours of
$\tilde{A}$ for the shell solution.

Another solution is that we call {\it flux rope solution}. 
As the magnetic loops expand, a current sheet is
formed inside the magnetic loops. It is suggested that the magnetic
reconnection taking place in the current sheet is responsible for the
SGR flares \citep{2001ApJ...552..748W, 2006MNRAS.367.1594L}.
When the magnetic reconnection takes place, flux ropes (namely plasmoids)
are formed inside magnetic loops. The flux function
should then have a local maximum inside the flux rope. Such a
solution can be constructed by
\begin{eqnarray}
 \tilde{A}(\eta,\theta)=\left\{\begin{array}{lll}
{\displaystyle \frac{A_0a^2}{\sqrt{1-\eta^2}}\sin^2\theta}, 
 &(\rm{region\ I:\ }\eta\lid \textit{b}), \\
{\displaystyle  \frac{A_0a^2}{\sqrt{1-\eta^2}}\Lambda(\eta)\sin^2\theta},  
			 &(\rm{region\ II:\ }\textit{b}< \eta \lid
			 \textit{a}),
		\end{array}\right. \label{sol3:A}
\end{eqnarray}
where $A_0$, $a$ are constants and $\Lambda(\eta)$ is given by equation
(\ref{sol2:Lambda}). This function is shown by a dot-dashed curve in
Fig.~\ref{fig:shellflux}.
It has a local maximum in the domain $b<\eta<a$
(see Fig.~\ref{fig:shellflux}). 
The contours of $\tilde{A}$ for the
flux rope solution is shown in the right panel of Fig.~\ref{fig:flux}.
Flux ropes appear behind the shell.

\subsection{Dipolar Solutions}\label{sss1}
\begin{figure}
 \begin{tabular}{ccc}
  \begin{minipage}{0.33\hsize}
   \includegraphics[width=4.8cm]{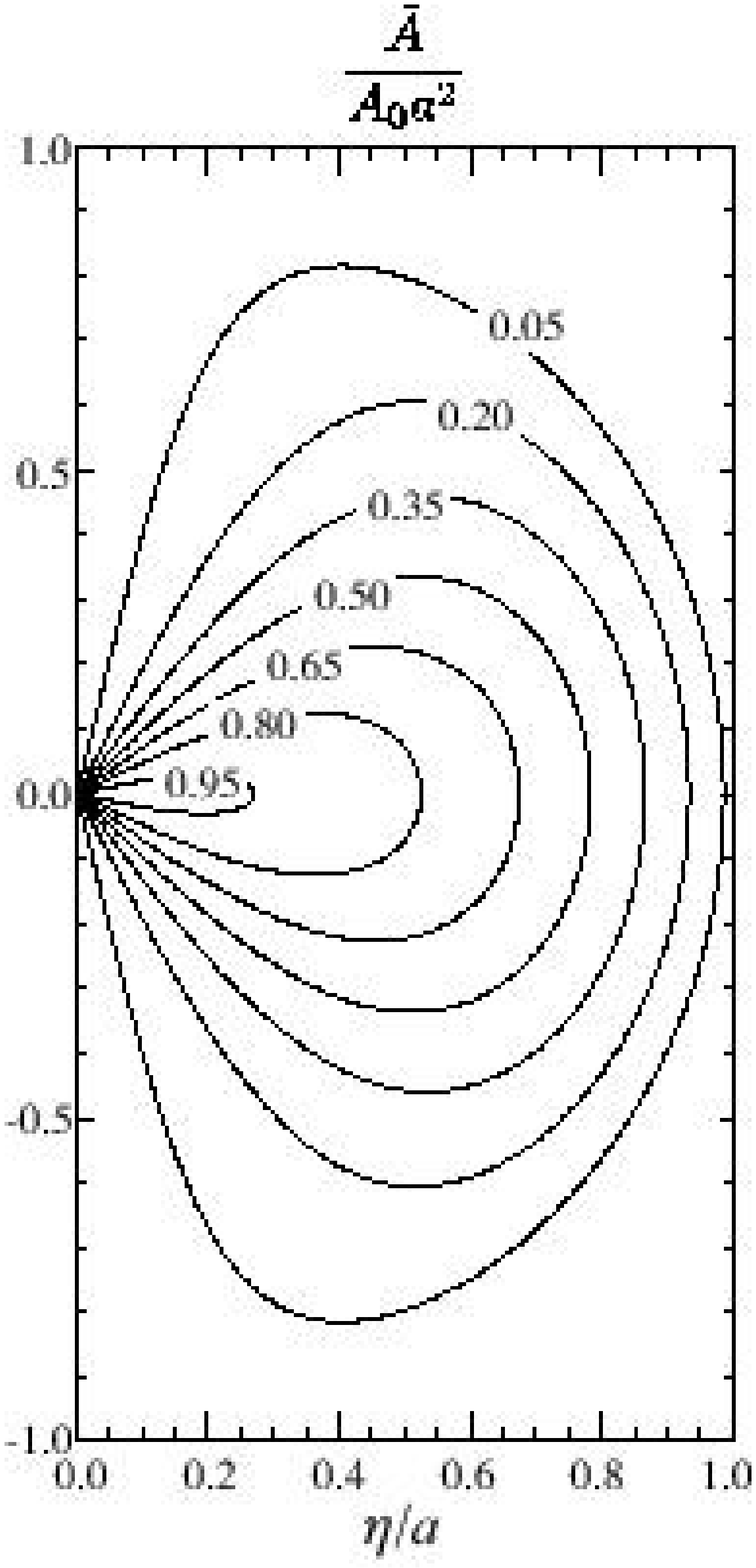}
  \end{minipage}
  \hspace*{1mm}
  \begin{minipage}{0.33\hsize}
   \includegraphics[width=4.8cm]{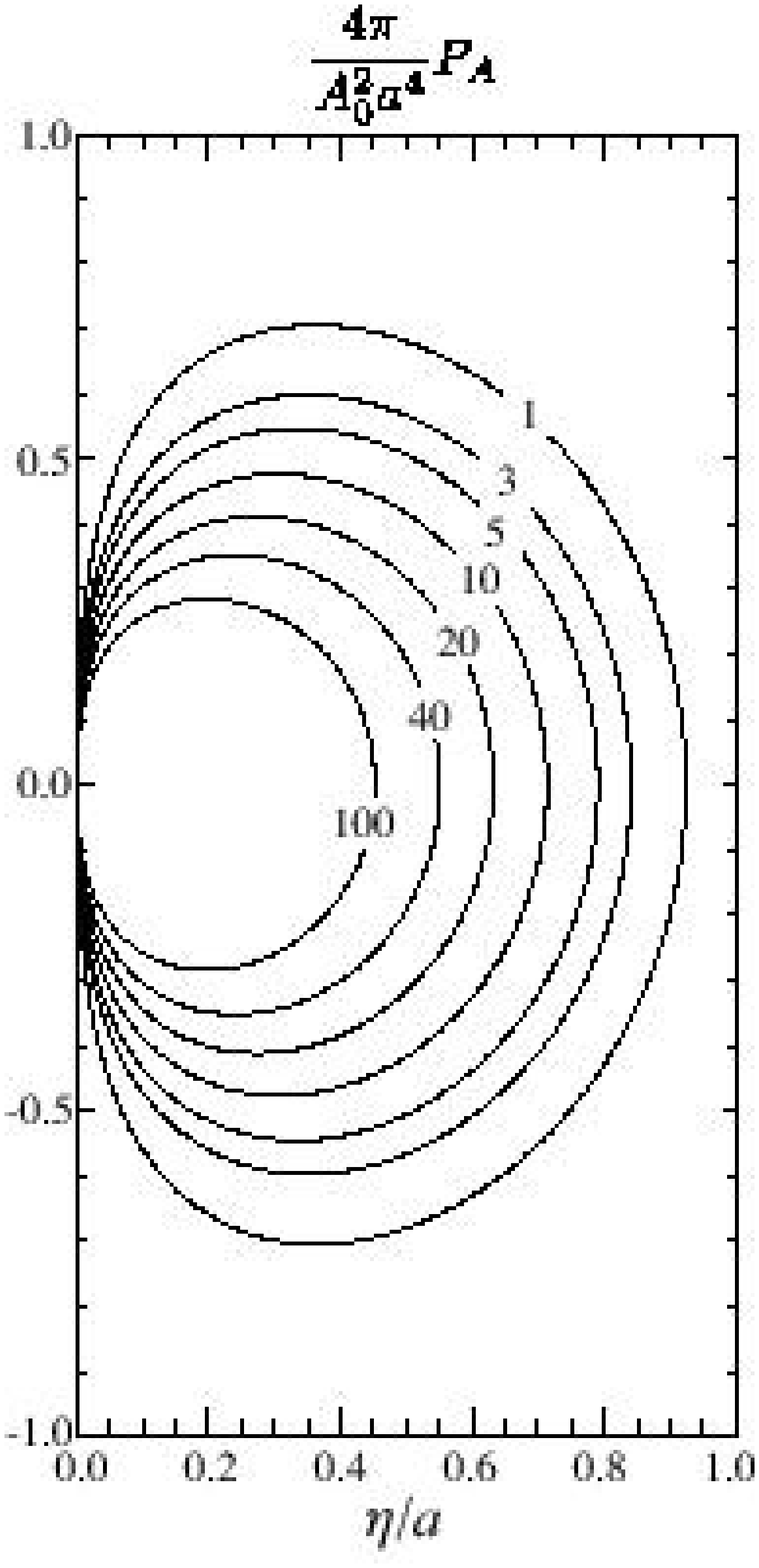}
  \end{minipage}
  \hspace*{1mm}
  \begin{minipage}{0.33\hsize}
  \includegraphics[width=4.8cm]{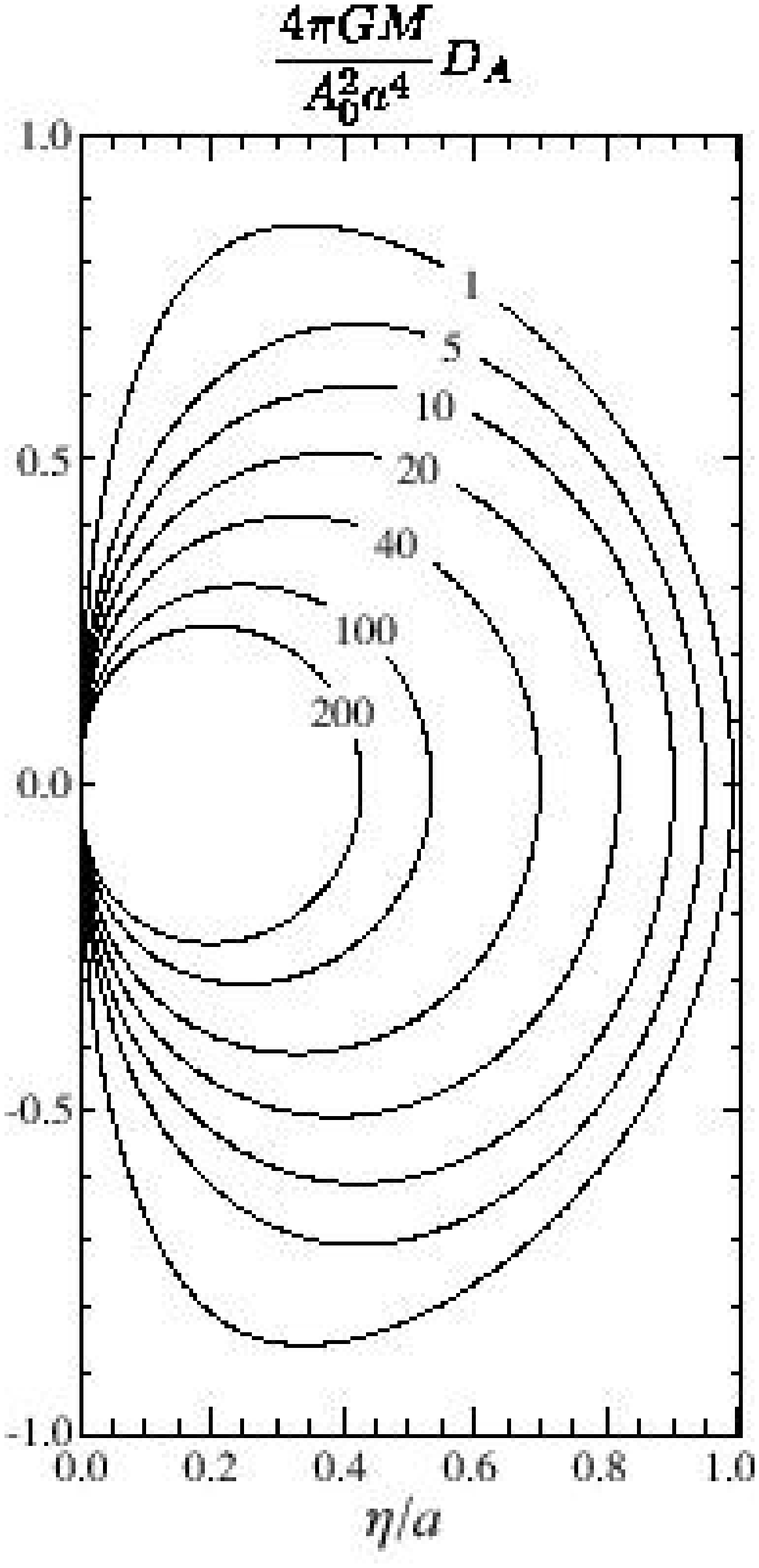}
  \end{minipage}
 \end{tabular}
 \caption{Contour plots of the magnetic flux $\tilde{A}$
 (left), the poloidal part of the pressure $P_A$ (centre), and that
 of the gas density $D_A$(right) for the dipolar solution in
 $\eta/a-\theta$ plane when $a=0.8$.}
 \label{fig:sol1A}
\end{figure}
Dipolar solutions are constructed by the flux function specified by
equation (\ref{sol1:A}).
 The azimuthal magnetic fields can be obtained
by substituting equation (\ref{sol1:A}) into equation
(\ref{eq:eomphi2}) as
\begin{equation} 
Q(\eta,\theta)=\sum_{n}Q_{0,n}
 \frac{(a^2-\eta^2)^{\frac{n}{2}}}{(1-\eta^2)^{1+\frac{n}{4}}}\sin^{n}\theta,
 \label{sol1:Q}
\end{equation}
where $Q_{0,n}$ are constants. Note that the solutions (\ref{sol1:A})
and (\ref{sol1:Q}) satisfy the formal solution given by equation
(\ref{sol:Qformal}). Substituting equations (\ref{sol1:A}) and
(\ref{sol1:Q}) into equation
(\ref{eq:eomtheta2}), we obtain the pressure function $P$:
\begin{equation}
 P(\eta,\theta)=P_0(\eta)+P_A(\eta,\theta)+P_Q(\eta,\theta),\label{sol1:P}
\end{equation}
where $P_0(\eta)$ is an arbitrary function arisen from the
integration and $P_A$ and $P_Q$ are given by
\begin{equation}
 P_A(\eta,\theta)=\frac{A_0^2}{4\pi\eta^4}\frac{a^2-\eta^2}{(1-\eta^2)^2}
  (2a^2-3 a^2 \eta^2-\eta^4+2\eta^6)\sin^2\theta,\label{sol1:PA}
\end{equation}
\begin{eqnarray}
 P_Q(\eta,\theta)&=&\left\{\begin{array}{ll}
 {\displaystyle  -\sum_{m+n\ne 2} \frac{n Q_{0,m}Q_{0,n}}{4\pi
  (m+n-2)}\frac{(a^2-\eta^2)^{\frac{m+n}{2}}}{\eta^2
  (1-\eta^2)^{1+\frac{m+n}{4}}} \sin^{m+n-2}\theta}, & \rmn{for\ \  }m+n\ne2,\\
 {\displaystyle -\sum_{m+n=2} \frac{n Q_{0,m}Q_{0,n}}{4 \pi
  }\frac{a^2-\eta^2}{\eta^2(1-\eta^2)^\frac{3}{2}}\log (\sin\theta)},
			     & \rmn{for\ \  }m+n=2.\\
 		      \end{array}\right. \label{sol1:PQ}
\end{eqnarray}
Substituting equations (\ref{sol1:A}), (\ref{sol1:Q}), (\ref{sol1:P}),
(\ref{sol1:PA}), and (\ref{sol1:PQ}) into (\ref{eq:eomr2}),
the density function $D$ can be determined as
\begin{figure}
 \begin{tabular}{ccc}
  \begin{minipage}{0.33\hsize}
   \includegraphics[width=4.8cm]{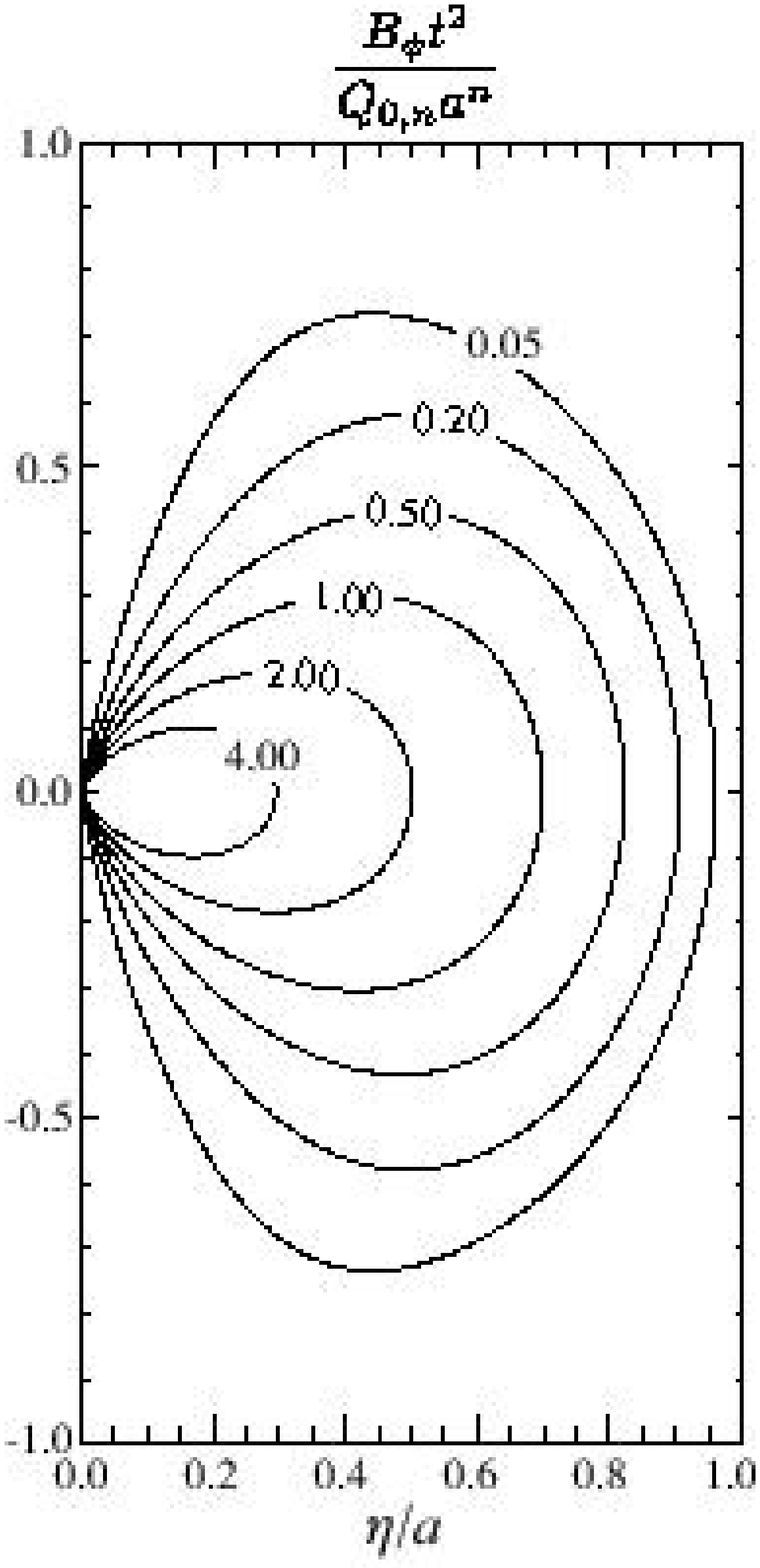}
  \end{minipage}
  \hspace*{1mm}
  \begin{minipage}{0.33\hsize}
   \includegraphics[width=4.8cm]{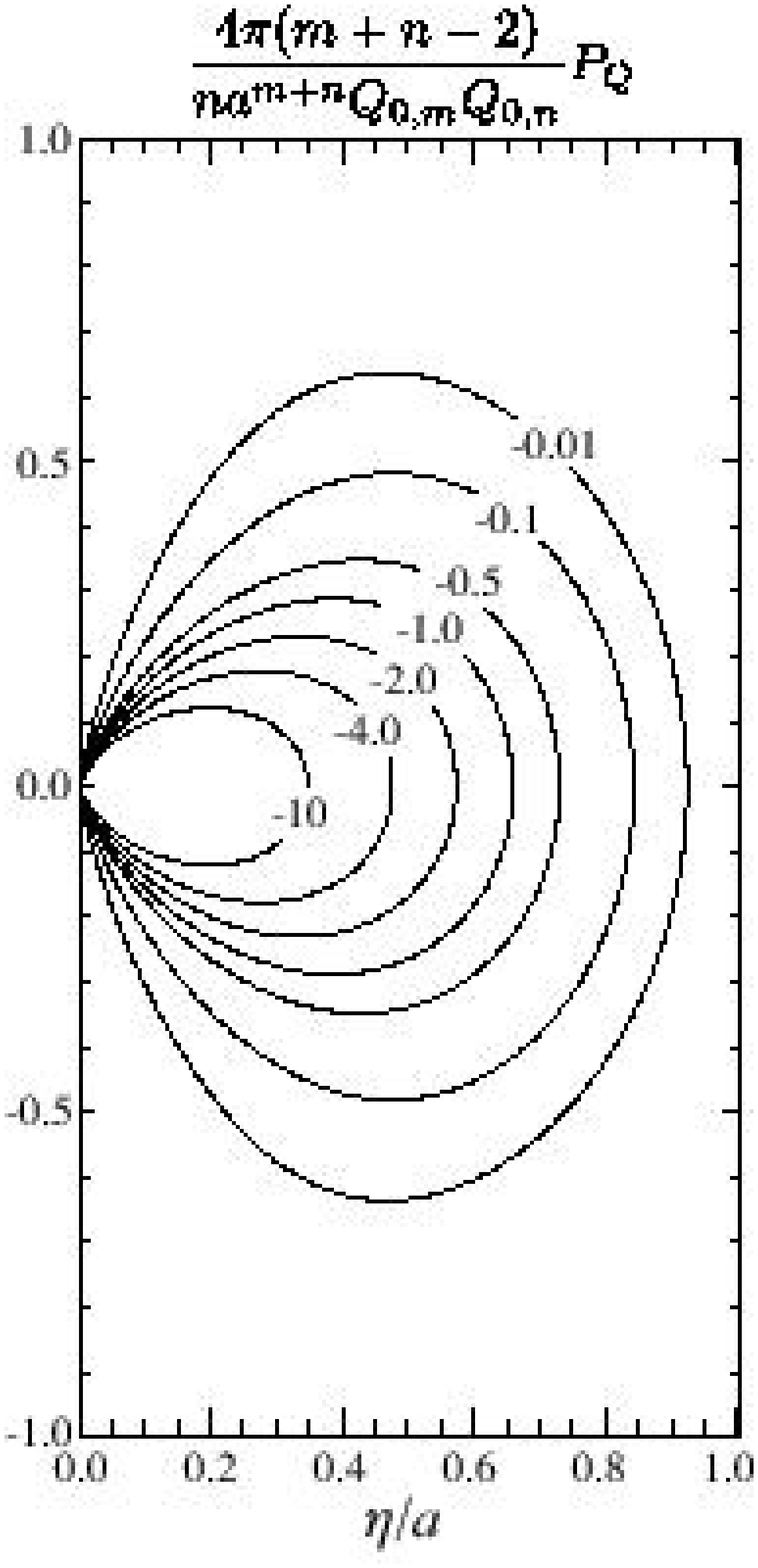}
  \end{minipage}
  \hspace*{1mm}
  \begin{minipage}{0.33\hsize}
  \includegraphics[width=4.8cm]{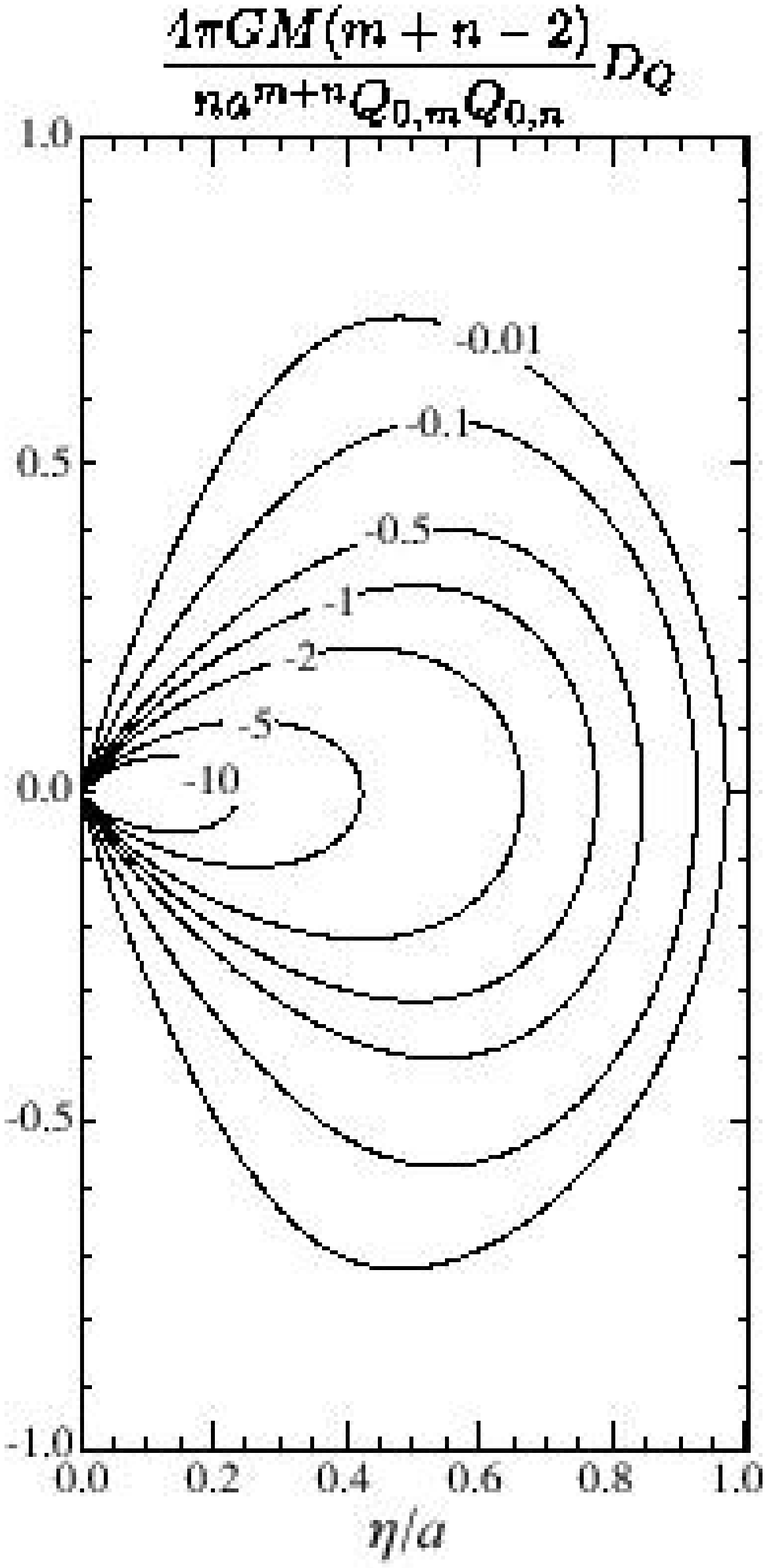}
  \end{minipage}
 \end{tabular}
 \caption{Contour plots of the toroidal magnetic field (left),
 the toroidal part of the pressure $P_{Q}$ (centre), and that of the
 density $D_Q$(right) for the dipolar solution in $\eta/a-\theta$ plane
 when $a=0.8$ and $m=n=4$.}
 \label{fig:sol1Q}
\end{figure}
\begin{equation}
 D(\eta,\theta)=D_0(\eta)+D_A(\eta,\theta)+D_Q(\eta,\theta),\label{sol1:D}
\end{equation}
where 
\begin{equation}
 D_0(\eta)=\frac{\eta^2}{GM}\left(\frac{4\eta
				   P_0}{1-\eta^2}-\frac{d
				   P_0}{d
				   \eta}\right),\label{sol1:D0}
\end{equation}
\begin{equation}
 D_A(\eta,\theta)=\frac{A_0^2}{4\pi G
  M\eta^3}\frac{(a^2-\eta^2)}{(1-\eta^2)^3}
  \left[a^2(8-12\eta^2+3\eta^4)-\eta^6(5-6\eta^2)\right]\sin^2\theta
  ,\label{sol1:DA}
\end{equation}
\begin{equation}
 D_Q(\eta,\theta)=\left\{\begin{array}{ll}
   {\displaystyle-\sum_{m+n\ne2}\frac{n Q_{0,m}Q_{0,n}}{4\pi G M
  (m+n-2)} \frac{(a^2-\eta^2)^{\frac{m+n-2}{2}}(2 a^2-a^2\eta^2-\eta^4)}{\eta
  (1-\eta^2)^{2+\frac{m+n}{4}}}\sin^{(m+n-2)}\theta},
		    & \rmn{for\ \  }m+n\ne 2,\\\\
  {\displaystyle-\sum_{m+n=2}\frac{n Q_{0,m}Q_{0,n}}{8 \pi G
  M}\frac{-\eta^2
  (2-a^2-\eta^2)+2(2a^2-a^2\eta^2-\eta^4)\log(\sin\theta)}
  {\eta (1-\eta^2)^\frac{5}{2}}}, & \rmn{for\ \  }m+n=2.\\
			 \end{array}\right.\label{sol1:DQ}
\end{equation}

The parameters $m$ and $n$ correspond to the Fourier modes in the $\theta$
direction.
These parameters should be determined by the boundary condition on the
surface of the central star where magnetic twist is injected. 

Equation (\ref{sol1:P}) and (\ref{sol1:D}) indicate that the solution
consists of three parts, $P_0$, $P_A$ and $P_Q$ (or
$D_0$, $D_A$ and $D_Q$). The arbitrary function $P_0(\eta)$ describes an
isotropic pressure in the region $r<R(t)$. The isotropic
density profile $D_0(\eta)$ is related to $P_0$ through equation
(\ref{sol1:D0}). This equation is similar to that in 
non-relativistic model \citep{1982ApJ...261..351L}. In the
non-relativistic model,
gravity is supported by the gradient of $P_0(\eta)$. In the relativistic case, 
relativistic correction of the plasma inertia cannot be ignored. 
This effect is included in the first term in the right hand
side of equation (\ref{sol1:D0}). Other functions $P_A$ and $P_Q$
(or $D_A$ and $D_Q$) come from the interaction with the
electromagnetic force. Note that the plasma pressure $P_Q$, which balances with the
electromagnetic force produced by the toroidal magnetic field, is always
negative. This suggests that the pressure is smaller for larger
toroidal magnetic fields.

Fig.~\ref{fig:sol1A} shows the contour plots of the magnetic flux
$\tilde{A}$ (left), the poloidal part of the pressure $P_A$ (centre),
and that of the gas density $D_A$ (right), while Fig.~\ref{fig:sol1Q}
shows the contour plots of the toroidal magnetic field $B_\phi$ (left),
the toroidal part of the pressure $P_Q$ (centre), and that of the gas
density $D_Q$ (right) in the $\eta/a-\theta$
plane for $m=n=4$ and $a=0.8$.

The magnetic field is explicitly expressed as
\begin{equation}
 \bmath
  B=\frac{2A_0}{r^2}\frac{a^2-(r/t)^2}{\sqrt{1-(r/t)^2}}\cos\theta\bmath
  e_r
  +\frac{A_0}{t^2}\frac{2-a^2-(r/t)^2}{\left[1-(r/t)^2\right]^{\frac{3}{2}}}
  \sin\theta \bmath e_\theta 
  +\sum_n \frac{Q_{0,n}}{r t}\frac{\left[a^2-(r/t)^2\right]^{\frac{n}{2}}} 
  {[1-(r/t)^2]^{1+\frac{n}{4}}}\sin^{n-1}\theta \bmath e_\phi,\label{sol1:Bs}
\end{equation}
where $\bmath e_r$, $\bmath e_\theta$, and $\bmath e_\phi$ are unit
vectors in $r$, $\theta$, and $\phi$ directions in the polar coordinate,
respectively.
Note that $B_r$ and $B_\phi$ are zero at $r=R(t)$ but $B_\theta$ is not zero
and it depends on time when $a \ne 1$. We will discuss the physical
meaning of this result later in \S \ref{sss2}.

In later stage, the magnetic field becomes stationary,
\begin{equation}
 \lim_{t\rightarrow\infty}\bmath B=\frac{2A_0a^2}{r^2}\cos\theta\bmath
  e_r, \label{sol1:Blimit}
\end{equation}
and the magnetic field becomes radial.
In the limit $t \gg r$, the pressure and the gas density inside the
magnetic loop are given by
\begin{equation}
 \lim_{t\rightarrow\infty}p=\frac{A_0^2a^4}{2\pi
  r^4}\sin^2\theta+\frac{1}{r^4}\left(P_0 \eta^4\right)\bigr|_{\eta=0},
  \label{sol1:plimit}
\end{equation}
\begin{equation}
 \lim_{t\rightarrow\infty}\rho=\frac{2A_0^2a^4}{\pi GM r^3}\sin^2\theta
  +\frac{1}{r^3}\left(D_0 \eta^3\right)\bigr|_{\eta=0}.
  \label{sol1:rholimit}
\end{equation}
Since the toroidal magnetic field tends to be zero in this limit,
the pressure and density do not depend on the amplitude of the
toroidal magnetic fields.

\begin{figure}
 \begin{tabular}{ccc}
  \begin{minipage}{0.33\hsize}
   \includegraphics[width=4.8cm]{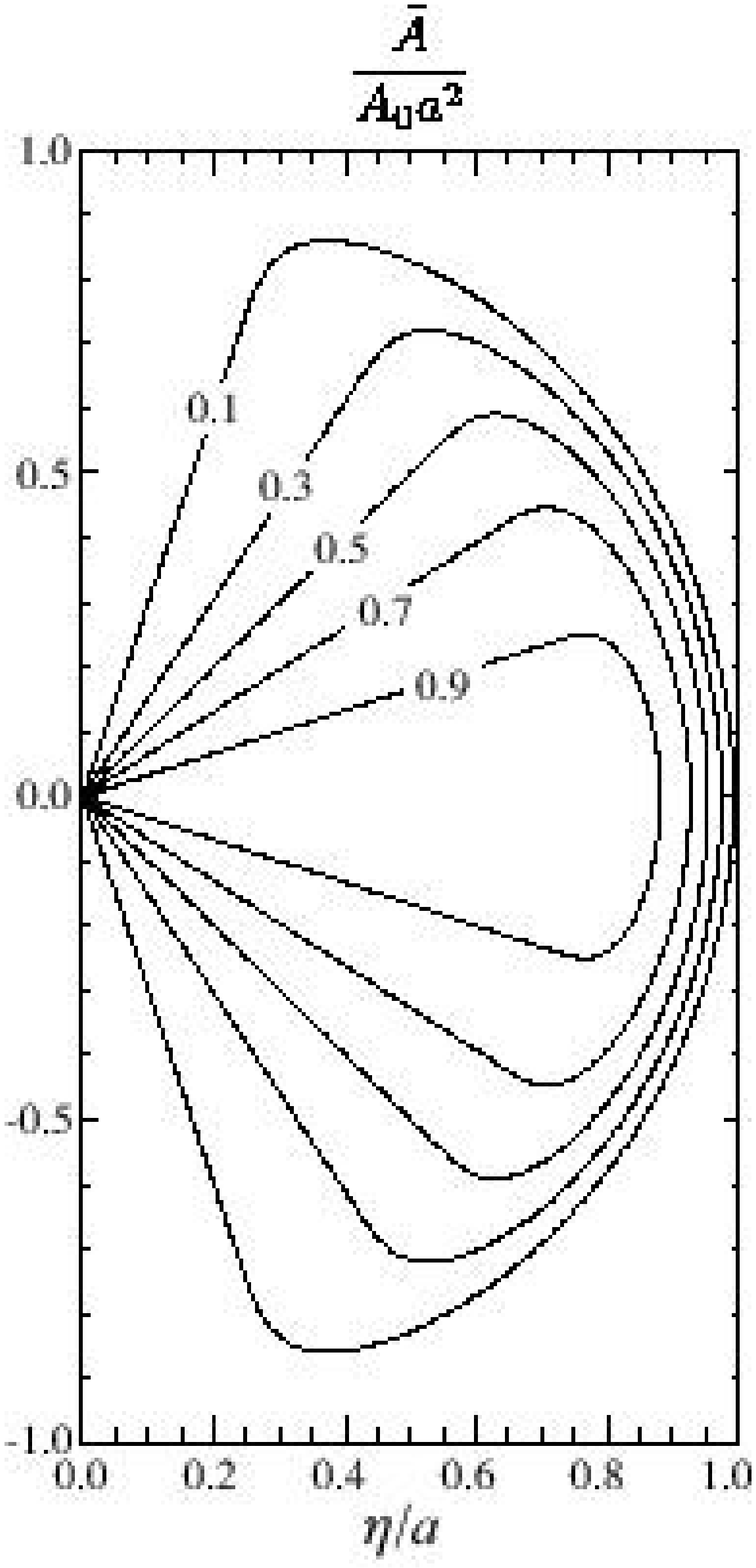}
  \end{minipage}
  \begin{minipage}{0.33\hsize}
   \includegraphics[width=4.8cm]{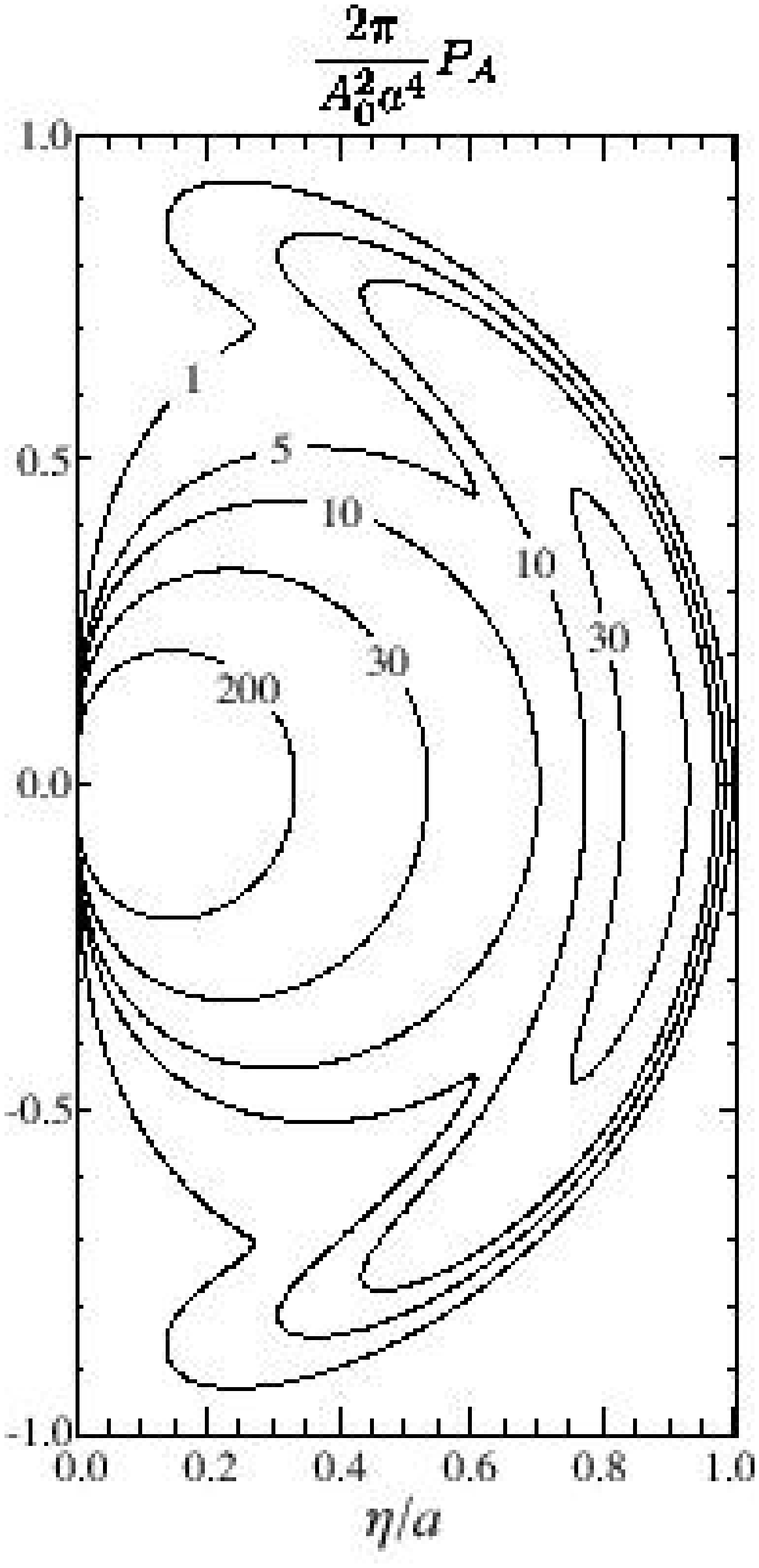}
  \end{minipage}
  \begin{minipage}{0.33\hsize}
   \includegraphics[width=4.8cm]{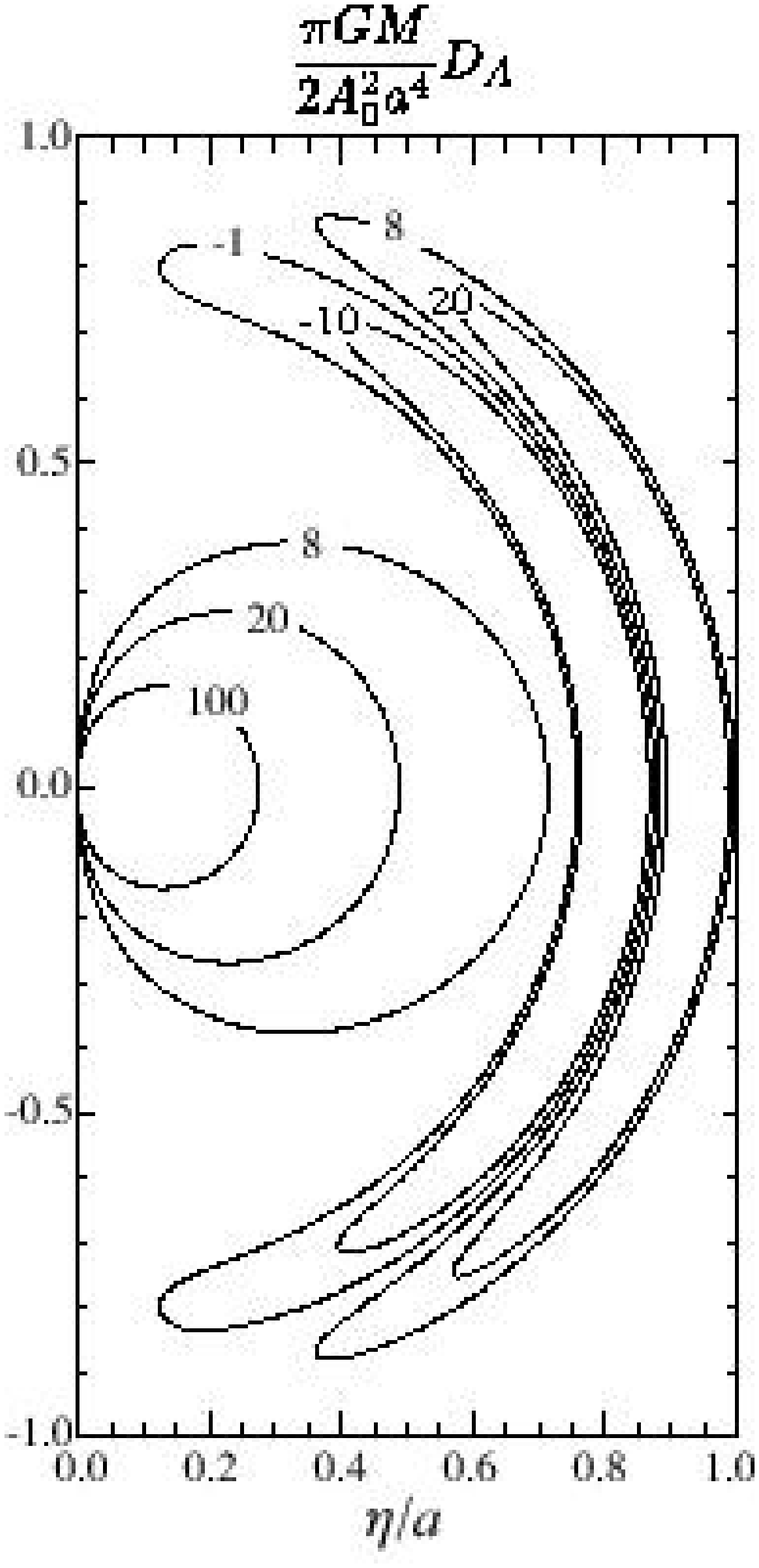}
  \end{minipage}
 \end{tabular}
 \caption{Contour plots of the magnetic flux $\tilde{A}$
 (left), the poloidal part of the pressure $P_{A}$ (centre), and that of
 the gas density $D_A$ (right) for the shell solution in $\eta/a-\theta$
 plane when $a=0.8$ and $b=0.75a$.}
 \label{fig:sol2A}
\end{figure}
\subsection{Shell Solutions}\label{sss2}
Shell solutions are constructed from the flux function
(\ref{sol2:A}). By substituting equation (\ref{sol2:A}) into equation
(\ref{eq:eomphi2}), the function $Q$ can be obtained as
\begin{equation}
Q^{I}(\eta,\theta)=Q^I_0 \frac{f(\theta)}{1-\eta^2},
\end{equation}
\begin{equation}
 Q^{II}(\eta,\theta)=\sum_{n}\frac{Q^{II}_{0,n}}{1-\eta^2} \left[\sin^4
    T(a)-\sin^4 T(\eta)\right]^{\frac{n}{2}}\sin^n\theta,\label{sol2:Q2t}
\end{equation}
where $f(\theta)$ is an arbitrary function of $\theta$, and $Q^I_{0}$
and $Q^{II}_{0,n}$ are constants. The subscripts $I$ and $II$
mean that the function is defined in region I and in region II,
respectively. 
The arbitrary function $f(\theta)$ can be determined by applying the
boundary condition that magnetic field should be connected smoothly at
$\eta=b$, 
\begin{equation}
Q^{I}(\eta=b,\theta)=Q^{II}(\eta=b,\theta).\label{sol2:boundaryab}
\end{equation}
By using the boundary condition, the function $f$ is given by
\begin{equation}
 f(\theta)=\sum_{n}\sin^n \theta.
\end{equation}
and the function $Q^{I}$ is obtained as
\begin{equation}
Q^{I}(\eta,\theta)=\sum_n Q^I_{0,n} \frac{\sin^n\theta}{1-\eta^2}. \label{sol2:Q1t}
\end{equation}
The constants $Q^{I}_{0,n}$ and $Q^{II}_{0,n}$ should be related by
\begin{equation}
Q_{0,n}\equiv Q^{I}_{0,n}=Q^{II}_{0,n}\sin^{2n}T(a) \label{sol2:Q1Q2bound}.
\end{equation}
from the boundary condition (\ref{sol2:boundaryab}).
Substituting equations (\ref{sol2:A}), (\ref{sol2:Q2t}),
(\ref{sol2:Q1t}) and  (\ref{sol2:Q1Q2bound})
into  equation (\ref{eq:eomtheta2}), we obtain the pressure function
$P(\eta, \theta)$. The density function $D(\eta, \theta)$ is obtained
from equation (\ref{eq:eomr2}). The functions $Q$, $P$, and $D$ obtained
in region I and region II are given in appendix \ref{ap-shell}.

\begin{figure}
 \begin{tabular}{ccc}
  \begin{minipage}{0.33\hsize}
   \includegraphics[width=4.8cm]{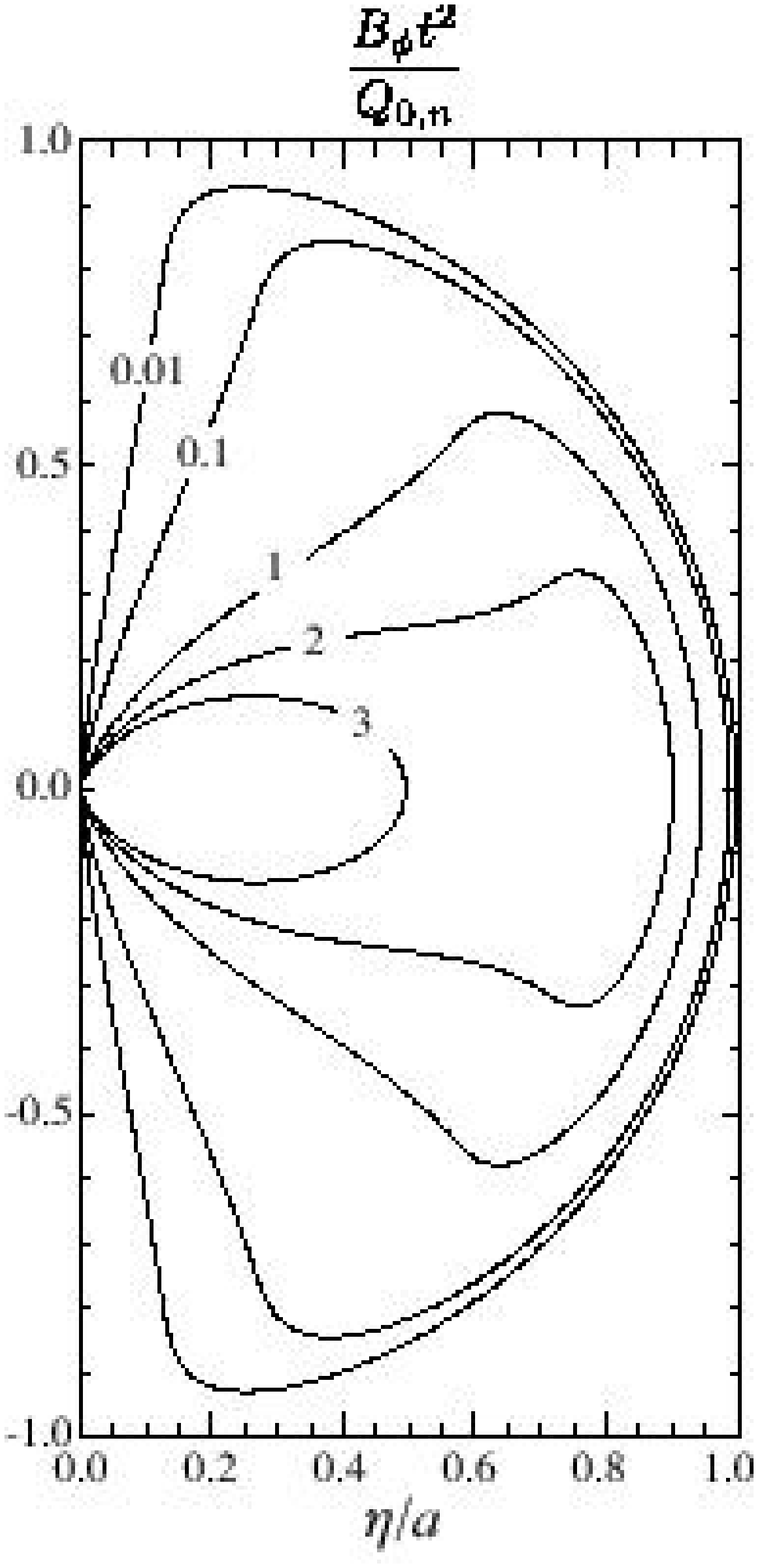}
  \end{minipage}
  \begin{minipage}{0.33\hsize}
   \includegraphics[width=4.8cm]{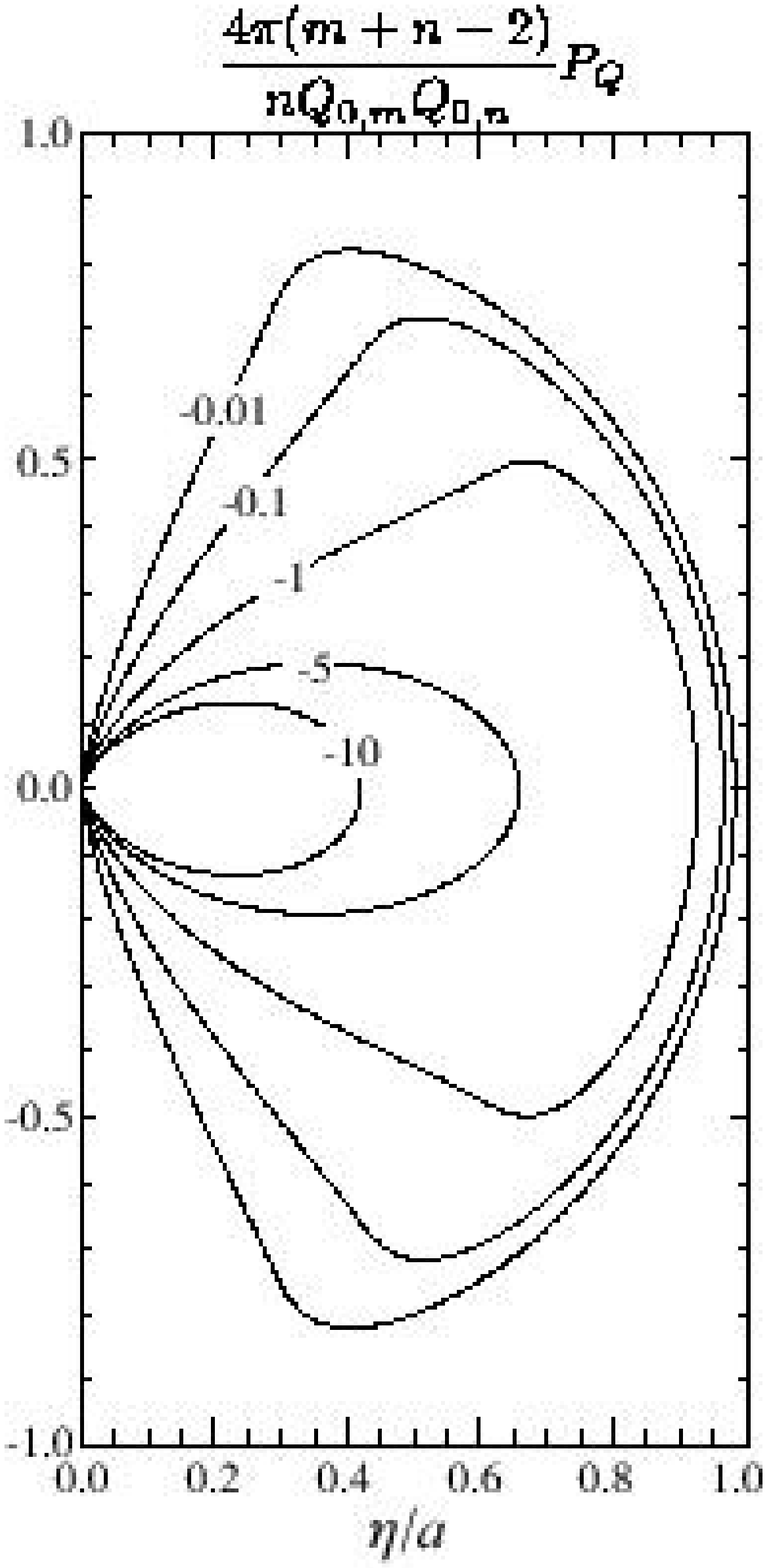}
  \end{minipage}
  \begin{minipage}{0.33\hsize}
   \includegraphics[width=4.8cm]{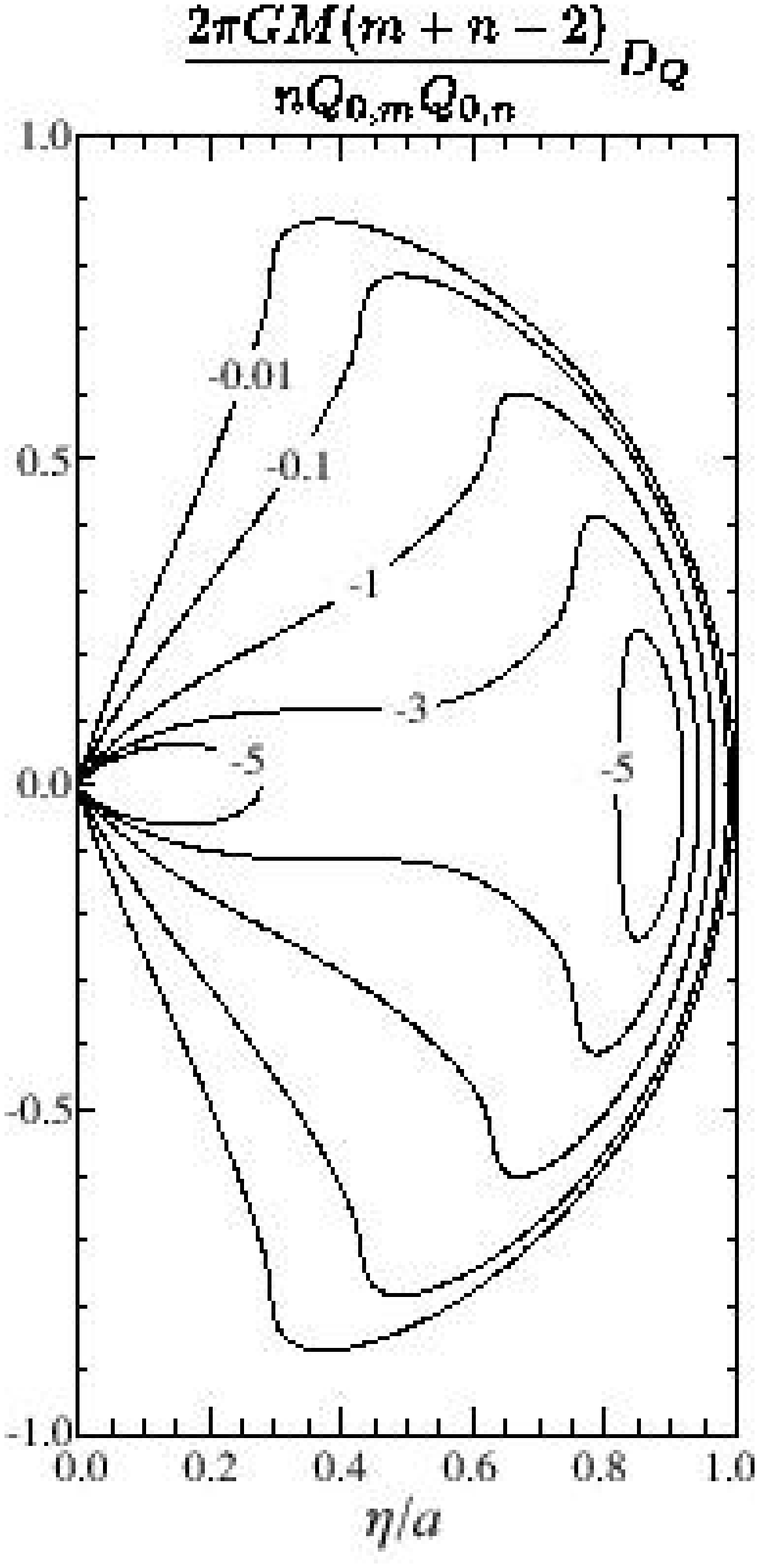}
  \end{minipage}
 \end{tabular}
 \caption{Contour plots of the toroidal magnetic field $B_\phi$ (left),
 the toroidal part of the pressure $P_Q$ (centre), and that of the gas
 density $D_Q$ (right) for the shell solution in $\eta/a-\theta$ plane
 when $a=0.8, b=0.75a$, and $n=m=4$.}
 \label{fig:sol2Q}
\end{figure}
The pressure and the gas density consist of three parts, the
isotropic part $P_0$ and parts representing the
interaction with the electromagnetic force by the poloidal and toroidal
components of the magnetic field, $P_A$ and $P_Q$, similarly to the
dipolar solutions (see equations (\ref{sol2:P1}) and (\ref{sol2:P2}),
for the pressure and equations (\ref{sol2:D1}) and (\ref{sol2:D2})
for the gas density). 

Fig.~\ref{fig:sol2A} shows the contour plots of the magnetic flux
$\tilde{A}$ (left), the poloidal part of the pressure $P_A$
(centre), and that of the gas density $D_A$ (right) in
$\eta/a-\theta$ plane. Fig.~\ref{fig:sol2Q} shows contour plots of the
toroidal magnetic field $B_\phi$ (left), the toroidal part of the
pressure $P_Q$ (centre), and that of the gas density $D_Q$ (right) in
$\eta/a-\theta$ plane. The parameters are taken to be $a=0.8$,
$b=0.75a$, and $m=n=4$ in both figures.
A shell structure appears behind the loop top.

The pressure $P_Q$ is always negative (see equations
(\ref{sol2:PQ1}) and (\ref{sol2:PQ2}) and the middle panel of
Fig.~\ref{fig:sol2Q}) and its amplitudes is
proportional to that of the toroidal magnetic fields, $Q_{0,n}$. This
indicates that the pressure is smaller for a larger toroidal
magnetic field.

The magnetic field is explicitly given by
\begin{equation}
\bmath B=\left\{\begin{array}{ll}
	  {\displaystyle\frac{2A_0a^2}{r^2}\cos\theta\bmath
	   e_r+\sum_{n}Q_{0,n}\frac{t}{r(t^2-r^2)}\sin^{n-1}\theta \bmath e_\phi}, &
	   (\eta \lid b),\\
		 {\displaystyle\frac{2A_0a^2}{r^2}\Lambda\left(r/t\right)\cos\theta\bmath
		  e_r+\frac{4 A_0 a^2}{r t}k \frac{\sin^3T(r/t)\cos
		  T(r/t)}{\sin^4 T(a)}\sin\theta\bmath
		  e_\theta+\sum_{n}Q_{0,n}\Lambda^{\frac{n}{2}}\left(r/t\right)\frac{t
		  \sin^{n-1}\theta}{r(t^2-r^2)}\bmath
		  e_\phi},
		  & (b < \eta \lid a).
		 \end{array}\right.\label{sol2:Blimit}
\end{equation}

Similarly to the dipolar solution, the shell solutions have the parameter
$m$ which corresponds to the Fourier modes in the polar angle
$\theta.$ These modes and the corresponding amplitude $Q_{0,m}$ of the
toroidal magnetic fields should be determined by the
boundary condition at the surface of the central star where the magnetic
twist is injected.

In contrast to the dipolar solution, the magnetic field lines do not
cross the equatorial plane in region I (see the left panel in
Fig.~\ref{fig:sol2A}). Note that in the limit that
$t\gg r$, the magnetic fields and plasma distribution approach those of
the dipolar solution, given by (\ref{sol1:Blimit}), (\ref{sol1:plimit})
and (\ref{sol1:rholimit}).

At the boundary $r=R(t)$, the field components $B_r$ and $B_\phi$ are
exactly zero, but $B_\theta$ is not zero. 
Since the Poynting flux $\bmath S = (\bmath E\times \bmath
B)/(4\pi)$ is not zero at $r=R(t)$, the energy flux will be transmitted
to the region outside the boundary at $r=R(t)$.
When $\cos T(a)=0$, since the magnetic field vanishes at $r=R(t)$, 
the energy is not transferred to $r>R(t)$. This happens
when the constant $k$ is given by
\begin{equation}
 k=\frac{(2l+1)}{2}\frac{\pi}{a-b},\label{sol2:knowave}
\end{equation}
where $l$ is an integer number.

\begin{figure}
 \begin{tabular}{ccc}
  \begin{minipage}{0.49\hsize}
    \includegraphics[width=8cm]{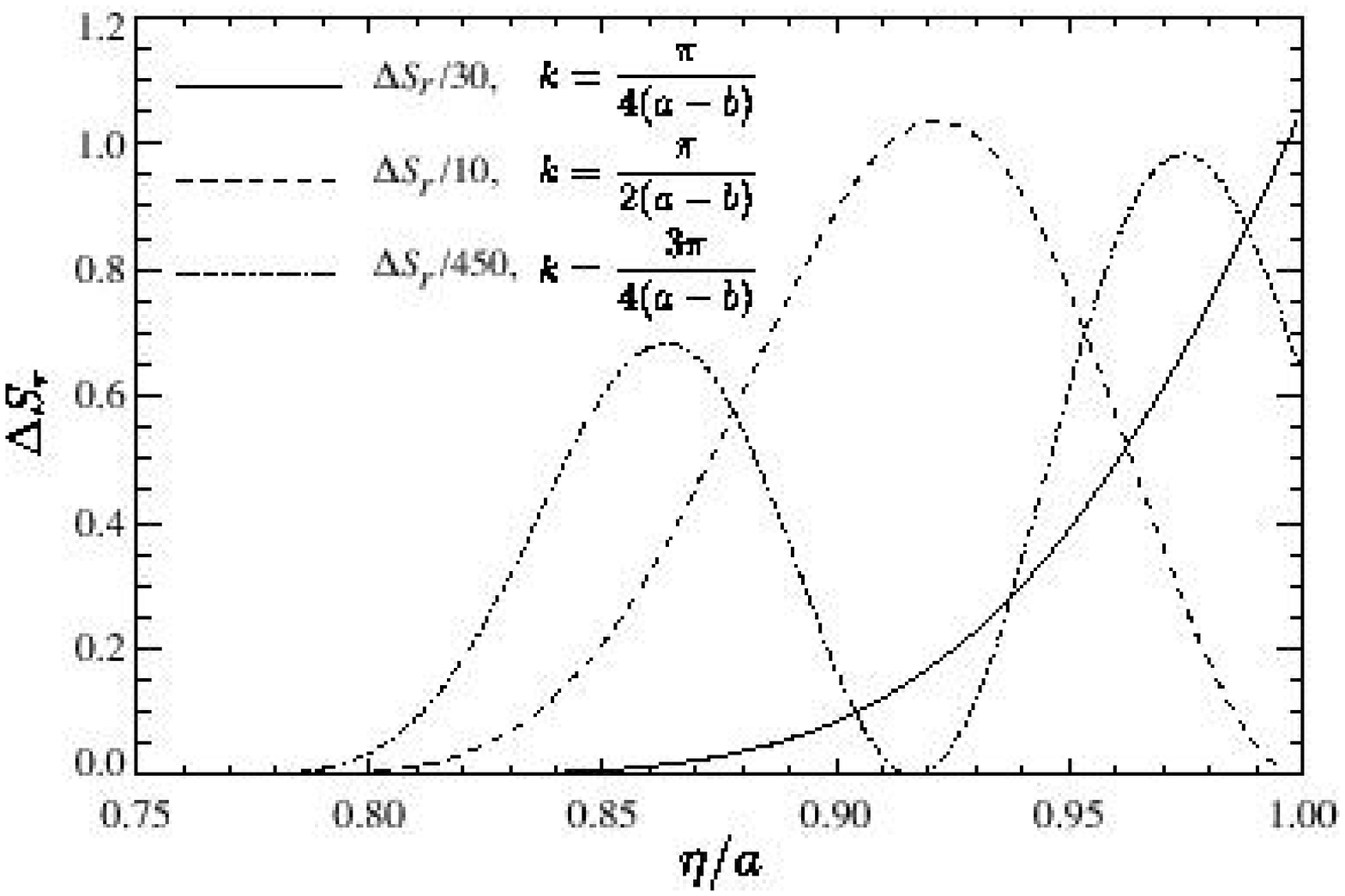}
  \end{minipage}
  \begin{minipage}{0.49\hsize}
    \includegraphics[width=8cm]{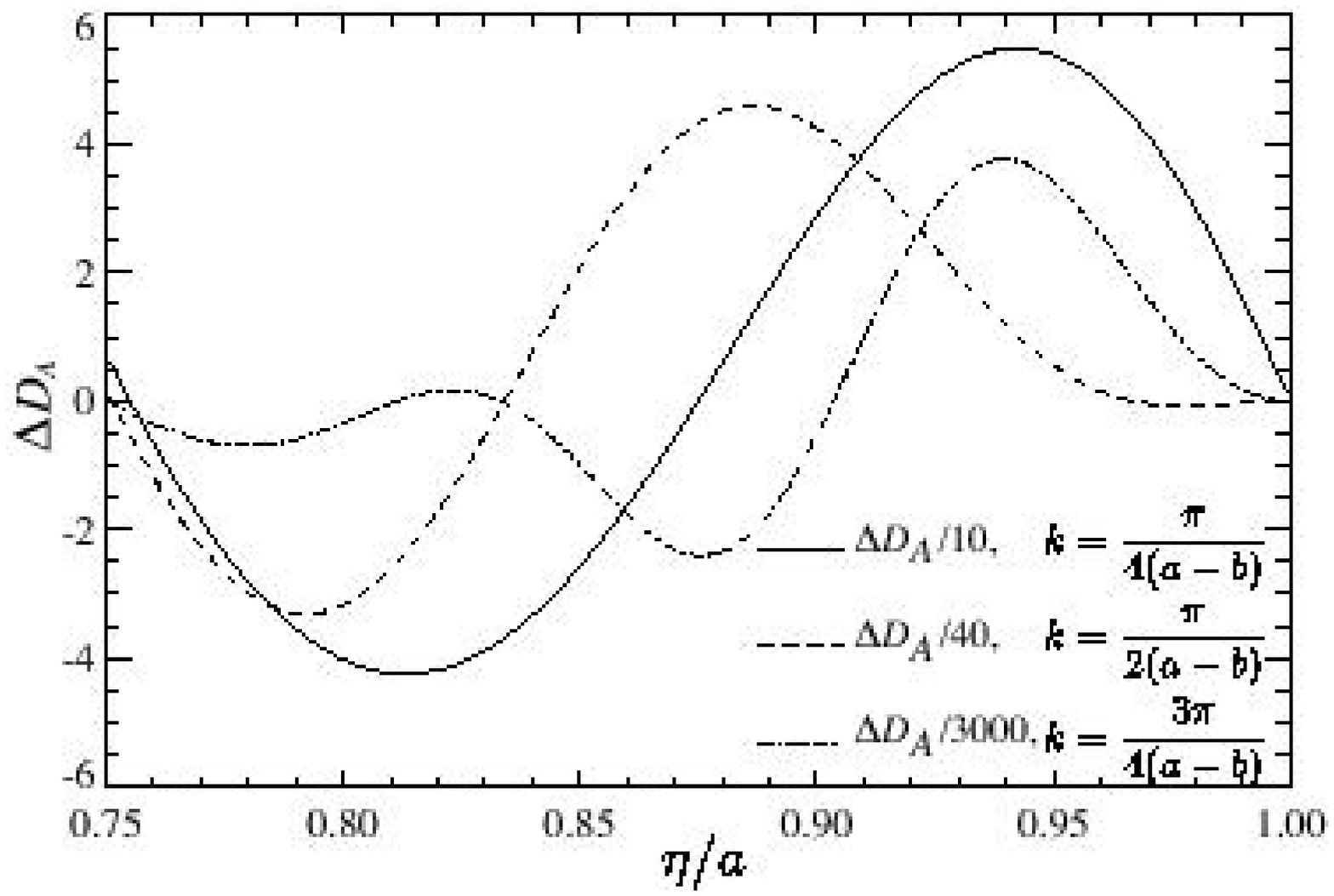}
  \end{minipage}
 \end{tabular}
 \caption{Distributions of the Poynting flux $\Delta S_r\equiv \pi
 r^4S_r/(A_0^2 a^4\sin^2\theta)$ (left) and $\Delta D_A \equiv D_A \pi G
 M \eta^3 (1-\eta^2)/[2 A_0^2 a^4 \sin^2\theta]$ (right) for the shell
 solutions,  where $\bmath S_r$ is the
 Poynting flux in the radial direction. Solid curves show for
 $k=\pi/[4(a-b)]$, while dashed and dot-dashed ones for
 $k=\pi/[2(a-b)]$, and $k=3\pi/[4(a-b)]$, respectively. Here we take
 $Q_{0,n}=0$.}\label{fig:sol2-Pr}
\end{figure}

Fig.~\ref{fig:sol2-Pr} shows the distributions of the Poynting flux
$\Delta S_r =r^4 \pi S_r/(A_0^2 a^4\sin^2\theta)$ (left panel) and
$\Delta D_A \equiv D_A \pi G M \eta^3 (1-\eta^2)/[2 A_0^2 a^4
\sin^2\theta]$ (right
panel) for $a=0.8$, $b/a=0.75$, and $B_\phi=0$ for shell
solutions. Solid curve denotes that for $k=\pi/[4(a-b)]$, while dashed and
dot-dashed ones do for $k=\pi/[2(a-b)]$ and $k=3\pi/[4(a-b)]$,
respectively. When $k$ satisfies equation (\ref{sol2:knowave}),
$B_\theta(t,r=at,\theta)=0$ and thus $S_r(t, r=at,
\theta)=0$. Electromagnetic energy is not transmitted ahead of the loop
top. When $k$ does not satisfy equation (\ref{sol2:knowave}), 
the Poynting flux $S_r$ at $r=at$ is not zero and 
the electromagnetic energy is transmitted to $r>at$. 
The physical interpretation of the condition given in equation
(\ref{sol2:knowave}) is as follows.

\begin{figure}
 \begin{tabular}{ccc}
  \begin{minipage}{0.4\hsize}
   \includegraphics[width=7cm]{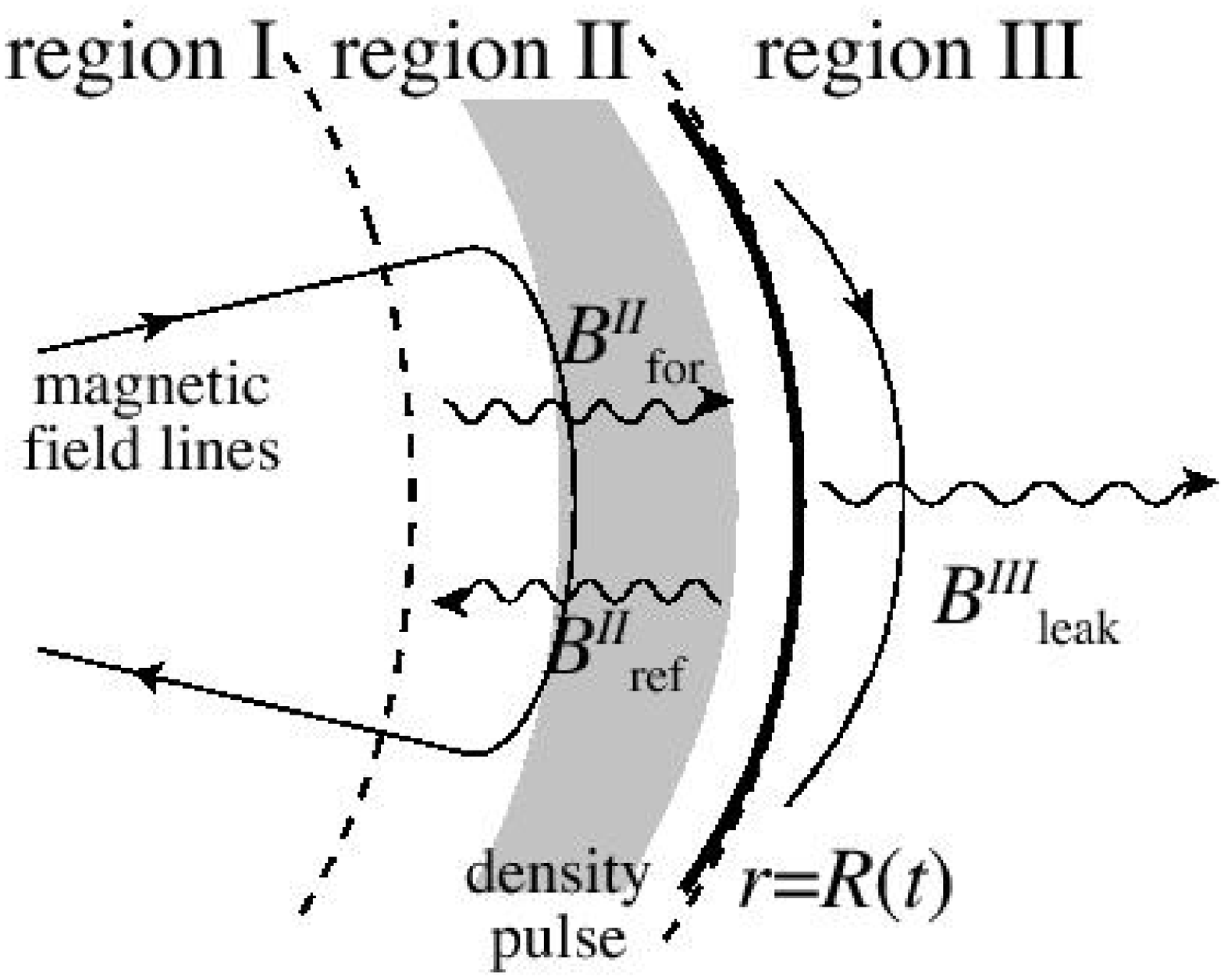} 
   \caption{Schematic picture of the propagating waves. The forward wave
   $B_\mathrm{for}$ and the reflected waves $B_\mathrm{ref}$ propagate
   inside the loops. Solid curves show the loop top at $r=R(t)$, while
   the thin curves show the magnetic field lines. The superposition of these waves determines
   $B_\theta$ in region II. The leak wave $B_\mathrm{leak}$ appears
   ahead of the loop top in region III.}
   \label{fig:leakingwave}
  \end{minipage}
  \hspace{2mm}
  \begin{minipage}{0.6\hsize}
   \includegraphics[width=4.8cm]{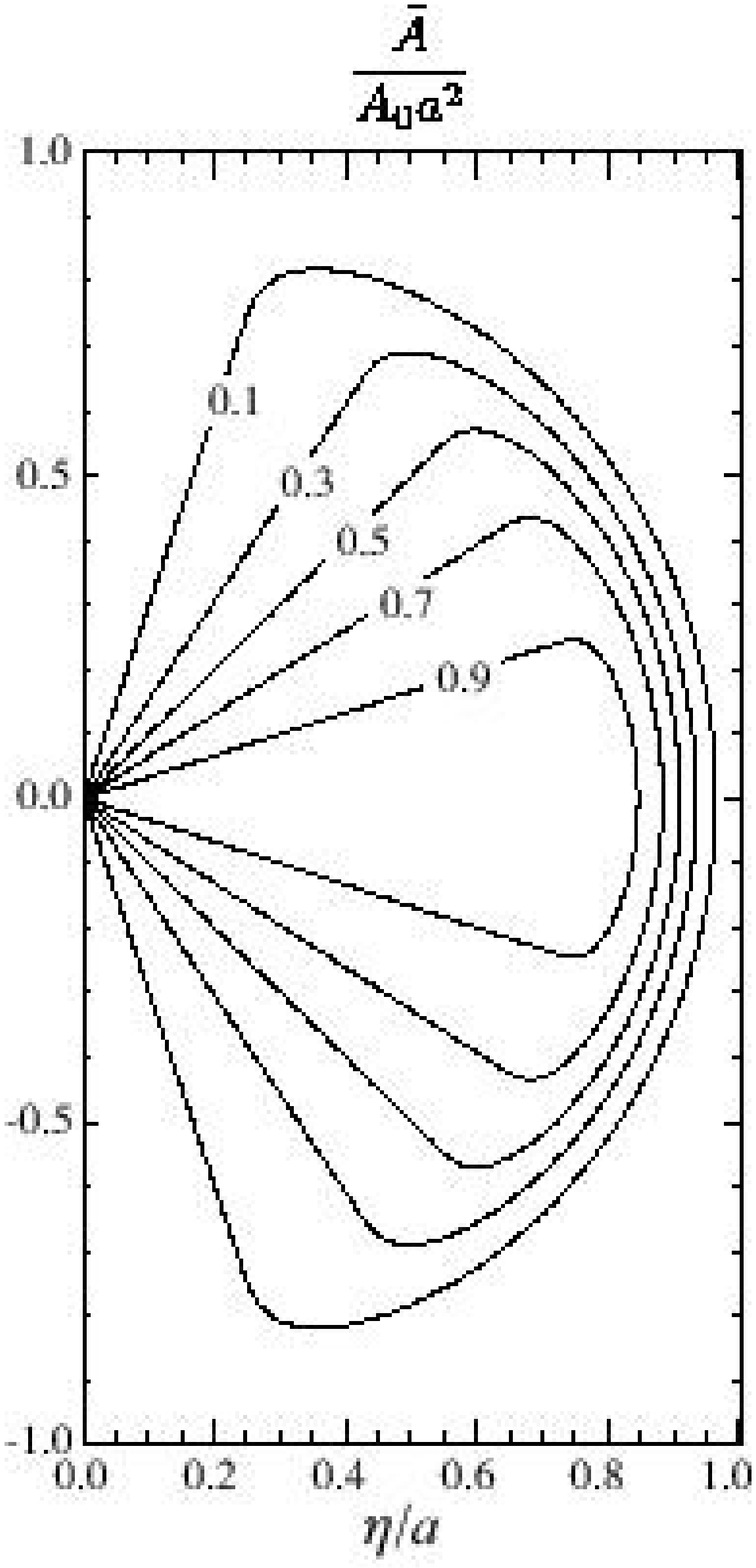} 
   \includegraphics[width=4.8cm]{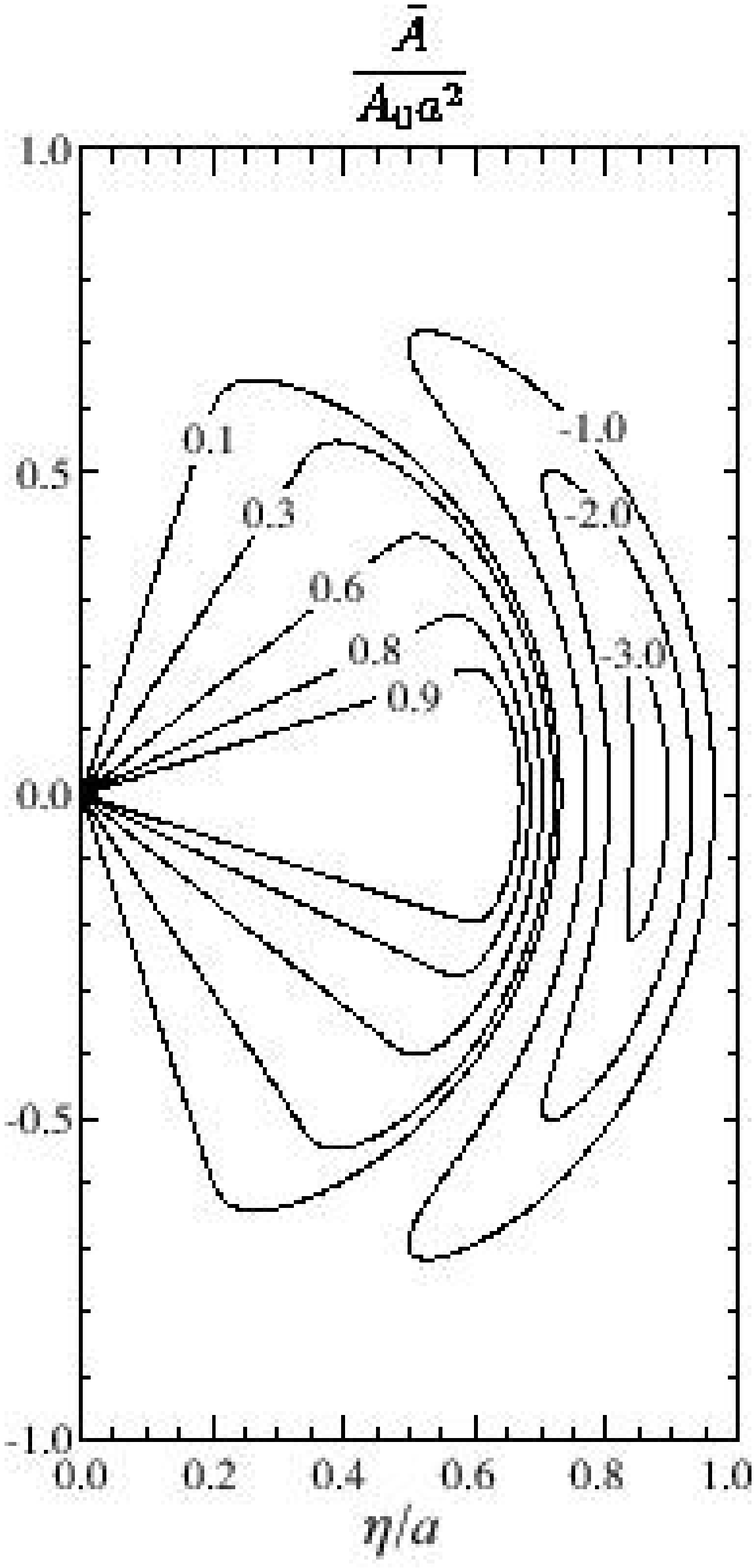} 
   \caption{Contour plots of the flux function $\tilde{A}/(A_0 a^2)$ of the
   shell solution for $k=\pi/[2(a-b)]$ (left panel) and
   $k=3\pi/[4(a-b)]$ (right panel).}
   \label{fig:sol2-klarge}
  \end{minipage}
 \end{tabular}
\end{figure}

Let us consider the MHD waves propagating inside the magnetic loops.
The MHD waves consist of the forward wave $B_\mathrm{for}$ and the
reflected wave $B_\mathrm{ref}$
($B_\mathrm{for}$ and $B_\mathrm{ref}$ are the magnetic fields in the
poloidal plane).
When the wave $B^{III}_\mathrm{leak}$ is transmitted to region III (see
Fig.~\ref{fig:leakingwave}), the electromagnetic energy
can be converted to the kinetic and thermal energies in region III. The
magnetic field $B_\theta$ given in equation (\ref{sol2:Blimit}) can be
expressed by the superposition of the forward and reflected waves.
When the density enhancement appears ahead of the magnetic loop in
region II, the forward waves can be partially reflected by it. The
condition for the perfect reflection should be determined by the
wavelength $\lambda$ and the thickness of the density enhancement $d\sim
(a-b)t$. This situation is analogous to the enhancement of the
reflection rate by coating a glass with dielectric medium. The
reflection rate becomes maximum when the width of the dielectric medium
$d$ satisfies $d=(2 l+1)\lambda/4$. When $\lambda=2\pi/k$, this
condition coincides with equation (\ref{sol2:knowave}). Note that the
parameter $k$ in equation (\ref{sol2:knowave}) is not exactly the
wave number but it determines the profile of the magnetic fields
(see equations (\ref{sol2:A}), (\ref{sol2:Lambda}), and (\ref{func:T})
for the definition of $k$). When the condition (\ref{sol2:knowave}) is
satisfied, the MHD waves propagating in the $+r$ direction are totally
reflected by the density enhancement produced by the loop expansion.
For the dipolar solution, the magnetic energy is transmitted to $r>R(t)$
because the density enhancement does not appear (see the right
panels of Fig.~\ref{fig:sol1A} and Fig.~\ref{fig:sol1Q}).

When $k>\pi/[2(a-b)]$, the magnetic shell recedes from $\eta \sim a$ to
the region $b < \eta < a$, and a flux rope appears around $\eta =a$ ahead of
the magnetic shell. Fig.~\ref{fig:sol2-klarge} shows the contour plots
of the magnetic flux $\tilde A$ of the shell solutions for
$k=\pi/[2(a-b)]$ (left panel) and $k=3\pi/[4(a-b)]$ (right panel).

\subsection{Flux Rope Solutions}\label{sss3}
Flux rope solutions which include flux ropes inside the expanding
magnetic loops are constructed by the flux function (\ref{sol3:A}).
\begin{figure}
 \begin{tabular}{ccc}
  \begin{minipage}{0.33\hsize}
   \includegraphics[width=4.8cm]{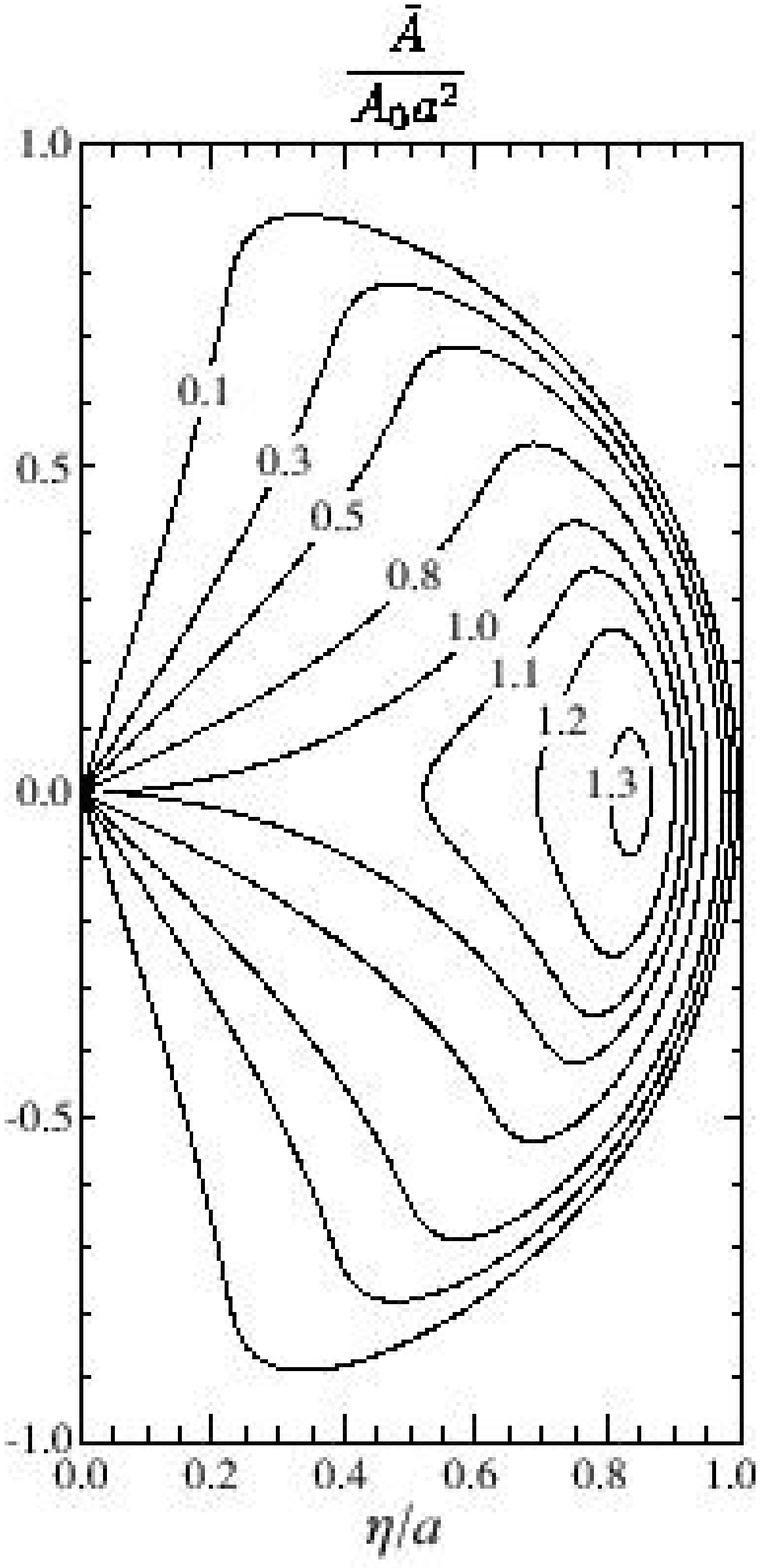}
  \end{minipage}
  \begin{minipage}{0.33\hsize}
   \includegraphics[width=4.8cm]{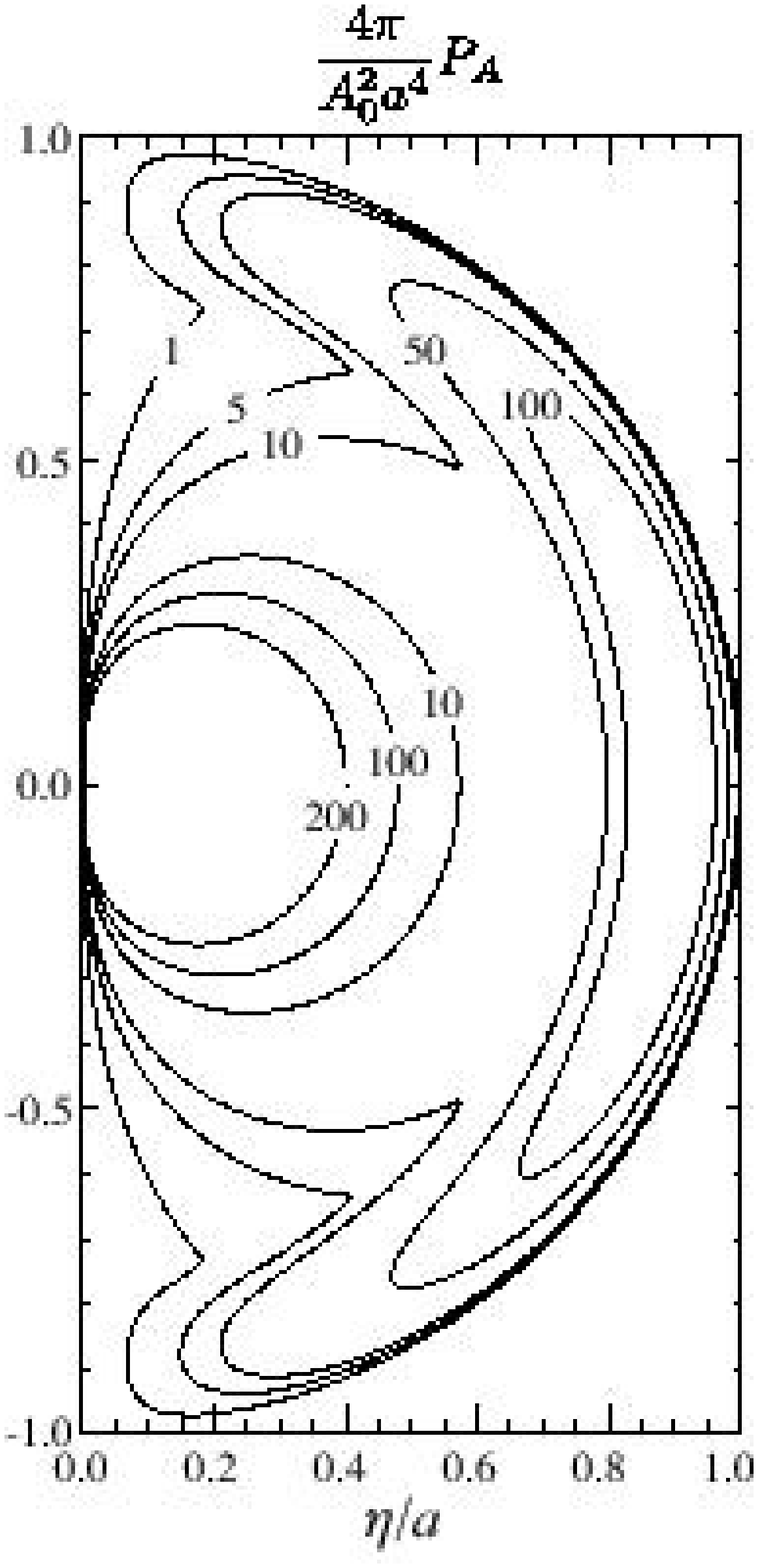}
  \end{minipage}
  \begin{minipage}{0.33\hsize}
   \includegraphics[width=4.8cm]{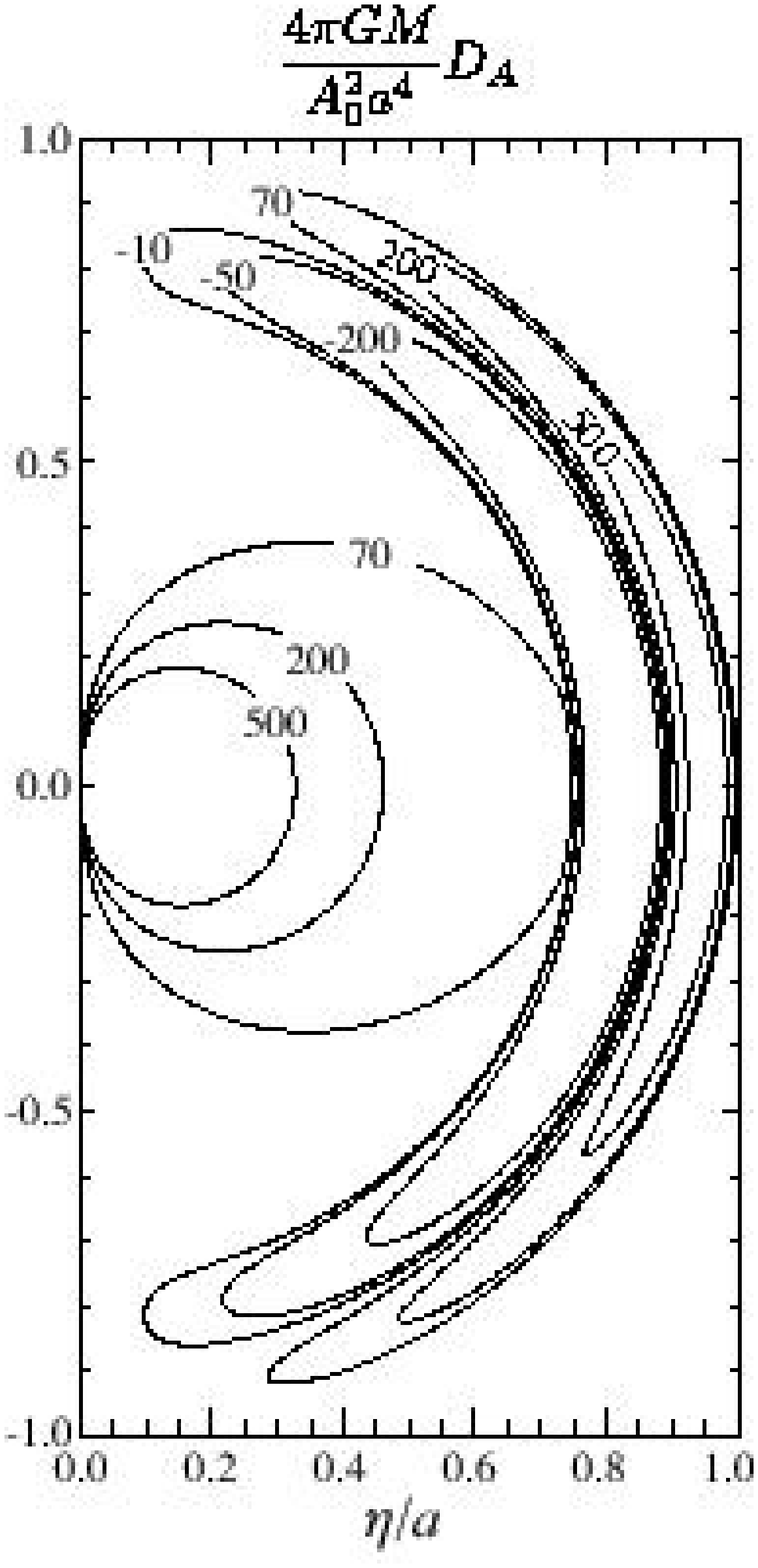}
  \end{minipage}
 \end{tabular}
 \caption{Contour plots of the magnetic flux $\tilde{A}$ (left), the poloidal
 part of the pressure $P_A$ (centre), and that of the gas density
 $D_A$ (right) for the flux rope solution in $\eta/a-\theta$ plane when
 $a=0.8, b=0.75a$.}
 \label{fig:sol3A}
\end{figure}

By substituting equation (\ref{sol3:A}) into equation
(\ref{eq:eomphi2}), the function $Q$ can be written as
\begin{equation}
 Q^I(\eta,\theta)=\sum_n \frac{Q_{0,n}}{(1-\eta^2)^{1+\frac{n}{4}}}
  \sin^n\theta, \label{sol3:Q1}
\end{equation}
\begin{equation}
 Q^{II}(\eta,\theta)=\sum_n \frac{Q_{0,n}}{(1-\eta^2)^{1+\frac{n}{4}}}
  \Lambda^{\frac{n}{2}}(\eta)\sin^n\theta, \label{sol3:Q2}
\end{equation}
where $Q_{0,n}$ is a constant and subscripts $I$ and $II$ denote region
I and region II, respectively.
The pressure and density functions (i.e., $P$ and $D$) can be obtained by
substituting equations (\ref{sol3:A}), (\ref{sol3:Q1}), (\ref{sol3:Q2})
into equations (\ref{eq:eomr2}) and (\ref{eq:eomtheta2}).
The functions $P$ and $D$ obtained in region I and region II are given
in appendix \ref{ap-fluxrope}.

The pressure and the gas density consist of three
parts, the isotropic part $P_0$ and parts representing the interaction
with the electromagnetic force by the poloidal and toroidal magnetic
fields, $P_A$ and $P_Q$ (see equations (\ref{sol3:P1}) and
(\ref{sol3:P2}) for the pressure and equations (\ref{sol3:D1}) and
(\ref{sol3:D2}) for the gas density).

\begin{figure}
 \begin{tabular}{ccc}
  \begin{minipage}{0.33\hsize}
   \includegraphics[width=4.8cm]{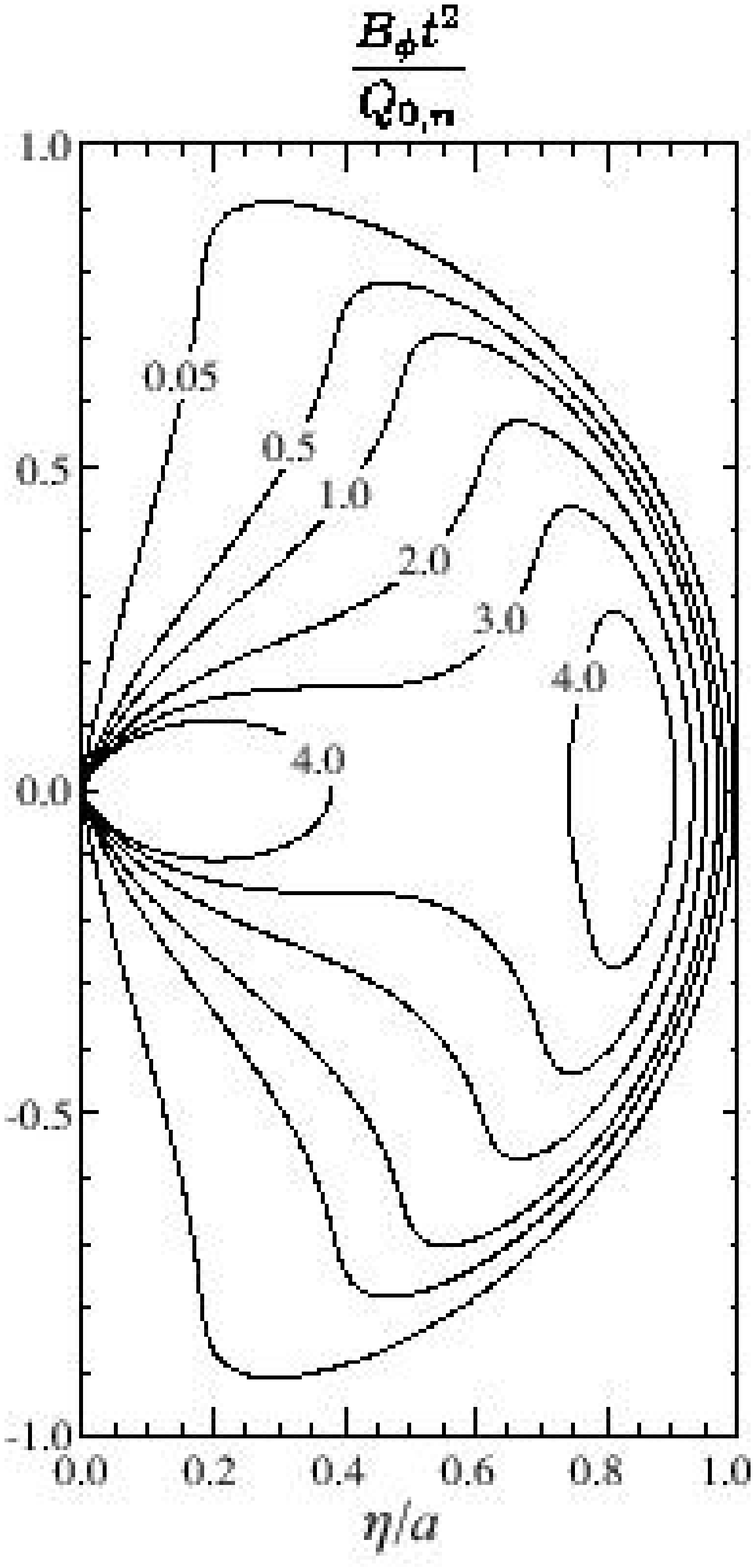}
  \end{minipage}
  \begin{minipage}{0.33\hsize}
   \includegraphics[width=4.8cm]{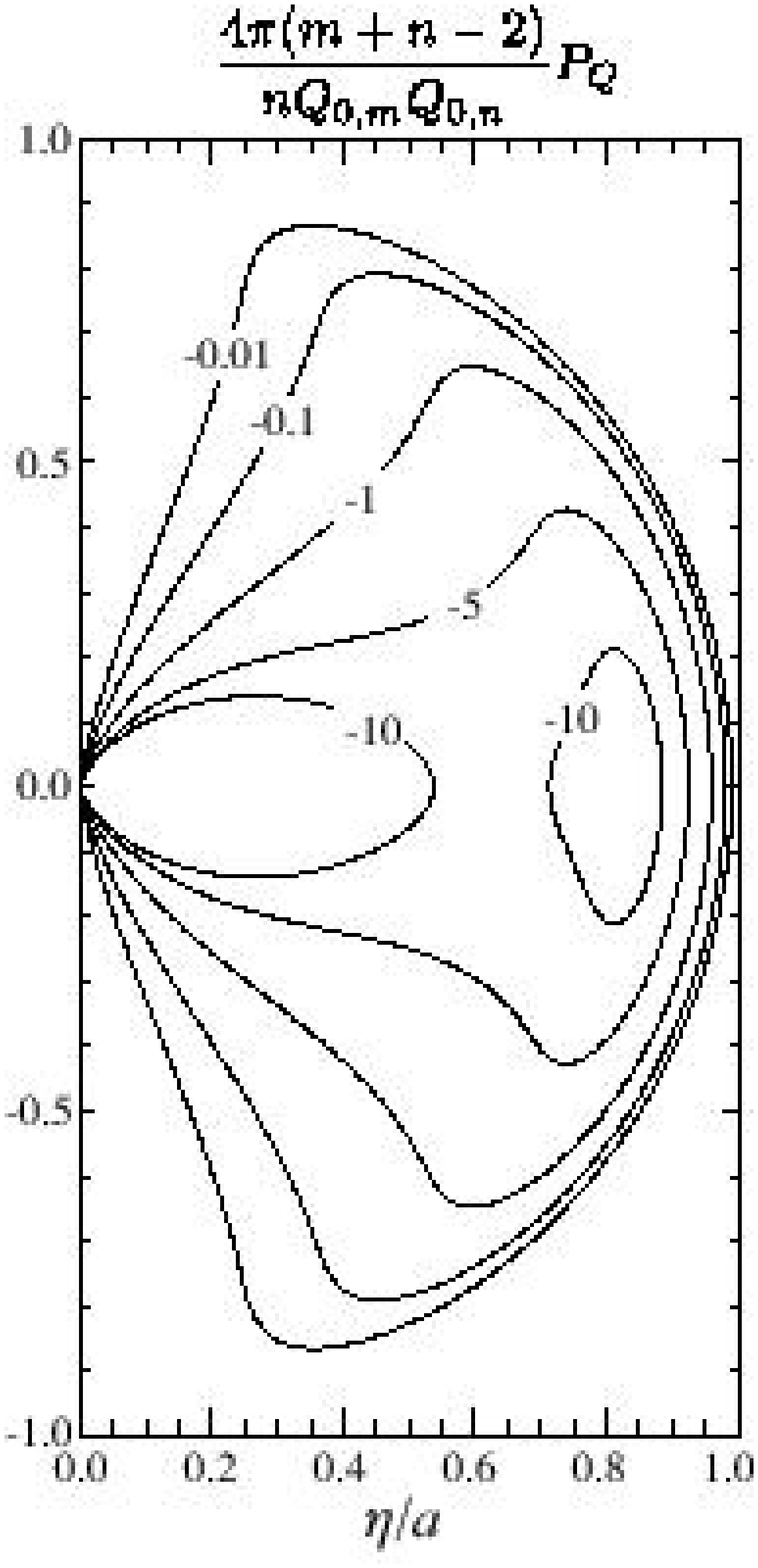}
  \end{minipage}
  \begin{minipage}{0.33\hsize}
   \includegraphics[width=4.8cm]{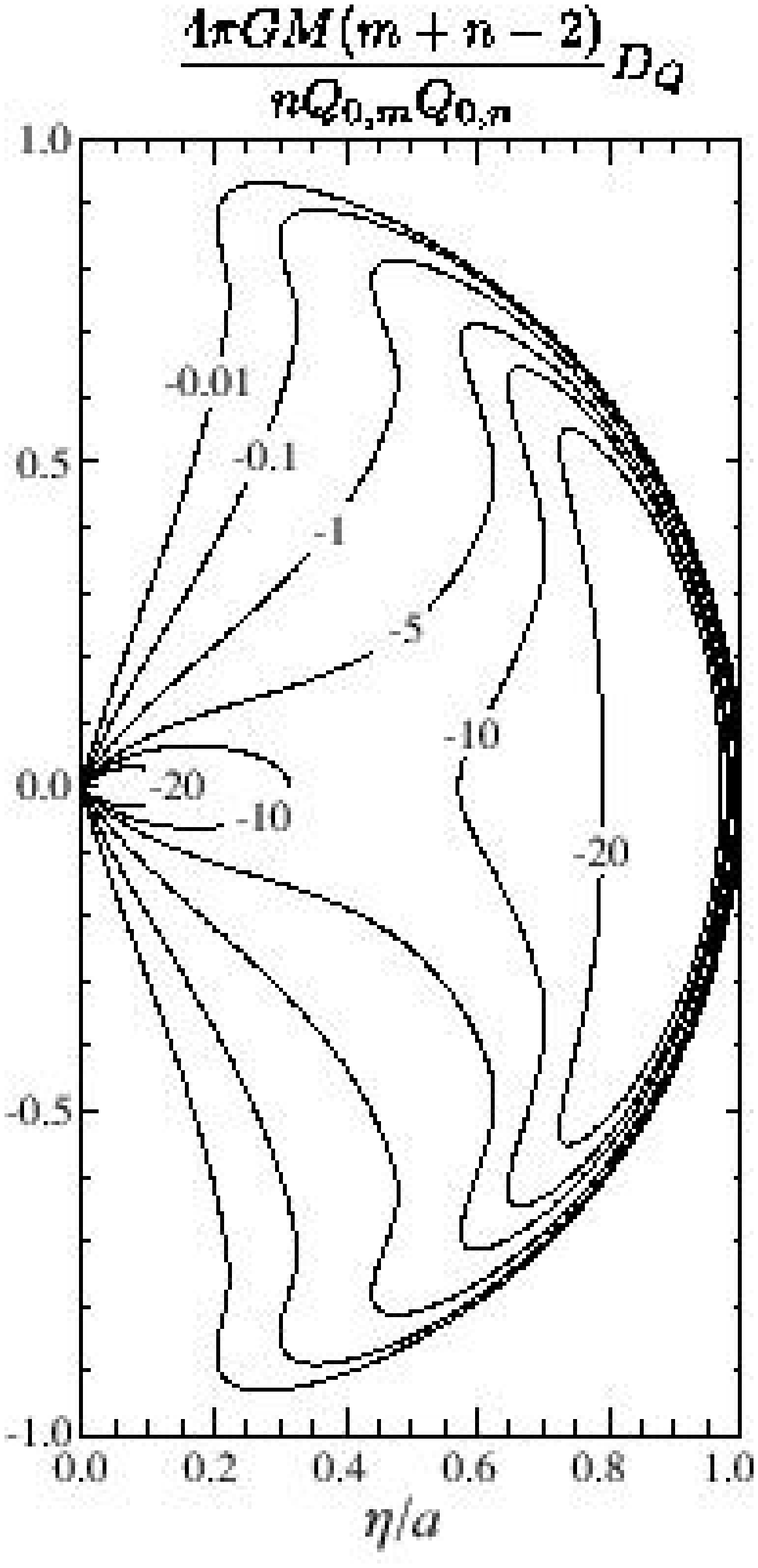}
  \end{minipage}
 \end{tabular}
 \caption{Contour plots of the toroidal magnetic field $B_\phi$ (left),
 the toroidal part of the pressure $P_Q$ (centre), and that of the gas
 density $D_Q$ (right) for the flux rope solution in $\eta/a-\theta$
 plane when $a=0.8, b=0.75a$, and $n=m=4$ .}
 \label{fig:sol3Q}
\end{figure}

Fig.~\ref{fig:sol3A} shows the contour plots of the magnetic flux
$\tilde{A}$ (left), the poloidal part of the pressure $P_A$
(centre), and that of the gas density $D_A$ (right) in
$\eta/a-\theta$ plane. Fig.~\ref{fig:sol3Q} shows contour plots of the
toroidal magnetic field $B_\phi$ (left), the toroidal part of the
pressure $P_Q$ (centre), and that of the gas density $D_Q$ (right) in
$\eta/a-\theta$ plane. The parameters are taken to be $a=0.8$,
$b=0.75a$, and $m=n=4$ in both figures. The flux ropes exist behind the
loop top (see the left panel of Fig.~\ref{fig:sol3A}).

The magnetic fields in region I and II are explicitly given by
\begin{equation}
 B^{I}_r=\frac{2A_0a^2}{r^2\sqrt{1-\left(r/t\right)^2}}\cos\theta,
\end{equation}
\begin{equation}
 B^{I}_\theta=-\frac{A_0a^2}{t^2 [1-(r/t)^2]^\frac{3}{2}}\sin\theta,
\end{equation}
\begin{equation}
 B^{I}_\phi=\sum_n\frac{Q_{0,n}}{rt[1-(r/t)^2]^{1+\frac{n}{4}}}\sin^{n-1}\theta,
\end{equation}
\begin{equation}
 B^{II}_r=\frac{2A_0a^2}{r^2}\frac{\Lambda\left(r/t\right)}
  {\sqrt{1-\left(r/t\right)^2}}\cos\theta, \label{sol3:Br2}
\end{equation}
\begin{equation}
 B^{II}_\theta=\frac{A_0
  a^2}{t^2}\frac{1}{\left[1-\left(r/t\right)^2\right]^{\frac{3}{2}}}\left\{4
		     \frac{tk}{r}\left[1-\left(r/t\right)^2\right]\frac{\sin^3(T(r/t))\cos(T(r/t))}{\sin^4
		     (T(a))}-\Lambda\left(r/t\right)\right\}\sin\theta,\label{sol3:Btheta2}
\end{equation}
\begin{equation}
 B^{II}_\phi=\sum_n\frac{Q_{0,n}}{rt}
  \frac{\Lambda^{\frac{n}{2}}\left(r/t\right)}
  {\left[1-\left(r/t\right)^2\right]^{1+\frac{n}{4}}}\sin^{n-1}\theta.\label{sol3:Bphi2}
\end{equation}
Here the subscripts $I$ and $II$ denote the magnetic fields in
region I and region II, respectively.
Similarly to the dipolar and shell solutions, the parameter $m$ 
represents the Fourier modes which specify where the magnetic twist is
injected.

In the limit that $t\gg r$, the flux rope solution reduces to the
dipolar solution given by equation (\ref{sol1:Blimit}),
(\ref{sol1:plimit}) and (\ref{sol1:rholimit}). 
In this limit, the magnetic field becomes stationary and radial.

Note that the field component $B_r$ and $B_\theta$ are exactly zero but
$B_\theta$ is not zero at $r=R(t)$ unless the condition
(\ref{sol2:knowave}) is satisfied. 
The electromagnetic energy is transmitted to $r>R(t)$ unless the
condition (\ref{sol2:knowave}) is satisfied
as discussed in the previous subsection. When equation
(\ref{sol2:knowave}) is satisfied, the Poynting flux is totally reflected at
$r=R(t)$ and the electromagnetic energy is not transmitted to $r \gid R(t)$.

\section{Physical Properties}\label{physprop}
Here we discuss physical properties of the three solutions we derived in
$\S$ \ref{sss}.
In this section, we organize our discussion into
four parts. First we consider the energetics. Second we show the shell and
flux rope structures derived in $\S$ \ref{sss2} and $\S$ \ref{sss3}
inside the magnetic loops. Third we study the relativistic effects,
especially the role of the displacement current. Finally, we apply our
solutions to SGR flares.

\subsection{Energetics}\label{energetics}
First let us consider the dipolar solution without the toroidal magnetic
field (i.e., $Q_{0,n}=0$) for simplicity.
Total energy $\mathcal{E}$ contained inside the expanding magnetic loops is
given as
\begin{equation}
 \mathcal{E}=K+U_\mathrm{in}+U_\mathrm{th}+U_\mathrm{E}+U_\mathrm{M}+W,\label{en:tot}
\end{equation}
where
\begin{equation}
 K=\int_{V} dV \rho \gamma^2,\label{en:kinetic}
\end{equation}
\begin{equation}
 U_{\mathrm{in}}=\int_V d V \frac{\Gamma}{\Gamma-1}\gamma^2 v^2
  p,\label{en:iner}
\end{equation}
\begin{equation}
 U_{\mathrm{th}}=\int_V d V \frac{p}{\Gamma-1}, \label{en:ther}
\end{equation}
\begin{equation}
 U_{\mathrm{E}}=\int_V d V \frac{\bmath E^2}{8\pi},\label{en:ele}
\end{equation}
\begin{equation}
 U_{\mathrm{M}}=\int_V d V \frac{\bmath B^2}{8\pi},\label{en:mag}
\end{equation}
\begin{equation}
 W=-\int_V d V \frac{G M\gamma\rho}{r},\label{en:grav}
\end{equation}
are kinetic, thermal inertial, thermal, electric, magnetic, and
gravitational potential energies, respectively.
Since the solutions we derived describe the freely expanding magnetic
loops, i.e., $Dv/Dt=0$, the total kinetic energy $K$ given by 
\begin{equation}
K=\int_V \rho\gamma^2 d V=\int^{a}_0 d \eta
 \int_0^{2\pi} d \theta \int_0^\pi d \phi\ 
 \frac{\eta^2 D(\eta,\theta)}{\sqrt{1-\eta^2}} \sin\theta,
 \label{en:kinetic2}
\end{equation}
does not change with time. Here the total kinetic energy is integrated
inside the spherical surface of $r = R(t)$.
Other energies can be evaluated by carrying out the integration
directly. The non-kinetic part of the total energy
$\mathcal{E}'\equiv
U_{\mathrm{in}}+U_\mathrm{th}+U_\mathrm{E}+U_\mathrm{M}+W$ contained
inside $r=R(t)$ is then given as
\begin{equation}
 \mathcal{E}'=\frac{4A_0^2 a^3}{3 t}.  \label{en:totdipole}
\end{equation}
\begin{figure}
 \begin{center}
 \includegraphics[width=12.0cm]{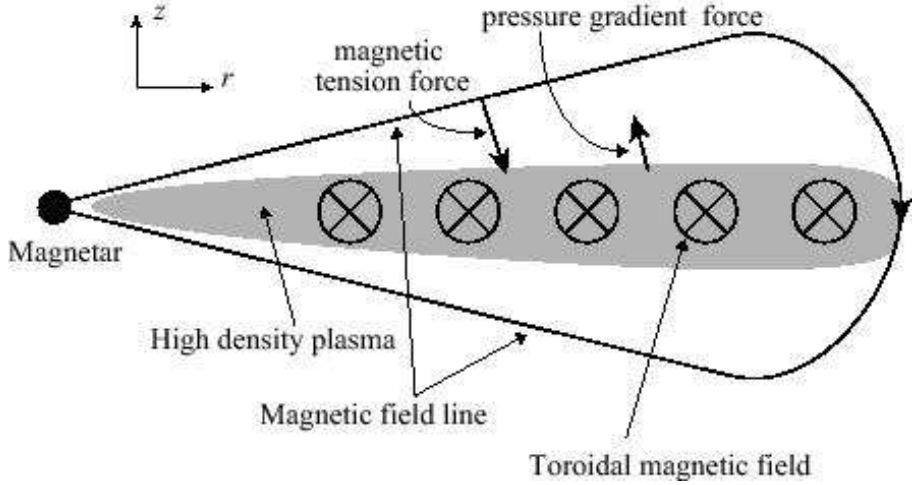}
 \caption{A schematic picture showing an expanding magnetic loop. 
  Toroidal magnetic field is created inside the magnetic loop due to
  the twist injection from the surface of the magnetar. The magnetic
  pressure gradient force plus the pressure gradient force balances with
  the magnetic tension force by the poloidal magnetic
  field.}\label{fig:cartoon1}
 \end{center}
\end{figure}

Since the thermal, gravitational potential, and magnetic energies
contain infinity due to the divergence of $p$, $\rho$, and $B_r$ at
$r=0$ (see equations (\ref{sol1:PA}), (\ref{sol1:DA}), and
(\ref{sol1:Bs}), respectively), we renormalized the infinite parts of
$U_\mathrm{th}$, $U_\mathrm{M}$, and $W$ to zero
\citep[see][]{1982ApJ...261..351L}.

The non-kinetic part of the total energy $\mathcal E'$ depends on the
amplitude of the
poloidal magnetic field $A_0$, but is independent of the isotropic
component (i.e., $P_0$ and $D_0$). The isotropic component does not
contribute
to the total energy because the thermal energy of the isotropic plasma
cancels with that of the gravitational potential energy.

The energy $\mathcal{E}'$ diverges at $t=0$ because we assumed a
point mass at the origin. 
In magnetars, since the magnetar has a finite radius $R_s$,
the self-similar expansion will take place when $r>r_0>R_s$ and $t>t_0$. Let us
denote the total energy and the non kinetic part of the total energy
inside the spherical surface of $r_0 \equiv R(t_0)$ as $\mathcal{E}_0$ and
$\mathcal{E}'_0$, respectively. The expansion takes place when
$\mathcal{E}_0>\mathcal{E}'_0$. Since the total kinetic energy
$K=\mathcal{E}_0 -\mathcal{E}'_0$ does not change with time and
$\mathcal{E}'(t)$ given in equation (\ref{en:totdipole}) decreases with time
for the dipolar solution,
$\mathcal{E}(t)=\mathcal{E}'(t)+K<\mathcal{E}'_0+K=\mathcal{E}_0$. The
released energy $\mathcal{E}_0-\mathcal{E}(t)$ is carried away to $r>R(t)$.
This can be confirmed by integrating the energy conservation
equation inside the spherical surface $r=R(t)$ as
\begin{equation}
 \int_{V(t)} \frac{\partial}{\partial t}\left[(\rho+4
       p)\gamma^2-p+\frac{\bmath{E}^2+\bmath{B}^2}{8\pi}-\frac{GM\rho\gamma}{r}\right]
d^3 \bmath r+\int_{V(t)} \nabla \cdot \left[(\rho+4p)\gamma^2 \bmath v+\frac{\bmath
		      E\times \bmath
		      B}{4\pi}-\frac{GM\rho\gamma}{r}\bmath
		      v\right]d^3 \bmath r=0.\label{eq:energycons}
\end{equation}
Since this integration is cumbersome, we do not show the details of the
calculation. We have to point out that the integration
with the volume $V$ cannot be exchanged with the time derivative in the first
term since the volume $V$ changes with time.

\begin{figure}
 \center
 \includegraphics[width=10.0cm]{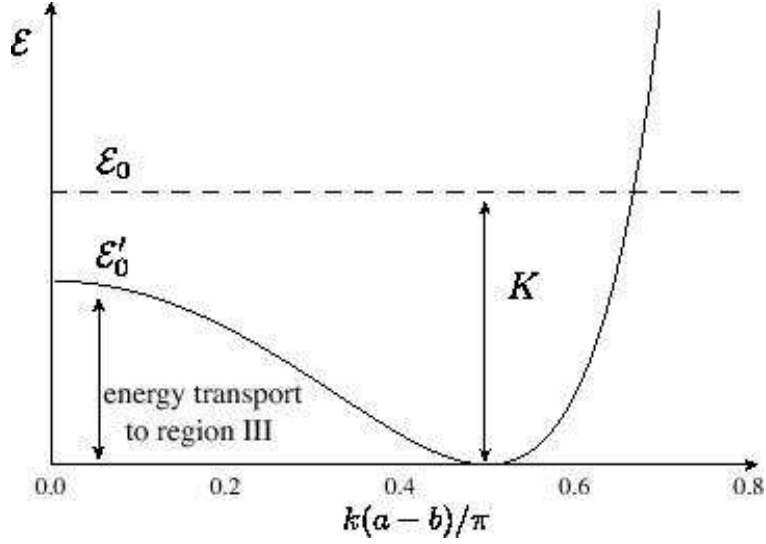}
 \caption{$k$ dependence of the non kinetic part of the total energy
 $\mathcal{E}'$ (solid curve) for the shell and flux rope solutions.}
 \label{fig:Ekdependence}
\end{figure}
Next let us consider the case $Q_{0,n}\ne 0$.
Since the integration of equations (\ref{en:kinetic})-(\ref{en:grav}) is
complex, we evaluate the total energy inside the closed boundary by
a different method.
Note that equation (\ref{eq:reducedeom}), which represents equations of
motion in the self-similar space, indicates that the self-similar
equations we derived are closely related to the
static relativistic MHD solutions except the existence of the
thermal inertial term and the electric field. We can derive the
virial theorem for the relativistic self-similar MHD (see appendix
\ref{ap1});
\begin{equation}
 3(\Gamma-1)U_{\mathrm{th}}+U_{\mathrm{in}}+U_{\mathrm{M}}+U_{\mathrm{E}}+W
  =\mathcal{H}+\mathcal{S},\label{eq:virial}
\end{equation}
where
\begin{equation}
\mathcal{H}=\int p \bmath r \cdot d \bmath{\mathcal{A}}
 -\frac{1}{8\pi}\int
 \left\{2\left[(\bmath r \cdot \bmath E)(\bmath E\cdot
	 d \bmath{\mathcal{A}})+(\bmath r \cdot \bmath B)(\bmath B\cdot
	 d \bmath{\mathcal{A}})
\right]-(\bmath E^2+\bmath B^2)(\bmath r\cdot d \bmath{\mathcal{A}})\right\},\label{en:surface}
\end{equation}
and
\begin{equation}
 \mathcal{S}=\int dV \frac{\partial }{\partial t}
  \left(\bmath r \cdot \frac{\bmath E\times \bmath B}{4\pi}\right).\label{en:poynting}
\end{equation}
Here $\bmath{\mathcal{A}}$ is the surface enclosing the volume $V$.
The non-kinetic part of the total energy $\mathcal{E}'$ can be written
from equation (\ref{eq:virial}) as
\begin{equation}
 \mathcal{E}'=-(3\Gamma-4)U_\mathrm{th}+\mathcal{H}+\mathcal{S}.\label{en:totalEvirial}
\end{equation}
We can evaluate $\mathcal{E}'$ inside the expanding spherical surface of
$r=R(t)$ by using the fact that $p=\rho=B_r=B_\phi=0$ at $r=R(t)$ as
\begin{equation}
 \mathcal{E}'=\left\{\begin{array}{lll}
{\displaystyle \frac{4 A_0^2 a^3}{3 t}}, &
    (\mathrm{dipolar \ solution}),\\
{\displaystyle \frac{16 A_0^2 a^5 k^2 (1-a^2) \cot^2 T(a)}{3t}}, &
    (\mathrm{shell\ solution}),\\
{\displaystyle \frac{16 A_0^2 a^5 k^2 \cot^2 T(a)}{3t}}, & 
    (\mathrm{flux \ rope\ solution}), \label{en:energy}
		      \end{array}\right.
\end{equation}
In all solutions, the non-kinetic part of the total energy does not
depend on the toroidal
magnetic field because the toroidal magnetic field does not change the
dynamics of the expanding magnetic loops in the
self-similar stage. This can be understood from the fact that when we
take $Q_{0,n}=0$, the
solutions we derived satisfy equations (\ref{eq:eomr2}),
(\ref{eq:eomphi2}) and (\ref{eq:eomtheta2}) without any modification
on the poloidal magnetic field. To understand
this reason, let us consider the
equation of motion in the $\theta$ direction. Since $\bmath v=v \bmath
e_r$, the force balance should be attained in the $\theta$ direction (see
Fig. \ref{fig:cartoon1}). 
The pressure $P_Q$ is smaller for 
larger toroidal magnetic fields because $P_Q$ is
proportional to $-Q_{0,n}^2$ (see equations
(\ref{sol1:PQ}) for the dipolar solution,
(\ref{sol2:PQ1}) and (\ref{sol2:PQ2}) for the shell solution, and
(\ref{sol3:PQ1}) and (\ref{sol3:PQ2}) for the flux rope solution).
For larger toroidal magnetic fields, the magnetic pressure gradient
force by the toroidal
magnetic field balances with the magnetic tension force from the poloidal
magnetic field. As a result, the existence of the toroidal magnetic
field modifies the plasma distribution, but does not change the dynamics.

Fig.~\ref{fig:Ekdependence} shows the $k$ dependence of $\mathcal{E}_0'$
for the shell and flux rope solutions (note that both solutions have the
same $ k $ dependence).
When the condition (\ref{sol2:knowave}) is satisfied, the
total energy contained inside the magnetic loops is
equal to $K$ and conserved for the shell and flux rope solutions because the
energy flux is zero at $r=R(t)$.
When (\ref{sol2:knowave}) is not satisfied, the total energy is larger
than $K$ by $\mathcal{E}'$ (see equation (\ref{sol2:knowave})). The
excess energy is carried away to the region III ($r>R(t)$) to attain the
free expansion, i.e., $Dv/Dt=0$.

\begin{figure}
 \begin{tabular}{ccc}
  \begin{minipage}{0.49\hsize}
   \includegraphics[width=8.0cm]{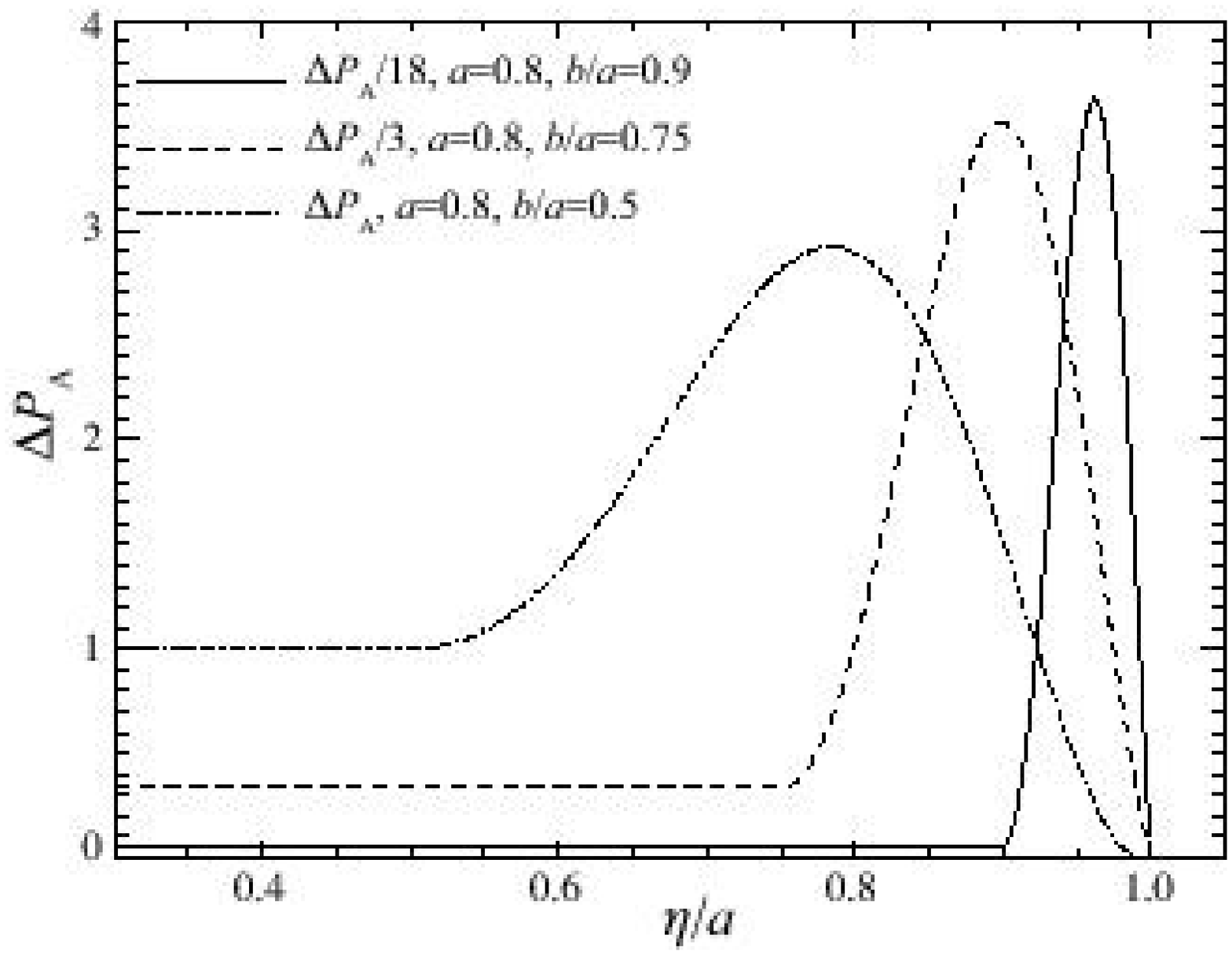}
  \end{minipage}
  \hspace*{1mm}
  \begin{minipage}{0.49\hsize}
   \includegraphics[width=8.0cm]{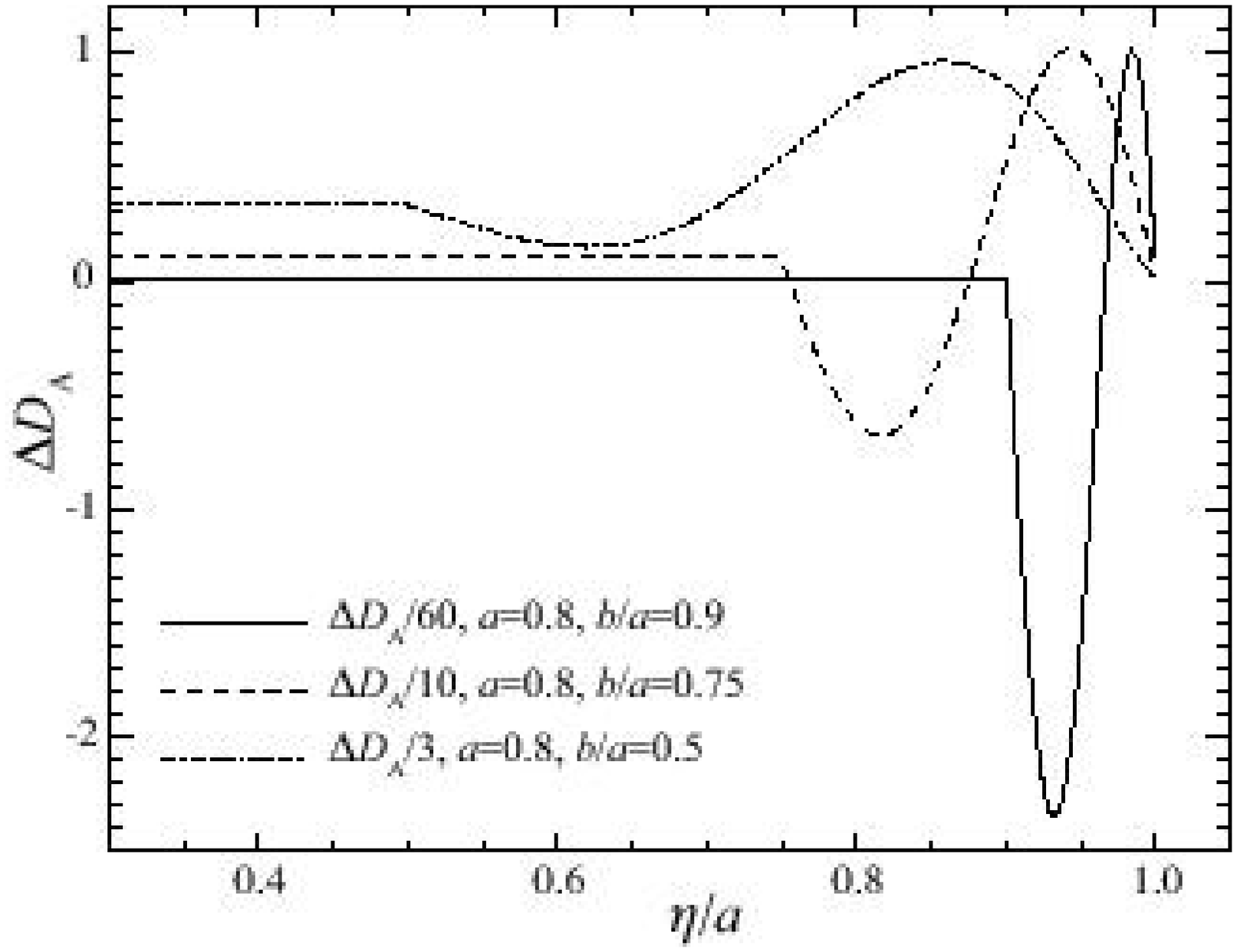}
  \end{minipage}
 \end{tabular}
 \caption{The pressure enhancement $\Delta P_A$ (left) and the density
 enhancement $\Delta D_A$ (right) for the shell solution are shown as a
 function of $\eta/a$.
 Solid curve is for $b=9a/10$, while the dashed and dot-dashed ones
 are for $b=3a/4$ and $b=a/2$, respectively. Other parameters are fixed at
 $a=0.8$ and $k=\pi/[4(a-b)]$.}\label{fig:sol2enhanceA}
\end{figure}

\subsection{Shell and Flux Rope Structures}\label{shellstructure}
Let us examine the density and pressure distribution for the shell
solution derived in $\S$ \ref{sss2}.
We define the density and pressure enhancements as 
\begin{equation}
 \Delta P_A=\left[\frac{2\pi \eta^4}{A_0^2 a^4 \sin^2\theta}\right]P_A,
  \label{sol2:pAenhance}
\end{equation}
\begin{equation}
 \Delta D_A = \left[\frac{\pi G M \eta^3 (1-\eta^2)}{2
		   A_0^2 a^4 \sin^2\theta} \right]D_A, \label{sol2:rhoAenhance}
\end{equation}
\begin{equation}
 \Delta P_Q=\left[\frac{4\pi (m+n-2)\eta^2(1-\eta^2)}{n Q_{0,m}Q_{0,n}
	     \sin^{m+n-2}\theta}\right] P_Q, \label{sol2:pQenhance}
\end{equation}
\begin{equation}
 \Delta D_Q =\left[\frac{2\pi G M (m+n-2)\eta
	      (1-\eta^2)^{2}}{n
	      Q_{0,m}Q_{0,n}\sin^{m+n-2}\theta}\right]D_Q,\label{sol2:rhoQenhance}
\end{equation}

These functions are normalized to be unity in region I, where the
poloidal magnetic field lines are radial for the shell solution.
 In Fig.~\ref{fig:sol2enhanceA}, the pressure and density
enhancements, $\Delta P_A$ and $\Delta D_A$ are plotted for
$b=a/2,~ 3a/4,~ 9a/10$ when $a=0.8$ and $k=\pi/[4(a-b)]$.  In all three
cases, the pressure and density pulses appear at the top of the magnetic
loops. Their amplitudes are larger for a thinner shell.
The peak of the pressure enhancement appears behind that of the
density enhancement. This structure comes from the requirement for the
force balance with the gravity. As mentioned in $\S$
\ref{energetics},
this relativistic self-similar solution is similar to the static
solution in which the force balance is attained. As plasma is swept up into
the shell, the density increases inside the shell. To support the
gravity by this excess density, the pressure gradient appears behind the
density enhancement.
The density decrease behind the pressure enhancement also comes from
the requirement for the force balance. Since the decrease of the density
enables the buoyancy force to push the plasma in the radial direction,
this buoyancy force maintains the pressure pulse. These structures are
identical to those in non-relativistic solution
\citep{1982ApJ...261..351L}.

\begin{figure}
 \begin{tabular}{ccc}
  \begin{minipage}{0.49\hsize}
   \includegraphics[width=8cm]{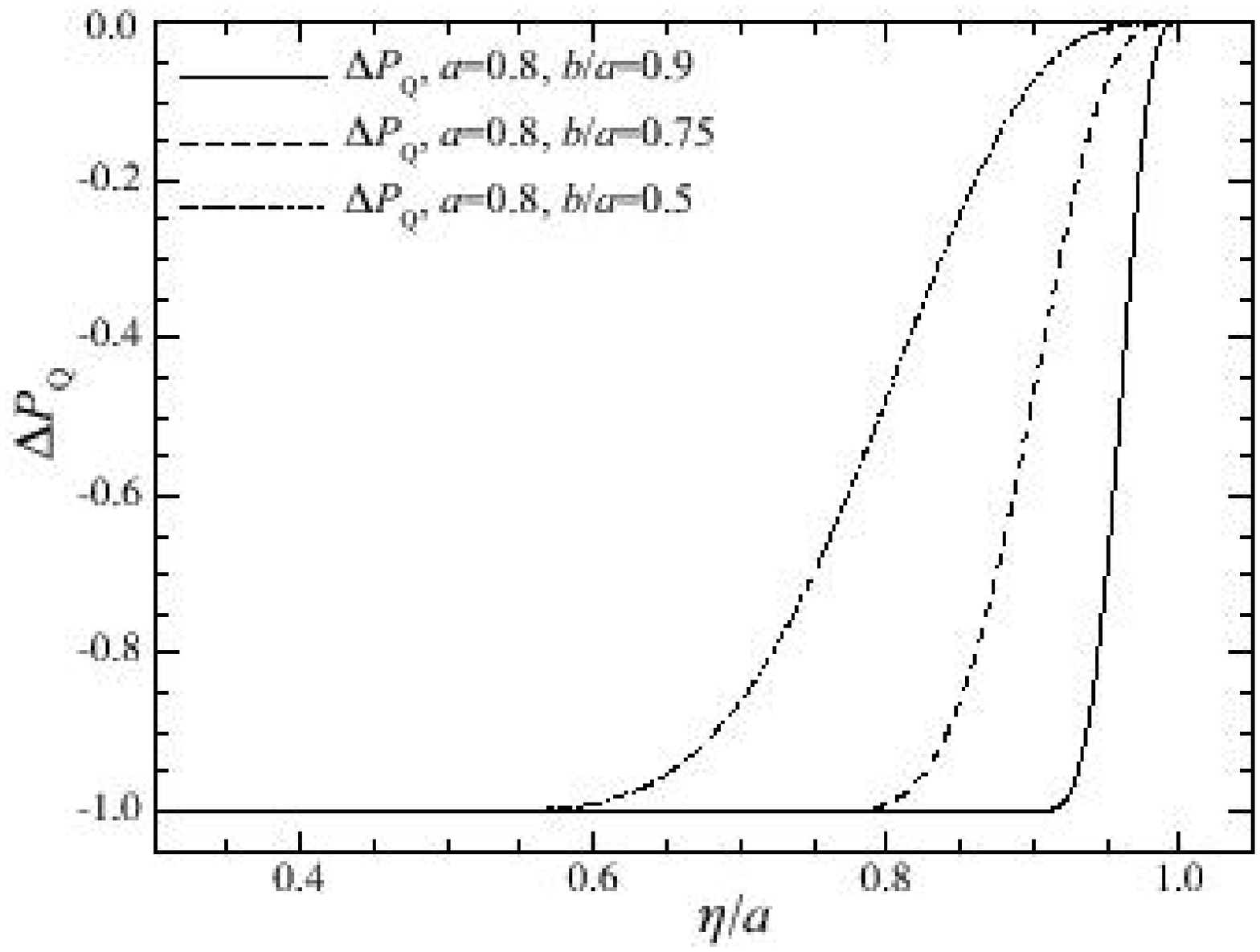}
  \end{minipage}
  \hspace*{1mm}
  \begin{minipage}{0.49\hsize}
   \includegraphics[width=8cm]{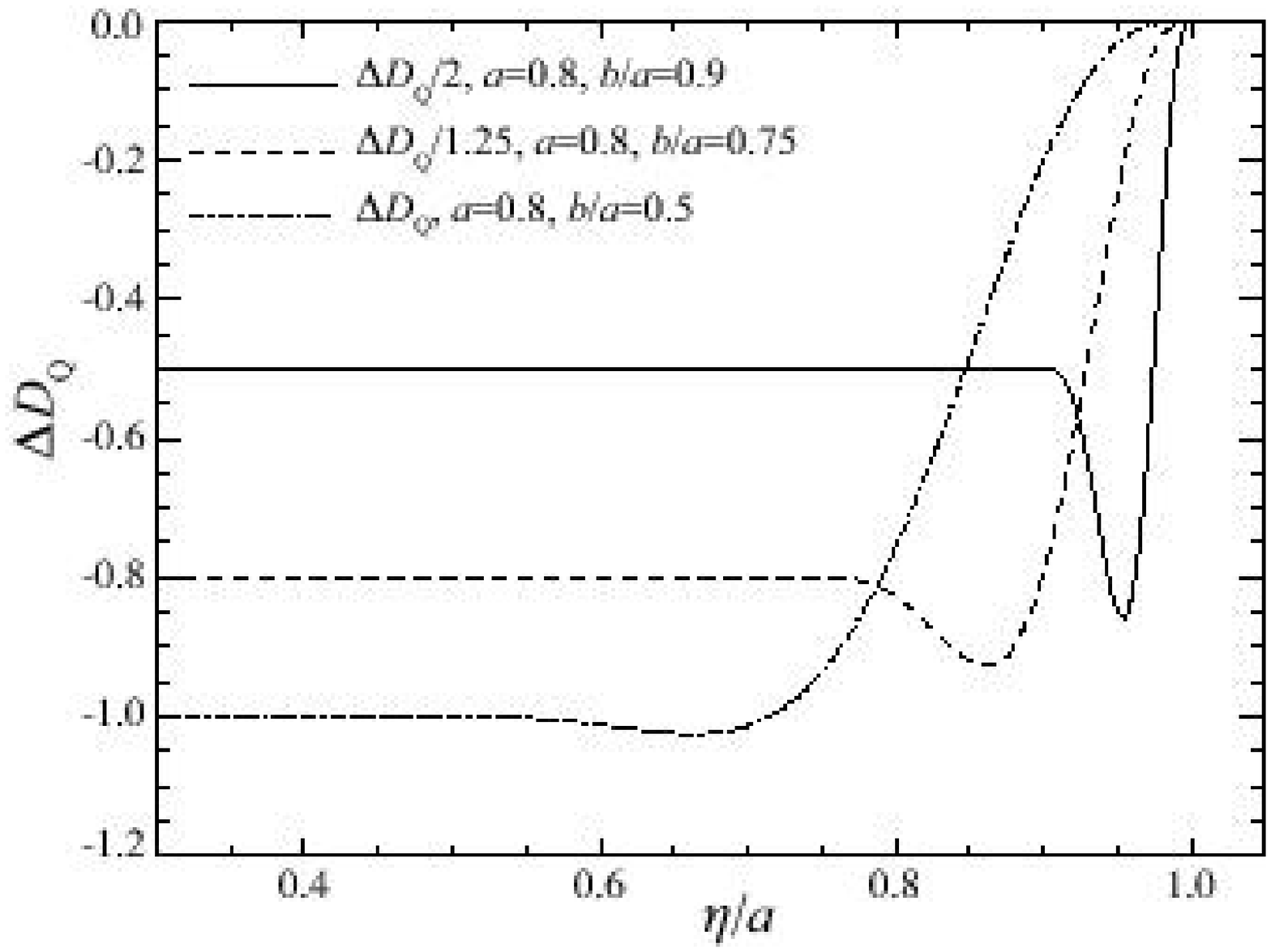}
  \end{minipage}
 \end{tabular}
 \caption{The pressure decrement $\Delta P_Q$ (left) and the density
 decrement $\Delta D_Q$ (right) by the azimuthal field are shown as a
 function of $\eta/a$ for the shell solution.
 Solid curve is for $b=9a/10$, while dashed and dot-dashed curves are for
 $b=3a/4$ and $b=a/2$, respectively. Other parameters are fixed at
 $a=0.8$ and $k=\pi/[4(a-b)]$. }\label{fig:sol2enhanceQ}
\end{figure}

Fig. \ref{fig:sol2enhanceQ} plots $\Delta P_Q$ and $\Delta D_Q$ for
$b=a/2,~3a/4,~9a/10$ when $a=0.8$ and $k=\pi/[4(a-b)]$. As
mentioned in $\S$ \ref{sss}, the Lorentz force exerted by the toroidal
magnetic fields always reduces the pressure. A local minimum of the
density enhancement $\Delta D_Q$ locates behind a local maximum of
$d\Delta P_Q/d\eta$. This structure also comes from the force balance. Pressure
gradient force balances with the buoyancy force in the rarefied region.

\begin{figure}
 \begin{tabular}{ccc}
  \begin{minipage}{0.49\hsize}
   \includegraphics[width=8cm]{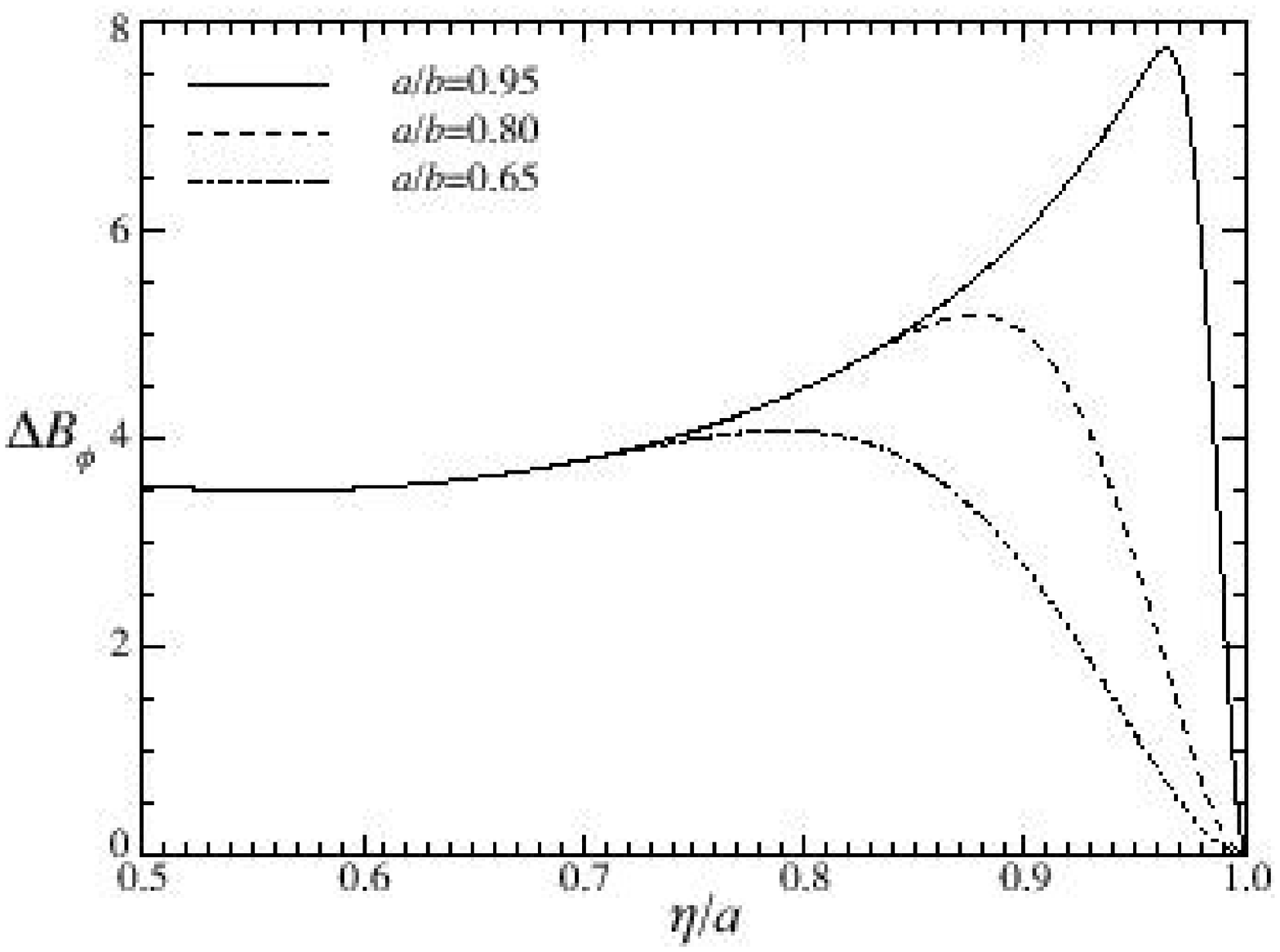}
   \caption{Distribution of $\Delta B_\phi$ for the flux rope solution
   as a function of $\eta/a$ for
   $a=0.8$. Solid curve is for $b/a=0.95$, while dashed and
 dot-dashed ones are for $b/a=0.8$ and $b/a=0.65$, respectively.}
   \label{fig:sol3Qenhance}
  \end{minipage}
  \hspace*{1mm}
  \begin{minipage}{0.49\hsize}
   \includegraphics[width=8cm]{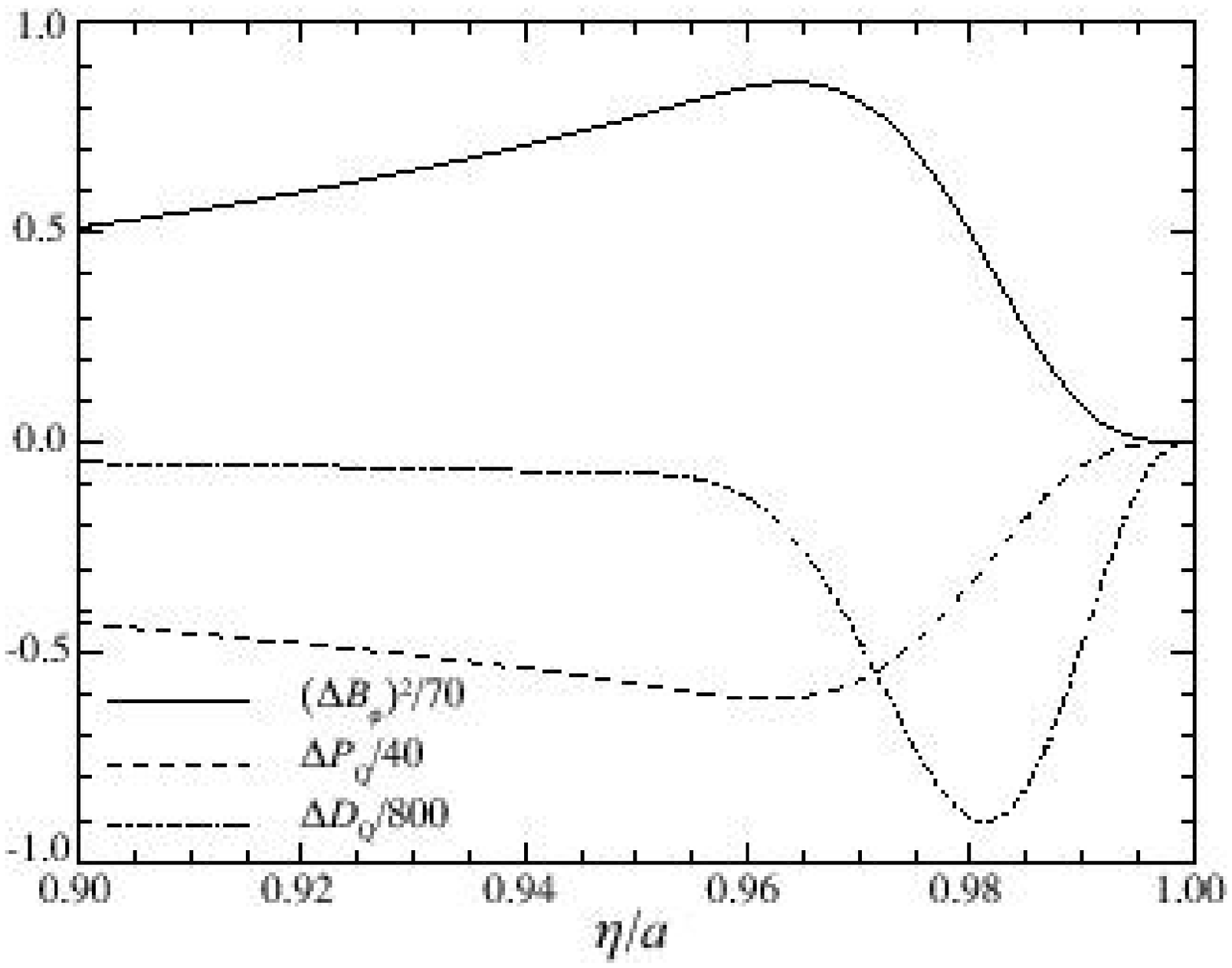}
   \caption{Distribution of $\Delta B_\phi^2$ for the flux rope solution
   as a function of $\eta/a$ for $a=0.8$ and $b/a=0.95$ (solid
   curve). Dashed and dot-dashed ones denote $\Delta P_Q$ and $\Delta
   D_Q$, respectively.}
   \label{fig:sol3QPDenhance}
  \end{minipage}
 \end{tabular}
\end{figure}

Next we examine the structure of the flux rope solution derived in $\S$
\ref{sss3}. We define the normalized toroidal magnetic field strength as
\begin{equation}
 \Delta B_\phi(\eta)=\frac{B_\phi t^2}{Q_{0,n} \sin^{n-1}\theta}.
\end{equation}
Solid curve in Fig. \ref{fig:sol3Qenhance} shows $\Delta B_\phi$ as a
function of $\eta$ for $b=0.95a$, while the dash and dot-dashed ones
show that for $b=0.8a$ and $b=0.65a$, respectively. Other parameters are
fixed at $a=0.8$ and $k=\pi/[4(a-b)]$.
The toroidal magnetic field has a peak inside the flux rope.
Its amplitude is larger for a larger $a$ and a thinner shell. The shell
structure also appears behind the loop top (see Fig.~\ref{fig:sol3A}
 and \ref{fig:sol3Q}).
Solid curve in Fig. \ref{fig:sol3QPDenhance} shows $\Delta B_\phi^2$,
which corresponds to the magnetic pressure by the toroidal magnetic
field, as a function of $\eta$ for $a=0.8$, $b=0.95a$, and
$k=\pi/[4(a-b)]$. Dashed and
dot-dashed curves show $\Delta P_Q$ and $\Delta D_Q$, respectively.
Plasma density decreases inside the shell.  The decrement of the plasma
density leads to the buoyancy force which balances with the pressure
gradient force in front of the shell. Behind the shell, the pressure
gradient force balances with that of the  magnetic pressure. This
effect is more prominent for the flux
rope solution than for the shell solution since the magnetic pressure is
enhanced inside the flux rope.

\subsection{The Role of the Displacement Current}\label{reffect}
We showed that $P_Q$ is always negative.
On the other hand, $P_A$ can have either positive or negative values. In
this subsection, we obtain the condition for $P_A<0$.
 
First let us consider the dipolar solution derived in $\S$ \ref{sss1}.
The condition that $P_A$ given by equation (\ref{sol1:PA})
is positive in $0\lid \eta \lid a$ is given by
\begin{equation}
 2 a^2-3 a^2 \eta^2-\eta^4+2\eta^6\gid 0.\label{sol1:pacond}
\end{equation}
The critical value $a_*$ for $P_A>0$ in $0\lid \eta \lid a$ is 
\begin{equation}
 a_*=\frac{\sqrt{69+11\sqrt{33}}}{12}\simeq 0.958. \label{eq:astar}
\end{equation}
When $a > a_*$, $P_A$ has negative values in the domain $0\lid \eta \lid a$.
Since $a$ denotes the expansion speed of the magnetic loops at $r=R(t)$,
the above condition indicates that $P_A$ can be negative for faster
expansion.

Next let us calculate the azimuthal component of the current density,
\begin{equation}
 j_\phi=j_\mathrm{rot}+j_\mathrm{disp},
\end{equation}
where
\begin{equation}
 j_\mathrm{rot}\equiv\frac{(\nabla\times\bmath B)_\phi}{4\pi}
  =\frac{A_0 a^2}{4\pi r^3}\frac{2 a^2-5 a^2\eta^2+5\eta^4-2\eta^6}{(1-\eta^2)^\frac{5}{2}}\sin\theta,\label{eq:jrotdipole}
\end{equation}
\begin{equation}
 j_\mathrm{disp}\equiv-\frac{1}{4\pi}\frac{\partial E_\phi}{\partial t}
  =-\frac{A_0 a^2}{4\pi r^3}\frac{\eta^4\left(6-3 a^2-5 \eta^2+2 \eta^4\right)}{(1-\eta^2)^\frac{5}{2}}\sin\theta,\label{eq:jdispdipole}
\end{equation}
and
\begin{equation}
 j_\phi=\frac{A_0 a^2}{4\pi r^3}\frac{2 a^2 -3 a^2\eta^2-\eta^4+2
  \eta^6}{(1-\eta^2)^\frac{3}{2}}\sin\theta. \label{eq:jphidipole}
\end{equation}
The current $j_\mathrm{rot}$ is always positive, while $j_\mathrm{disp}$
has negative values for a larger $a$ in $0 \lid \eta \lid a$.
The displacement current $j_\mathrm{disp}$ cannot be ignored for a
larger $a$ and it reduces the
azimuthal current $j_\phi$. Thus the current $j_\phi$
changes its sign for a larger $a$. 
Remember that the pressure $P$ is determined by the $\theta$ component
of the equation of motion given by
\begin{equation}
 \left(-\nabla p+\bmath j \times \bmath B+\rho_e \bmath E\right)_\theta=0. \label{eq:thetaeom}
\end{equation}
According to the definition of $P_A$ and $P_Q$, the poloidal component of
equation (\ref{eq:thetaeom}) is given by
\begin{equation}
 \frac{1}{t^4 r}\frac{\partial P_A}{\partial \theta}=j_\phi B_r.
\end{equation}
Since both $\partial P_A/\partial \theta$ and $j_\phi B_r$ depend on
$\theta$ by $\sin\theta \cos\theta$, and $B_r/\cos \theta$ is positive,
the sign of $P_A$ is determined by that of $j_\phi/\sin \theta$. Thus
$P_A$ can be
negative when the displacement current $j_\mathrm{disp}$
dominates the current $j_\mathrm{rot}$.
The condition that $j_\phi \lid 0$ coincides with the condition that
$P_A \lid 0$ (i.e. $a \gid a_*$, where $a_*$ is given by equation
(\ref{eq:astar})).

\begin{figure}
 \begin{center}
   \includegraphics[width=8cm]{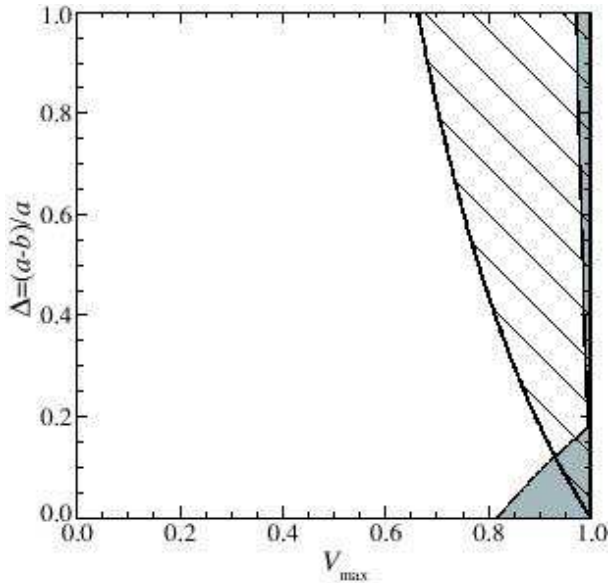}
 \end{center}
 \caption{The diagram showing where the solution $P_A=0$ exists in the
 parameter space $V_\mathrm{max}$ and  $\Delta=(a-b)/a$ when
 $k=\pi/[4(a-b)]$ for the shell solution (the shaded area) and for the
 flux the rope solution (grey area).} \label{fig:diagram}
\end{figure}

Next let us consider the shell and flux rope solutions. Since these solutions
are more complex,
the equation $P_A=0$ is solved numerically. Instead of using parameters
$a$ and $b$, we introduce the following parameters,
\begin{equation}
 V_\mathrm{max}=a, \label{def:Vmax}
\end{equation}
\begin{equation}
 \Delta=\frac{a-b}{a}.\label{def:Delta}
\end{equation}
These parameters denote the maximum speed of the expanding loops and
thickness of the shell behind the loop top, respectively.
Fig.~\ref{fig:diagram} is a diagram showing whether the solution that
$P_A=0$ exists for the shell and flux rope solutions in the parameter space of
$V_\mathrm{max}$ and $\Delta$ for $k=\pi/[4(a-b)]$. Solutions where
$P_A=0$ exist in the shaded area for the shell solution and in the grey
area for the flux rope solution.
Similarly to the dipolar solution, the effect of the displacement current is
more prominent for a larger $V_\mathrm{max}$. Generally, $P_A$ is smaller
for a larger $V_\mathrm{max}$ and thicker shells in parallel that 
the displacement current becomes important for larger $V_\mathrm{max}$
and $\Delta$.
For the flux rope solution, the displacement current is
important not only in region II but in region I (right bottom region in
Fig.~\ref{fig:diagram}). In this case, $P_A$ is negative for $b\gid
0.817$.

\subsection{Application to SGR Explosions}\label{estimate}
SGR flares can be triggered by energy injection into  magnetic loops at
the surface of a strongly magnetized neutron star \cite[e.g.,][]{2006MNRAS.367.1594L}.
When sufficiently large energy is injected, the magnetic loops will
become dynamically unstable, and expand relativistically. Magnetic
energy release in the expanding magnetic loops can be the origin of SGR
flares. The expanding magnetic loops will also produce magnetosonic
waves propagating ahead of the loops. High energy particles can be
produced in the magnetic reconnection inside the loops, and in shock
fronts formed ahead of the loops. 

In this paper, we did not solve the structure of the region ahead of the
magnetic loops ($r > R(t)$). When the outer region is a vacuum,
electromagnetic waves will be emitted from the boundary at
$r=R(t)$. When the plasma density is much larger than the
Goldreich-Julian density \citep{1969ApJ...157..869G}
and the wave frequency is much smaller than the plasma frequency, 
the outer plasma can be studied by using MHD equations. It will be our
future work to connect the self-similar solutions inside $r=R(t)$ and
the solutions in $r\gid R(t)$.

Now let us estimate the energy for the SGR explosion based on the
self-similar solutions. 
Let us take the field strength to be $10^{15}$ Gauss
(\citealt{1998Natur.393..235K}; \citealt{2002ApJ...574L..51I};
\citealt{2003ApJ...584L..17I}) at the stellar radius $R_s=10^6$ cm.
This leads to 
\begin{equation}
 \frac{2A_0 a^2}{R_s^2}=10^{15}~\rmn{Gauss}.
\end{equation}
By assuming that the self-similar expansion begins when $t_0=R_s/a$, the
released energy from the expanding magnetic loops can be estimated from
equation (\ref{en:energy}) as
\begin{equation}
\mathcal{E}'=\left\{\begin{array}{lll}
{\displaystyle 6\times 10^{46} ~\rmn{erg}}, &
    (\mathrm{dipolar \ solution}),\\
{\displaystyle 2\times 10^{47}\Delta^{-2} ~\rmn{erg}}, &
    (\mathrm{shell\ solution}),\\
{\displaystyle 8\times 10^{47}\Delta^{-2}~\rmn{erg}}, & 
    (\mathrm{flux \ rope \ solution}).\label{sol1:estenergy}
			   \end{array}\right.
\end{equation}
Here we take $a=V_\mathrm{max}\simeq
0.9$ and $k=\pi/[4(a-b)]$. These results agree with
the observed energy of SGR giant flares
(\citealt{2005Natur.434.1098H}; \citealt{2005Natur.434.1107P};
\citealt{2005Natur.434.1110T}).
Note that the total energy contained in the expanding magnetic loops is
more energetic for thinner shells.
The non-kinetic part of the total energy $\mathcal{E'}$ is inversely
proportional to the square of the shell thickness.
When some fraction of the kinetic energy $K$ is converted to the
electromagnetic energy, the released energy can be
larger than that estimated by equation (\ref{sol1:estenergy}).
\section{Summary \& Discussions}\label{summary}
By extending the self-similar solutions derived by
\citet{1982ApJ...261..351L}, we
derived self-similar solutions of relativistically
expanding magnetic loops taking into account the toroidal magnetic fields. 
The dipolar solution derived in $\S$ \ref{sss1} gives us an insight into
the relativistic expansion of the magnetic loops because of its
simplicity. However, the shell and flux rope solutions derived in $\S$
\ref{sss2}, \ref{sss3} have more physically interesting properties such
as an enhanced magnetic pressure at the shells and flux rope
structures. Such configurations might be more probable for SGR flares.

The equations of motion in the self-similar stage are similar to those
of the static equilibrium state except the existence of the relativistic
thermal inertial term and the electric field. This fact allows us to evaluate
the non-kinetic part of the total energy in the magnetic loops by using
the virial theorem.
The magnetic loops with shell or flux rope structures carry more energy
than the simple dipole solution. 
The energy is comparable to the observed energy of the SGR giant flares. 

In relativistically expanding magnetic loops, the effect of the
displacement current becomes important. In dipolar solution, the displacement
current becomes larger than the real current $\nabla \times \bmath B/(4\pi)$
for faster expansion speed ($V_\mathrm{max}>a_*$). This effect reduces the
toroidal current and weakens the magnetic tension force. To balance the
reduced magnetic tension force, the pressure decreases.

We found that the energy is transferred to $r>R(t)$ in
dipolar solutions. In the shell and flux rope solutions, the energy 
is transferred to $r>R(t)$ unless condition (\ref{sol2:knowave}) is satisfied.
The condition can be interpreted as that for the total reflection of the
MHD waves in the shell. Dipolar solutions always have leakage (transmission of
Poynting flux to the region $r\gid R(t)$) because $B_\theta \ne 0$ at
$r=R(t)$.
The shell and flux rope solutions have perfectly reflecting solutions in
which the total energy in $r<R(t)$ is conserved. It means that the
solutions are energy eigenstates of the system. The eigenstates can be
obtained by adjusting the parameter $k$.

In this paper, we obtained solutions for freely expanding magnetic
loops, i.e., $Dv / Dt = 0$. We assumed that the magnetic loops have
sufficiently large energy to drive the expansion. When the flux function
$\tilde{A}$ increases with time, the toroidal magnetic fields
will also increase with time. The toroidal magnetic fields will then
affect the dynamics through the magnetic pressure. Such solutions can
describe the accelerating magnetic loops.

Magnetic fields can be expressed as the sum of the Fourier
modes in the polar angle. The modes and their amplitudes should be
determined at the boundary where
the magnetic twist is injected on the surface of the star. 
It is not shown but we can construct more complex solutions that the
poloidal magnetic fields are expressed by the sum of the Fourier modes,
i.e., $\tilde A \propto \sin^n \theta$. In actual
explosion, the opening angle of the expanding magnetic loops depends on
the location at which the magnetic twist is injected on the surface of
the central star. Such a solution may be expressed as the sum of the
Fourier modes for the poloidal and toroidal magnetic fields.
We should note that SGR flares are not necessarily
axisymmetric. Models including the non-axisymmetrically expanding
magnetic loops will be a subject of future works.

\section*{Acknowledgments}
We are grateful to the anonymous referee for constructive comments
improving the paper.
Fruitful discussions with Tomoyuki Hanawa, Akira Mizuta, and
Tomohisa Kawashima at Chiba University are greatly appreciated. This
work was supported by the
Grants-in-Aid for Scientific Research of Ministry of Education, Culture,
Sports, Science, and Technology (RM:20340040).

\appendix
\section[]{Construction of the Shell Solutions}\label{ap-shell}
The functions $Q$ and $P$ in region I are given by
\begin{equation}
 Q^I(\eta,\theta)=\sum_{n}\frac{Q_{0,n}}{1-\eta^2}\sin^n\theta,\label{sol2:Q1}
\end{equation}
\begin{equation}
 P^I(\eta,\theta)=P_0(\eta)+P^I_A(\eta,\theta)+P^I_Q(\eta,\theta),\label{sol2:P1}
\end{equation}
where $P_A^I$ and $P_Q^I$ are given by
\begin{equation}
 P^I_A(\eta,\theta)=\frac{A_0^2a^4}{2\pi\eta^4}\sin^2\theta,\label{sol2:PA1}
\end{equation}
\begin{equation}
 P^I_Q(\eta,\theta)=\left\{\begin{array}{ll}
		  {\displaystyle -\sum_{m+n\ne 2}\frac{n
		   Q_{0,m} Q_{0,n}}{ 4\pi (m+n-2)\eta^2 (1-\eta^2)}
		   \sin^{m+ n-2}}  \theta, & \rmn{for\ \  }m+n\ne2,\\
			{\displaystyle -\sum_{m+n=2}\frac{n
			 Q_{0,m}Q_{0,m}}{4\pi \eta^2 (1-\eta^2)}
			 \log(\sin\theta)}, & \rmn{for\ \  }m+n=2.\\
			\end{array}\right.  \label{sol2:PQ1}
\end{equation}
The function $D$ in region I is given by
\begin{equation}
 D^I(\eta,\theta)=D_0(\eta)+D^I_A(\eta,\theta)+D^I_Q(\eta,\theta),
  \label{sol2:D1}
\end{equation}
where $D^I_A$ and $D^I_Q$ are given by
\begin{equation}
 D^I_A(\eta,\theta)=\frac{2A_0^2a^4}{\pi G M \eta^3 (1-\eta^2)}\sin^2\theta,\label{sol2:DA1}
\end{equation}
\begin{equation}
 D^I_Q(\eta,\theta)=\left\{\begin{array}{ll}
		  {\displaystyle -\sum_{m+n\ne 2}\frac{n
		   Q_{0,m}Q_{0,n}}{2\pi G M (m+n-2)\eta
		   (1-\eta^2)^{2}}\sin^{m+n-2}\theta}, & \rmn{for\ \  }m+n\ne2,\\
		  {\displaystyle
		   -\sum_{m+n=2}\frac{nQ_{0,m}Q_{0,m}}{2\pi G M \eta
		   (1-\eta^2)^2}\log(\sin\theta)},& \rmn{for\ \  }m+n=2.
			\end{array}\right.\label{sol2:DQ1}
\end{equation}

The functions $Q$ and $P$ in region II are given by
\begin{equation}
 Q^{II}(\eta,\theta)=\sum_{n}\frac{Q_{0,n}}{1-\eta^2}
  \Lambda^{\frac{n}{2}}(\eta)\sin^n\theta,\label{sol2:Q2}
\end{equation}
\begin{equation}
 P^{II}(\eta,\theta)=P_0(\eta)+P^{II}_A(\eta,\theta)+P^{II}_Q(\eta,\theta),\label{sol2:P2}
\end{equation}
where $P_A^{II}$ and $P_Q^{II}$ are given by
\begin{equation}
P^{II}_A(\eta,\theta)=\frac{A_0^2a^4}{2\pi\eta^4}\Lambda(\eta)
\left\{1-\frac{\sin^2 T(\eta)}{\sin^4T(a)}\left[\sin^2T(\eta)
   +4k\eta^3\sin T(\eta)\cos T(\eta)
   -2k^2\eta^2(1-\eta^2)\left(3-4\sin^2 T(\eta)\right)\right]\right\}
\sin^2\theta,\label{sol2:PA2}
\end{equation}
\begin{equation}
P^{II}_Q(\eta,\theta)=\left\{\begin{array}{ll}
		 {\displaystyle -\sum_{m+n\ne 2}\frac{n
		  Q_{0,m}Q_{0,n}\Lambda^\frac{(m+n)}{2}(\eta)}{4 \pi (m + n-2) \eta^2 (1-\eta^2)}
		   \sin^{m+n-2}\theta}, & \rmn{for\ \  }m+n\ne2,\\
			 {\displaystyle-\sum_{m+n=2}\frac{n Q_{0,m}Q_{0,n}\Lambda(\eta)}{4 \pi
			 \eta^2 (1-\eta^2)} \log(\sin\theta)},& \rmn{for\ \  }m+n=2.\\
		       \end{array}\right.\label{sol2:PQ2}
\end{equation}
The function $D^{II}$ is given by
\begin{equation}
 D^{II}(\eta,\theta)=D_0(\eta)+D^{II}_A(\eta,\theta)+D^{II}_Q(\eta,\theta),\label{sol2:D2}
\end{equation}
where $D^{II}_A$ and $D^{II}_Q$ are given by
\begin{equation}
 D^{II}_A(\eta,\theta)=\frac{2A_0^2a^4}{\pi G M \eta^3
  (1-\eta^2)}\Lambda(\eta)\left[1-\frac{\sin^4T(\eta)}{\sin^4T(a)}\Xi(\eta)\right]\sin^2\theta,\label{sol2:DA2}
\end{equation}
\begin{equation}
 D^{II}_Q(\eta,\theta)=\left\{\begin{array}{ll}
		 {\displaystyle -\sum_{m+n\ne2}\frac{n Q_{0,m}
		  Q_{0,n}\Lambda^{\frac{m+n-2}{2}}(\eta)}{2
		  \pi G M (m+n-2) \eta (1-\eta^2)^2}
		   \left\{ 1-\frac{\sin^4
			   T(\eta)}{\sin^4
			   T(a)}\left[1-2k\eta(1-\eta^2) \cot
				 T(\eta)\right]\right\}}
			   \sin^{m+n-2}\theta,\\
			       \hspace{12cm} \rmn{for\ \  }m+n\ne2,\\
			  {\displaystyle -\sum_{m+n=2}\frac{n Q_{0,m} Q_{0,n}}{2 \pi G
			   M \eta (1 -\eta^2)^2}\left\{
							   \log(\sin
							   \theta)-\frac{\sin^4
							   T(\eta)}{\sin^4
							   T(a)}\left[\log(\sin
								 \theta)+k
								\eta
								(1-\eta^2)
								\cot
								T(\eta)
								(1-2\log
								(\sin\theta))\right]\right\}
			   },\\
			       \hspace{12cm} \rmn{for\ \  }m+n=2.\\
			 \end{array}\right.\label{sol2:DQ2}
\end{equation}
Here $\Xi(\eta)$ is a function of $\eta$ given by
\begin{equation}
 \Xi(\eta)=1 -k \eta (1+\eta^2) (1-3\eta^2) \cot T(\eta) 
	   + k^2 \eta^2 (1-\eta^2)(1+3\eta^2) \left[ 1-3\cot^2
					       T(\eta)\right]
	   +k^3 \eta^3 (1 - \eta^2)^2 \left[3\cot^3T(\eta) - 5\cot
				       T(\eta)\right].\label{sol2:Xi}
\end{equation}
$P_0(\eta)$ is an arbitrary function of $\eta$ and is related to the
function $D_0(\eta)$ through equation (\ref{sol1:D0}).

\section[]{Construction of the Flux Rope Solutions}\label{ap-fluxrope}
The function $P$ in region I is given by
\begin{equation}
 P^I(\eta,\theta)=P_0(\eta)+P_A^I(\eta,\theta)+P_Q^I(\eta,\theta),
  \label{sol3:P1}
\end{equation}
where $P_A^{I}$ and $P_Q^I$ are 
\begin{equation}
 P_A^I(\eta,\theta)=\frac{A_0^2
  a^4}{4\pi\eta^4}\frac{2-3\eta^2}{(1-\eta^2)^2}
  \sin^2\theta,\label{sol3:PA}
\end{equation}
\begin{equation}
P^{I}_Q(\eta,\theta)=\left\{\begin{array}{ll}
{\displaystyle -\sum_{m+n\ne 2}\frac{n Q_{0,m}Q_{0,n}}{4\pi (m+n-2)}
 \frac{\sin^{m+n-2}\theta} {\eta^2(1-\eta^2)^{1+(m+n)/4}}}, 
 & \rmn{for\ \  }m+n\ne2,\\
 {\displaystyle-\sum_{m+n=2}\frac{n Q_{0,m}Q_{0,n}}{4
  \pi}\frac{\log(\sin\theta)}{\eta^2 (1-\eta^2)^\frac{3}{2}}}, 
  & \rmn{for\ \  }m+n=2.\\
			     \end{array}\right.\label{sol3:PQ1}
\end{equation}
The function $D$ in region I is given by
\begin{equation}
D^{I}(\eta,\theta)=D_0(\eta)+D_A^{I}(\eta,\theta)+D_Q^{I}(\eta,\theta),\label{sol3:D1}
\end{equation}
where $D^{I}_A$ and $D^{I}_Q$ are
\begin{equation}
 D^{I}_A(\eta,\theta)=\frac{A_0^2 a^4}{4\pi G M \eta^3}
  \frac{8-12\eta^2+3\eta^4}{(1-\eta^2)^3}\sin^2\theta,\label{sol3:DA}
\end{equation}
\begin{equation}
 D^{I}_Q(\eta,\theta)=\left\{\begin{array}{ll}
{\displaystyle -\sum_{m+n\ne 2}\frac{n Q_{0,m}Q_{0,n}}{4\pi G M (m+n-2)}
 \frac{2-\eta^2}{\eta(1-\eta^2)^{2+(m+n)/4}}  \sin^{m+n-2}\theta}, 
 & \rmn{for\ \  }m+n\ne2,\\
 {\displaystyle-\sum_{m+n=2}\frac{n Q_{0,m}Q_{0,n}}{8\pi G M}
  \frac{\eta^2+2(2-\eta^2)\log(\sin\theta)}{\eta (1-\eta^2)^\frac{5}{2}}}, 
  & \rmn{for\ \  }m+n=2.\\
			     \end{array}\right.\label{sol3:DQ1}
\end{equation}

The function $P$ in region II is described as
\begin{equation}
P^{II}(\eta,\theta)=P_0(\eta)+P_A^{II}(\eta,\theta)+P_Q^{II}(\eta,\theta),\label{sol3:P2}
\end{equation}
where $P_A^{II}$ and $P_Q^{II}$ are 
\begin{equation}
 P_A^{II}=\frac{A_0^2 a^4}{4\pi
  \eta^4}\Lambda(\eta)\left[\frac{2-3\eta^2}{(1-\eta^2)^2}\Lambda(\eta)
		      +4k^2\eta^2\frac{\sin^2T(\eta)(3-4\sin^2T(\eta))}{\sin^4
		      T(a)}\right]\sin^2\theta, \label{sol3:PA2}
\end{equation}
\begin{equation}
P^{II}_Q(\eta,\theta)=\left\{\begin{array}{ll}
{\displaystyle -\sum_{m+n\ne 2}\frac{n Q_{0,m}Q_{0,n}}{4\pi (m+n-2)}
 \frac{\Lambda^{\frac{m+n}{2}}(\eta)}{\eta^2(1-\eta^2)^{1+(m+n)/4}}
 \sin^{m+n-2}\theta}, 
 & \rmn{for\ \  }m+n\ne2,\\
 {\displaystyle-\sum_{m+n=2}\frac{n Q_{0,m}Q_{0,n}}{4\pi}
  \frac{\Lambda(\eta)}{\eta^2 (1-\eta^2)^\frac{3}{2}} \log(\sin\theta)}, 
  & \rmn{for\ \  }m+n=2.\\
			     \end{array}\right.\label{sol3:PQ2}
\end{equation}
The function $D$ is given by
\begin{equation}
D^{II}(\eta,\theta)=D_0(\eta)+D_A^{II}(\eta,\theta)+D_Q^{II}(\eta,\theta),\label{sol3:D2}
\end{equation}
where $D^{II}_A$ and $D_Q^{II}$ are
\begin{equation}
 D^{II}_A(\eta,\theta)=\frac{A_0^2 a^4}{4\pi G M \eta^3}\Lambda(\eta)\left[d_0(\eta)+d_1(\eta)+d_2(\eta)+d_3(\eta)\right]\sin^2\theta,\label{sol3:DA2}
\end{equation}
\begin{equation}
 D^{II}_Q(\eta,\theta)=-\sum_{m,n}\frac{n Q_{0,m}Q_{0,m}}{4\pi G M
  \eta(1-\eta^2)^{2+\frac{m+n}{4}}} Y(\eta,\theta).\label{sol3:DQ2}
\end{equation}
The functions $d_0$, $d_1$, $d_2$, $d_3$, and $Y$ are given by
\begin{equation}
 d_0(\eta)=\frac{3\eta^4-12 \eta^2+8}{(1-\eta^2)^3}\Lambda (\eta),\label{sol3:d0}
\end{equation}
\begin{equation}
 d_1(\eta)=\frac{4k\eta (2-3\eta^2)}{(1-\eta^2)^2}\frac{\sin^3
  T(\eta)\cos T(\eta)}{\sin^4 T(a)},\label{sol3:d1}
\end{equation}
\begin{equation}
 d_2(\eta)=\frac{4k^2\eta^2(2+3\eta^2)}{1-\eta^2}\frac{\sin^2
  T(\eta)(3-4\sin^2  T(\eta))}{\sin^4 T(a)},\label{sol3:d2}
\end{equation}
\begin{equation}
 d_3(\eta)=-8k^3\eta^3 \frac{\sin T(\eta)\cos T(\eta)(3-8\sin^2T(\eta))}{\sin^4T(a)},\label{sol3:d3}
\end{equation}
\begin{equation}
Y(\eta,\theta)=\left\{\begin{array}{ll}
{\displaystyle \frac{\Lambda^{\frac{m+n-2}{2}}(\eta)}{m+n-2}
 \left[(2-\eta^2)\Lambda(\eta)+4k \eta (1-\eta^2) \frac{\sin^3 T(\eta) \cos
  T(\eta)}{\sin^4 T(a)}\right]\sin^{m+n-2}\theta}, 
 & \rmn{for\ \  }m+n\ne2,\\
 {\displaystyle \frac{1}{2}\left[\eta^2+2(2-\eta^2)\log
		 (\sin\theta)\right]\Lambda(\eta)
			+2k\eta
			(1-\eta^2)(-1+2\log(\sin\theta))\frac{\sin^3
			T(\eta) \cos T(\eta)}{\sin^4 T(a)}}, 
  & \rmn{for\ \  }m+n=2.\\
			     \end{array}\right.\label{sol3:Y}
\end{equation}
The functions $P_0$ and $D_0$ are related each other through equation
(\ref{sol1:D0}).

\section[]{The Virial Theorem of the self-similar Relativistic MHD}\label{ap1}
The virial theorem in non-relativistic MHD was derived by
\citet{1953ApJ...118..116C}.
\citet{1982ApJ...261..351L} applied it to the expanding magnetic loops
by evaluating the surface term. 
\cite{1975ctf..book.....L} derived the theorem for a relativistic case
in elegant way by
integrating the energy
momentum tensor. In this appendix, we derive the virial
theorem for a relativistic self-similar MHD.

We start from the equations of motion in self-similar stage given by
(\ref{eq:reducedeom}). Taking the inner product with 
$\bmath r$ and integrating it within a volume $V$, we obtain
\begin{equation}
 \int d V \left[\frac{\Gamma}{\Gamma-1}\frac{p\gamma^2v^2(\bmath r\cdot \bmath
  e_r)}{r}-\bmath r \cdot \nabla p+\bmath r\cdot \left(\rho_e\bmath
						  E+\bmath j \times
						  \bmath B\right)-\frac{GM\gamma
  \rho}{r^2}(\bmath r\cdot \bmath e_r)\right]=0. \label{ap:eomint}
\end{equation}
The first term is the thermal inertial term $U_\mathrm{in}$ and the
forth term is the gravitational potential energy $W$.
Integrating the second term by parts, we obtain
\begin{equation}
 \int d V \bmath r\cdot \nabla p=-3(\Gamma-1)U_{\mathrm{th}}+\int p\bmath r\cdot
  d \bmath{\mathcal{A}}
\end{equation}
where $\bmath{\mathcal{A}}$ is a closed surface of the volume $V$. 
The third term can be rewritten by using the Maxwell equations as follows, 
\begin{equation}
 \int_V dV \left[\bmath r\cdot \left(\rho_e\bmath E+\bmath j\times \bmath
		 B\right)\right]=-\int d V
 \frac{\partial}{\partial t}(\bmath r \cdot \bmath S)-\int d V \bmath r
 \cdot (\nabla \cdot \bmath \sigma), \label{ap:emint}
\end{equation}
where $\bmath S$ and $\bmath \sigma$ are the Poynting flux and the
Maxwell's stress tensor, respectively.
Note that the time derivative cannot be exchanged with the integration
with volume in the first term since the volume $V$ changes with time.

The second term on the right hand side of equation (\ref{ap:emint}) has a form
\begin{equation}
 -\int d V \bmath r\cdot (\nabla\cdot\bmath
  \sigma)=U_{\mathrm{E}}+U_{\mathrm{M}}+\frac{1}{8\pi}\int\left\{2\left[(\bmath r\cdot \bmath
					    E)(\bmath E\cdot
					    d\bmath{\mathcal{A}})+(\bmath r\cdot
					    \bmath
					    B)(\bmath B\cdot
					    d\bmath{\mathcal{A}})\right]-(\bmath
  E^2+\bmath B^2)(\bmath r\cdot d\bmath{\mathcal{A}})\right\},
\end{equation}
where $U_\mathrm{E}$ and $U_\mathrm{M}$ are the electric and magnetic
energies given by equations (\ref{en:ele}) and (\ref{en:mag}), respectively.
By using these results, we obtain the virial theorem for the
relativistic self-similar MHD;
\begin{equation}
 3(\Gamma-1)U_{\mathrm{th}}+U_{\mathrm{in}}+U_{\mathrm{M}}+U_{\mathrm{E}}+W
  =\mathcal{H}+\mathcal{S},\label{ap:virial}
\end{equation}
where
\begin{equation}
\mathcal{H}=\int p \bmath r \cdot d \bmath{\mathcal{A}}
 -\frac{1}{8\pi}\int
 \left\{2\left[(\bmath r \cdot \bmath E)(\bmath E\cdot
	 d \bmath{\mathcal{A}})+(\bmath r \cdot \bmath B)(\bmath B\cdot
	 d \bmath{\mathcal{A}})
\right]-(\bmath E^2+\bmath B^2)(\bmath r\cdot d \bmath{\mathcal{A}})\right\},\label{ap:surface}
\end{equation}
and
\begin{equation}
 \mathcal{S}=\int dV \frac{\partial }{\partial t}
  \left(\bmath r \cdot \frac{\bmath E\times \bmath B}{4\pi}\right).\label{ap:poynting}
\end{equation}
Here, $K$ is the kinetic energy given by
equation (\ref{en:kinetic}).

Readers may wonder why the thermal inertial term appears. Actually,
the kinetic, thermal, and thermal inertial energies should
not be considered separately because they depend on the frame of
reference. Even so, we used this definition through the paper to make clear
the difference between the non-relativistic and relativistic
expansions. We can easily figure out that the total plasma energy
can be described as the sum of these energies, as
\begin{equation}
 K+U_\mathrm{th}+U_\mathrm{in}=\int dV \left[\left(\rho+\frac{\Gamma}{\Gamma-1}p\right)\gamma^2-p\right].
\end{equation}

\label{lastpage}

\begin{thebibliography}{27}
\expandafter\ifx\csname natexlab\endcsname\relax\def\natexlab#1{#1}\fi

\bibitem[{{Asano}(2007)}]{2007Chiba...1..1}
{Asano} E., 2007, PhD. thesis, Chiba Univ.

\bibitem[{{Asano} {et~al.}(2005){Asano}, {Uchida}, \&
  {Matsumoto}}]{2005PASJ...57..409A}
{Asano} E., {Uchida} T., {Matsumoto} R., 2005, \pasj, 57, 409

\bibitem[{{Chandrasekhar} \& {Fermi}(1953)}]{1953ApJ...118..116C}
{Chandrasekhar} S., {Fermi} E., 1953, \apj, 118, 116

\bibitem[{{Goldreich} \& {Julian}(1969)}]{1969ApJ...157..869G}
{Goldreich} P., {Julian} W.~H., 1969, \apj, 157, 869

\bibitem[{{Hurley} {et~al.}(2005){Hurley}, {Boggs}, {Smith}, {Duncan}, {Lin},
  {Zoglauer}, {Krucker}, {Hurford}, {Hudson}, {Wigger}, {Hajdas}, {Thompson},
  {Mitrofanov}, {Sanin}, {Boynton}, {Fellows}, {von Kienlin}, {Lichti}, {Rau},
  \& {Cline}}]{2005Natur.434.1098H}
{Hurley} K., {Boggs} S.~E., {Smith} D.~M., {Duncan} R.~C., {Lin} R., {Zoglauer}
  A., {Krucker} S., {Hurford} G., {Hudson} H., {Wigger} C., {Hajdas} W.,
  {Thompson} C., {Mitrofanov} I., {Sanin} A., {Boynton} W., {Fellows} C., {von
  Kienlin} A., {Lichti} G., {Rau} A., {Cline} T., 2005, \nat, 434, 1098

\bibitem[{{Ibrahim} {et~al.}(2002){Ibrahim}, {Safi-Harb}, {Swank}, {Parke},
  {Zane}, \& {Turolla}}]{2002ApJ...574L..51I}
{Ibrahim} A.~I., {Safi-Harb} S., {Swank} J.~H., {Parke} W., {Zane} S.,
  {Turolla} R., 2002, \apjl, 574, L51

\bibitem[{{Ibrahim} {et~al.}(2003){Ibrahim}, {Swank}, \&
  {Parke}}]{2003ApJ...584L..17I}
{Ibrahim} A.~I., {Swank} J.~H., {Parke} W., 2003, \apjl, 584, L17

\bibitem[{{Komissarov}(2002)}]{2002MNRAS.336..759K}
{Komissarov} S.~S., 2002, \mnras, 336, 759

\bibitem[{{Komissarov}(2006)}]{2006MNRAS.367...19K}
---, 2006, \mnras, 367, 19

\bibitem[{{Kouveliotou} {et~al.}(1998){Kouveliotou}, {Dieters}, {Strohmayer},
  {van Paradijs}, {Fishman}, {Meegan}, {Hurley}, {Kommers}, {Smith}, {Frail},
  \& {Murakami}}]{1998Natur.393..235K}
{Kouveliotou} C., {Dieters} S., {Strohmayer} T., {van Paradijs} J., {Fishman}
  G.~J., {Meegan} C.~A., {Hurley} K., {Kommers} J., {Smith} I., {Frail} D.,
  {Murakami} T., 1998, \nat, 393, 235

\bibitem[{{Landau} \& {Lifshitz}(1975)}]{1975ctf..book.....L}
{Landau} L.~D., {Lifshitz} E.~M., 1975, {The classical theory of fields}.
  4th edition: Volume 2 (Course of theoretical physics), p.~90,
		Betterworth-Heinemann
\bibitem[{{Low}(1982)}]{1982ApJ...261..351L}
{Low} B.~C., 1982, \apj, 261, 351

\bibitem[{{Low}(1984)}]{1984ApJ...281..392L}
---, 1984, \apj, 281, 392

\bibitem[{{Lyutikov}(2002)}]{2002PhFl...14..963L}
{Lyutikov} M., 2002, Physics of Fluids, 14, 963

\bibitem[{{Lyutikov}(2006)}]{2006MNRAS.367.1594L}
---, 2006, \mnras, 367, 1594

\bibitem[{{Lyutikov} \& {Blandford}(2003)}]{2003astro.ph.12347L}
{Lyutikov} M., {Blandford} R., 2003, astro-ph/0312347

\bibitem[{{Mereghetti}(2008)}]{2008A&ARv..15..225M}
{Mereghetti} S., 2008, \aapr, 15, 225

\bibitem[{{Palmer} {et~al.}(2005){Palmer}, {Barthelmy}, {Gehrels}, {Kippen},
  {Cayton}, {Kouveliotou}, {Eichler}, {Wijers}, {Woods}, {Granot}, {Lyubarsky},
  {Ramirez-Ruiz}, {Barbier}, {Chester}, {Cummings}, {Fenimore}, {Finger},
  {Gaensler}, {Hullinger}, {Krimm}, {Markwardt}, {Nousek}, {Parsons}, {Patel},
  {Sakamoto}, {Sato}, {Suzuki}, \& {Tueller}}]{2005Natur.434.1107P}
{Palmer} D.~M., {Barthelmy} S., {Gehrels} N., {Kippen} R.~M., {Cayton} T.,
  {Kouveliotou} C., {Eichler} D., {Wijers} R.~A.~M.~J., {Woods} P.~M., {Granot}
  J., {Lyubarsky} Y.~E., {Ramirez-Ruiz} E., {Barbier} L., {Chester} M.,
  {Cummings} J., {Fenimore} E.~E., {Finger} M.~H., {Gaensler} B.~M.,
  {Hullinger} D., {Krimm} H., {Markwardt} C.~B., {Nousek} J.~A., {Parsons} A.,
  {Patel} S., {Sakamoto} T., {Sato} G., {Suzuki} M., {Tueller} J., 2005, \nat,
  434, 1107

\bibitem[{{Prendergast}(2005)}]{2005MNRAS.359..725P}
{Prendergast} K.~H., 2005, \mnras, 359, 725

\bibitem[{{Spitkovsky}(2005)}]{2005KITP...CONF..HP}
{Spitkovsky} A., 2005, in KITP Program: Physics of Astrophysical
		outflows and Accretion disks
		(http://online.kitp.ucsb.edu/online/)

\bibitem[{{Spitkovsky}(2006)}]{2006ApJ...648L..51S}
---, 2006, \apjl, 648, L51

\bibitem[{{Stone} {et~al.}(1992){Stone}, {Hawley}, {Evans}, \&
  {Norman}}]{1992ApJ...388..415S}
{Stone} J.~M., {Hawley} J.~F., {Evans} C.~R., {Norman} M.~L., 1992, \apj, 388,
  415

\bibitem[{{Terasawa} {et~al.}(2005){Terasawa}, {Tanaka}, {Takei}, {Kawai},
  {Yoshida}, {Nomoto}, {Yoshikawa}, {Saito}, {Kasaba}, {Takashima}, {Mukai},
  {Noda}, {Murakami}, {Watanabe}, {Muraki}, {Yokoyama}, \&
  {Hoshino}}]{2005Natur.434.1110T}
{Terasawa} T., {Tanaka} Y.~T., {Takei} Y., {Kawai} N., {Yoshida} A., {Nomoto}
  K., {Yoshikawa} I., {Saito} Y., {Kasaba} Y., {Takashima} T., {Mukai} T.,
  {Noda} H., {Murakami} T., {Watanabe} K., {Muraki} Y., {Yokoyama} T.,
  {Hoshino} M., 2005, \nat, 434, 1110

\bibitem[{{Thompson} \& {Duncan}(2001)}]{2001ApJ...561..980T}
{Thompson} C., {Duncan} R.~C., 2001, \apj, 561, 980

\bibitem[{{Uchida}(1997)}]{1997PhRvE..56.2181U}
{Uchida} T., 1997, \pre, 56, 2181

\bibitem[{{Woods} {et~al.}(2001){Woods}, {Kouveliotou}, {G{\"o}{\u g}{\"u}{\c
  s}}, {Finger}, {Swank}, {Smith}, {Hurley}, \&
  {Thompson}}]{2001ApJ...552..748W}
{Woods} P.~M., {Kouveliotou} C., {G{\"o}{\u g}{\"u}{\c s}} E., {Finger} M.~H.,
  {Swank} J., {Smith} D.~A., {Hurley} K., {Thompson} C., 2001, \apj, 552, 748

\bibitem[{{Woods} \& {Thompson}(2006)}]{2006csxs.book..547W}
{Woods} P.~M., {Thompson} C., 2006, in Lewin W., van der Klis M., eds,
  Cambridge Astrophys. Ser. Vol. 39, p.~547, Compact stellar X-ray sources, Cambridge
  Univ. Press, Cambridge

\end{thebibliography}
\end{document}